\documentclass[12pt]{article}
\pdfoutput=1    %for arxiv
\usepackage{arxiv}

\usepackage{amsmath, amsfonts, amssymb}
\usepackage{amsthm}%proofs.

\usepackage[mathlines]{lineno}% Enable numbering of text and display math
\usepackage{pdflscape}
\usepackage{array}
\usepackage{rotating}

\usepackage{hyperref}
\usepackage{graphicx}
\usepackage{subcaption}
\usepackage{verbatim}
\usepackage{array}
\usepackage{xcolor}

\newtheorem{theorem}{Theorem}[section]
\newtheorem{proposition}{Proposition}[section]
\newtheorem{remark}{Remark}[section]
\newtheorem*{conjecture*}{Conjecture}

\title{Water flow model on vegetated hillslopes with erosion}
\date{July 09, 2026}

\usepackage{authblk}

\author[1]{Stelian Ion\thanks{\texttt{ro\_diff@yahoo.com}}}
\author[1]{Dorin Marinescu \thanks{\texttt{marinescu.dorin@ismma.ro}}}
\author[1,2]{Stefan-Gicu Cruceanu\thanks{\texttt{stefan.cruceanu@ismma.ro}}}
\affil[1]{``Gheorghe Mihoc-Caius Iacob'' Institute of Mathematical Statistics and Applied Mathematics of the ROMANIAN ACADEMY, Calea 13 Septembrie No. 13, PO Box 1-24, 050711 Bucharest, Romania}
\affil[2]{Corresponding author}

\hypersetup
{
  pdftitle={Water flow model on vegetated hillslopes with erosion},
  pdfsubject={Physics - Fluid Dynamics},
  pdfauthor={Stelian Ion, Dorin Marinescu, Stefan-Gicu Cruceanu},
  pdfkeywords={extended Saint-Venant model, porosity, numerical scheme, hydrographic basin, sediment transport, suspension and sedimentation}
}

\begin{document}
%==========================================================
%==========================================================

\maketitle

\begin{abstract}
  The water circulation in the {\bf S}oil-{\bf P}lant-{\bf
    A}tmosphere continuum and particularly the soil erosion
  induced by water are problems of main concern in the new
  era of climate change.  The present paper aims to provide
  a mathematical tool to investigate the water-soil and
  water-plant interactions involved in the complex process
  of water flow on plant-covered soil surfaces.  Basically,
  the mathematical model consists of an extended
  Saint-Venant system of equations for water flow coupled
  with Hairsine-Rose equations for soil erosion.  The
  classical Saint-Venant model is thus modified in order to
  take into account the presence of plants on the soil
  surface.

  A numerical approximation of the solution of our model is
  built using a Finite Volume Method for the discretization
  in space and a fractional time-step scheme to discretize
  the time variable and resulting time derivatives.  Several
  properties of the scheme with physical relevance are also
  discussed and investigated.

  In order to validate both the model and the numerical
  method, and to see if essence of the reality is adequately
  reflected, a series of qualitative and quantitative tests
  are performed.  Given that the mathematical model is
  flexible enough to reflect the variability of the
  environmental variables such as soil structure, soil
  surface roughness, or plant cover structure, each
  numerical experiment is constructed as an image of a
  target hydrological context.  The dam break problem, flash
  floods, water-induced soil erosion in a catchment basin
  are all subjects of numerical analysis.  It is shown that
  the presence of the plant cover drastically modifies the
  water dynamics and the distribution of the soil eroded
  particles and one can quantitatively evaluate such
  effects.  The methods described in the paper can also help
  one to manage the environmental resources in order to
  avoid the water induced disasters.
\end{abstract}



%{\bfseries \emph{ACM:}} G.1.8; G.4
%==========================================================
%==========================================================

%==========================================================
%==========================================================
\section{Introduction}
\label{section_intro}
\begin{figure}[!htbp]
  \centering
  \includegraphics[width=0.8\textwidth,height=5.5cm]{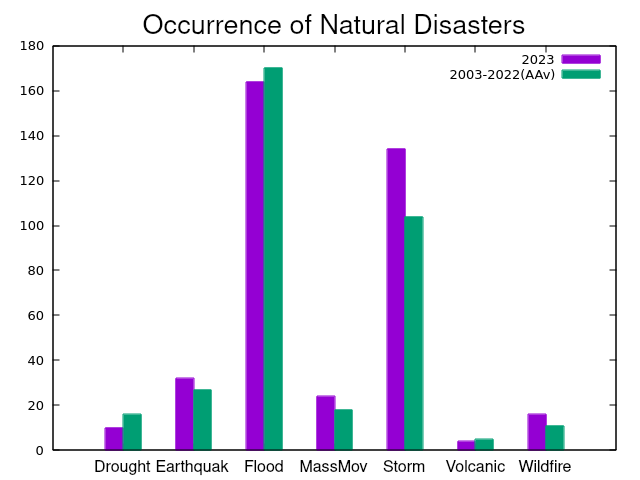}
  \caption{Number of water-related and non-water-related
    disasters in the world.  Source: EM-DAT.}
  \label{fig_disasters}
\end{figure}

Besides being one of the most basic and important
necessities for plants, animals, and human beings, water is
an essential element for life on Earth, covering about 71\%
of its surface.  The continuous circulation of water in the
Earth-Atmosphere system, the frequency and the intensity of
the storms were strongly affected by the increased climate
change over the last years with widespread effects on the
entire environment.  Knowing also that over 90\% of world's
disasters are weather-related (including floods, pollution,
wildfire, aridification)
\cite{unep_disasters_and_climate_change}, it is important
for all of us to understand the hydrological cycles, to keep
water in balance on earth, to proper manage the water
resources in order to protect our life and material
well-being.  Fig.~\ref{fig_disasters} illustrates a report
regarding the water-related and non-water-related disasters
in OECD countries \cite{oecd_book}.

Problems with water distribution can be easily raised due to
the nowadays increasing existence of dry or flooded areas.
The dams, these barriers across the water flow, are most of
the time made by humans with great purposes: to suppress
floods, to provide water for human consumption, irrigation
or navigability, to generate electricity, etc.  The
reservoirs behind these ``walls'' contain dangerous forces
and although a dam failure is rare, it can be catastrophic,
with heavy destructions on the environment as well as on the
civilian population (e.g. $1975$ Banqiao and Shimantan Dams,
$1979$ Machhu II Dam, $1985$ Val di Stava Dam failures); the
reader is referred to \cite{RCEM} for a comprehensive list
of Dam Failures around the world.  In this sense, having a
mathematical system to model the hydrodynamic processes in a
hydrographic basin and/or in a Dam Break Problem is
important and needed, can help estimating the risks
associated to various hazards and can provide important
information regarding an economical water usage.

Before describing the model, let us say a few words about
mathematical modeling and the difficulties we encounter
along this process.  A very brief answer to the question
``What is mathematical modeling?'' can be formulated as
follows \cite{HainesCrouch, VerschaffelGreerDeCorte}: {\it
  mathematical modeling is a cyclic process in which
  real-life problems are translated into mathematical
  language, solved within a symbolic system, and the
  solutions tested on real-world data}.  Thus, the
mathematical model is a result of the application of general
principles of natural sciences, logic and mathematics.
Modeling hydrological processes is a very challenging and
difficult task.  One can imagine for a moment the complexity
of the phenomena given by the various involved watershed
processes (e.g. runoff, soil erosion, precipitation,
evaporation, infiltration, root uptake, plant transpiration,
etc), the multitude of the factors that are involved in,
only a few being quantifiable.  We can add the environmental
heterogeneity (diversity of vegetation and soil, variations
in altitude, curvature of the soil surface etc) and it
becomes obvious that we cannot take everything into account
into one single model.  Moreover, one can use different
mathematical models for the same physical problem, depending
on
\begin{itemize}
\item the complexity of the physical phenomena,
\item the complexity of the mathematical description,
\item the possibility of solving the problem (analytic,
  numerical),
\end{itemize}
but almost every time one must deal with the {\bf modeller's
  dilemma}:
\begin{itemize}
\item if the mathematical description is too complicated,
  then we might end up with a mathematical system with no
  solution;
\item if the mathematical description is too simple, then we
  may end up not catching the essential or different aspects
  of the phenomenon.
\end{itemize}
Therefore, it would be ideal to use a mathematical model as
simple as possible that is as suitable as possible for the
physical phenomena, i.e. we do not want to kill the physical
problem for the sake of mathematics.  {\it The best models
  are the ones which give results closest to reality and at
  the same time require least number of parameters and model
  complexity}, \cite{wheater_book}.  We emphasize that the
following minimal requirements should always hold for any
mathematical model:

{\centering {\it solvable}\hspace{1cm} and \hspace{1cm}{\it
    physically relevant}.\par}

The studies for models covering various hydrological
applications as rainfall-runoff, flood and desertification,
Dam Break Problem, echo or agricultural hydrology, etc
continue to enrich the research literature.  Regardless of
their classification (e.g. metric, conceptual and
physics-based in \cite{wheater_book}; empirical, conceptual
or physical in \cite{epa_rainfall_runoff}), a model can cast
into one or more classes, depending on its task, scale or
structure.  We mention SHE \cite{abbott_SHE}, MIKE-SHE
\cite{mike_she}, KINEROS \cite{woolhiser_kineros}, VIC
\cite{liang_VIC}, PRMS \cite{singh_book}, SWASHES
\cite{delestre} among the most known physical models in
hydrology.

A general physical model describing the fluid motion is
given by the Navier-Stokes equations - a system of partial
differential equations very difficult to treat theoretically
and for which numerical approaches require huge
computational effort.  A simplified (derived from
Navier-Stokes) and widely used version for modelling surface
water flow is given by Saint-Venant equations
\cite{SaintVenant_theorie}.  These shallow-water equations
can be slightly modified by introducing porosity in order to
include the presence of the plant cover or the building
distribution in urban flood studies \cite{sds_apnum,
  cozolino_urban_flood, GUINOT2017133}.

It was observed that the water depth increases and its
velocity decreases on a given surface soil in the presence
of plants.  The Saint-Venant model with porosity described
later in this paper confirms these observations and thus it
is feasible to use it for study the hydrological process in
the presence of vegetation.  Before Saint-Venant with
vegetation, the previous models had considered the plants by
modifying Manning's coefficient \cite{manning,
  manning_strickler}.  Unlike these, the Saint-Venant model
with vegetation highlights the effects given by the
variability of the water depth and its velocity in the mass
and momentum balance equations with respect to the
vegetation heterogeneity.

It is known that the numerical approach to solve the porous
shallow-water equations must be carefully chosen such that a
good balance between the computational effort and the
accuracy of the obtained solution, between the precision of
the measured data and the numerical accuracy is created.
Moreover, the volume of processed data and the computing
effort given by high-order schemes increases excessively
when working at hydrographic basin scales.  In this context,
we will present here a mathematical model and a discrete
numerical scheme characterized by low computational
complexity and reduced memory requirements.  This study
extends the work in \cite{sds_apnum} by incorporating
erosion and sedimentation processes into the modeling
process.  The resulting model and its associated discrete
scheme generalize those introduced in \cite{sds_apnum},
which also form the foundation of the \textsc{ASTERIX}
software for simulating water flow over vegetated surfaces.
We should also mention that the model and our open-source
software can be used from a laboratory or teaching level up
to a hydrographic basin scale for various applications
related to studying water dynamics, as

$\succ$ propagation of floods produced by torrential rains,

$\succ$ propagation of the flood caused by a dam break,

$\succ$ studies on Riemann Problems,

$\succ$ studies on the influence of plants on surface
runoffs, estimating their role in the continuum
{\bf S}oil-{\bf P}lant-{\bf A}tmosphere,

$\succ$ studies on landscape water flows,

$\succ$ studies on vegetated bed rivers, etc.

This paper is organized as follows.  In
Section~\ref{section_math_model} we present the
partial-differential-equation (PDE) system we use for
modelling the water flow, erosion and sedimentation
processes, while Section~\ref{section_num_approx} is
dedicated to the numerical scheme built to approximate the
solution of this model.
Section~\ref{section_stationary_sol_sed} presents the exact
stationary solution found for a particular configuration of
the soil surface and vegetation density distribution.  The
convergence of the numerical solution in this case and a
stability result are formulated and proven.  We are able to
find some analytic solutions for the case of a single-class
sediment in Section~\ref{section_analytic_sols} and discuss
several mathematical properties of them.  Some numerical
applications for validating both the model and the numerical
method are considered in Section~\ref{section_num_app}.  One
test is performed to compare the numerical results with the
data obtained in the laboratory, while other tests
(qualitative ones) are performed to emphasize the effects of
vegetation, erosion and deposition processes on the
asymptotic behavior of dynamical systems defined by the
numerical scheme.
Section~\ref{section_conclusions_and_remarks} presents final
remarks and conclusions.
%==========================================================
%==========================================================

%==========================================================
%==========================================================
\section{Mathematical model}
\label{section_math_model}
The model we shortly present here for water flowing over a
soil surface is based on balance equations, closure
empirical relations and some simplifying assumptions that do
not alter the essence of the phenomena.  This model couples
an extended version of Saint-Venant equations (given by a
mass balance and two momentum balance equations) with the
Hairsine-Rose model for soil erosion and takes into account
the presence of the plants on the soil surface,
\cite{sds_apnum, hairrose}.

Assuming that the soil surface is modeled by the altitude
function $z$ defined on a bounded domain
$\mathfrak{D} \subset \mathbb{R}^2$
\begin{equation*}
  x^3 = z(\boldsymbol{x}), \quad \boldsymbol{x} = (x^1, x^2) \in \mathfrak{D},
\end{equation*}
that the sediment is partitioned into $M$ size classes, and
using Einstein summation notation, our model reads as
\begin{align}
  \partial_t(\theta h)+\partial_b(\theta h v^b) 
  &=\mathfrak{M},
    \label{eq_swe1}\\
  \partial_t(\theta hv^a)+\partial_b(\theta hv^av^b)+\theta h\partial_a{w} 
  &=\mathfrak{\tau}_v^a+\mathfrak{\tau}_s^a,
    \label{eq_swe2}
\end{align}
\begin{align}
  \partial_t(\theta h\rho_\alpha) + \partial_b(\theta \rho_\alpha h v^b)
  &=\theta( e_\alpha + e^r_\alpha - d_\alpha),
    \label{eq_eros1}\\
  \partial_t m_\alpha 
  &= \theta (d_\alpha- e^r_\alpha),
    \label{eq_eros2}
\end{align}
for $a={1,2}$ and $\alpha=\overline{1,M}$, where 

$\succ$ the water depth $h(t,\boldsymbol{x})$,

$\succ$ the two components $v^a(t,\boldsymbol{x})$ of the
water velocity $\boldsymbol{v}=(v^1,v^2)$,

$\succ$ the mass density (mass per unit volume)
$\rho_\alpha(t,\boldsymbol{x})$ of the suspended sediment of
the size class $\alpha$,

$\succ$ the mass density (mass per unit area)
$m_\alpha(t,\boldsymbol{x})$ of the deposited sediment of
the size class $\alpha$

\noindent are the unknown variables.  The potential $w$ of
the water level is given by
\begin{equation}
  w(t,\boldsymbol{x}) = g \left( z(\boldsymbol{x}) + h(t,\boldsymbol{x}) \right),
  \label{eq_potential_w}
\end{equation}
where $g$ denotes the gravitational acceleration and $z+h$
is the free water surface level.  The flow is influenced by
the presence of vegetation which is quantified by the
porosity function $\theta:\mathfrak{D}\rightarrow [0,1]$
defined as the volume of empty space among plant stems
(volume which can be filled by with water) present in a unit
volume.  With this notation, a bare soil is represented by
\begin{equation}
  \theta (\boldsymbol{x}) = 1, \quad \forall \boldsymbol{x} \in \mathfrak{D},
  \label{eq_theta_1}
\end{equation}
while a complete sealant plant cover is represented by
\begin{equation}
  \theta (\boldsymbol{x}) = 0, \quad \forall \boldsymbol{x} \in \mathfrak{D}.
  \label{eq_theta_0}
\end{equation}
The right hand side terms $\mathfrak{M}$,
$\mathfrak{\tau}_v^a$ and $\mathfrak{\tau}_s^a$ represent
the rate of water production (due to rain gain and
infiltration loss) and the rates of momentum production (due
to the plant cover resistance and the fluid-soil friction),
respectively.  The erosion and sedimentation processes are
taken into account through

$\circ$ $d_\alpha$ - deposition rate of the suspended
sediment from the size class $\alpha$,

$\circ$ $e_\alpha$ - entrainment rate of the size class
$\alpha$ sediment from the soil, and

$\circ$ $e^r_\alpha$ - re-entrainment rate of the size class
$\alpha$ deposited sediment, respectively.

The partial differential equations (\ref{eq_swe1} -
\ref{eq_eros2}) of our model are quite general and we need
some empirical relations in order to quantify the flow
resistance due to plant and soil frictions, as well as the
erosion and deposition rates.

The resistance opposed by plants to the water flow
\cite{baptist, nepf} and the water-soil frictional forces
\cite{rouse} are quantified by
\begin{equation}
  \mathfrak{\tau}^a_p = -\alpha_p h \left(1-\theta\right)\|\boldsymbol{v}\|v^a,\quad\quad
  \mathfrak{\tau}^a_s = -\theta \alpha_s (h) \|\boldsymbol{v}\|v^a,
  \label{eq_tau_p_s}
\end{equation}
respectively, where $\alpha_p$ and $\alpha_s$ are material
parameters.  The non-negative coefficient $\alpha_p$ depends
on the geometry of the plants from the vegetation cover,
while the non-negative function $\alpha_s(h)$ depends on the
given soil surface.  Some of the most used formulas for
$\alpha_s$ in the literature come from the experimental
relations of Manning, Ch{\'e}zy, or the Darcy-Weisbach.  The
Darcy-Weisbach expression
\begin{equation}
  \mathfrak{\tau}^a_s = -\theta \alpha_s \|\boldsymbol{v}\|v^a \quad\quad \rm{(Darcy-Weisbach)}
  \label{eq_tsa_darcy}
\end{equation}
has the advantage of being non-singular if the water depth
becomes zero.

The cumulative effects of the water-soil and water-plants
interactions is assumed to be additive in this model, and
we can thus introduce the resistance term
\begin{equation}
  \mathfrak{\tau}_v^a+\mathfrak{\tau}_s^a := -\mathcal{K}(h,\theta)\|\boldsymbol{v}\|v^a,
  \label{eq_tau_suma}
\end{equation}
where
\begin{equation}
  \mathcal{K}(h,\theta) = \alpha_p h \left(1-\theta\right) + \theta \alpha_s (h)
  \label{eq_K_h_theta}
\end{equation}
is the coefficient function of the frictional force of the
water-soil-plant system.

For erosion and sedimentation processes, things are more
complicated because the most of the existing data in the
literature relate specifically to flows into channels or
laboratory experiments.  Flows on sloping surfaces have
properties which are different from the ones of flows into
channels, and therefore, one must work with caution when
extrapolating data from one case to another.  In this
article, we will use the following set of empirical
relations \cite{hairrose,kim,sander} based on the ``power
stream'' concept:
\begin{equation}
  \begin{array}{l}
    d_\alpha = \nu_{s, \alpha}{\rho_\alpha},\\
    e_\alpha = p_\alpha(1-H) 
    \displaystyle\frac{F\left(\Omega-\Omega_{cr}\right)_{+}}{J},\\
    e^r_\alpha = \displaystyle H\frac{m_\alpha}{m_t}
    \frac{\gamma_s}{\gamma_s-1}\frac{F\left(\Omega-\Omega_{cr}\right)_{+}}{ gh},
  \end{array}
  \label{eq_da_ea_era}
\end{equation}
where $p_\alpha$ is the proportion of the sediment in size
class $\alpha$ in the original soil, 
\begin{equation*}
  0 < p_\alpha \leq 1, \quad\quad \displaystyle\sum_{\alpha=1}^{M}{p_\alpha} = 1,
\end{equation*}
$\nu_{s, \alpha}$ is the settling (falling) velocity of the
sediment in the size class $\alpha$, and $\gamma_s$ is the
relative density (with respect to water) of the sediment.
The parameters $F$ - effective fraction of power stream, $J$
- energy of soil particles detachment and $\Omega_{cr}$ -
critical power stream are specific to a given type of soil.
The erosion processes are controlled by the water flow
through the stream power $\Omega$ for which we will use the
law
\begin{equation}
  \Omega = \theta\rho_{\rm w}\|\boldsymbol{\mathfrak{\tau}}_s\|\|\boldsymbol{v}\|.
  \label{eq_Omega}
\end{equation}
The function
\begin{equation}
  H = \displaystyle\min\left\{\frac{m_t}{m_t^{\star}},1\right\}
  \label{eq_fcnH}
\end{equation}
plays the role of a protecting factor of the original soil
to the erosion process.  The terms
\begin{equation*}
  m_t=\displaystyle\sum_{a=1}^M m_a
\end{equation*}
and $m^{\star}_t$ from (\ref{eq_fcnH}) are the total mass of
sediment deposited on the soil and the mass required to
protect the original soil from erosion, respectively.
%==========================================================
%==========================================================

%==========================================================
%==========================================================
\section{Numerical approximation}
\label{section_num_approx}
In what follows, we will describe a numerical scheme to
approximate the solution of our model
(\ref{eq_swe1}-\ref{eq_eros2}).  This scheme casts into the
general class of methods of lines.

There are several ways to discretize a hyperbolic system of
partial differential equations.  One of them is to generate
an ODE system using a discretization method for the spatial
derivative operator (method of line), which is then
integrated.  One gets
\begin{equation}
  \partial_t U + \mathcal{F}(U) = \mathcal{R}(U),
  \label{eq_ode_U_FU_RU}
\end{equation}
where $U$ is the vector of the unknowns, $\mathcal{F}$ is
the ``flux'' term generated by the derivative of the spatial
variable, and $\mathcal{R}$ is the ``source'' term.

To obtain the numerical scheme that approximates the
equations (\ref{eq_swe1}-\ref{eq_eros2}), we apply the
Finite Volume Method (FVM) to discretize the spatial
variable and then we introduce a fractional time step method
to integrate the ODE system.

%----------------------------------------------------------
%----------------------------------------------------------
\subsection{FVM approximation}
\label{subsection_fvm_approx}
Let $\{\omega_i\}_{i=\overline{1,N}}$ to be an admissible
polygonal partition \cite{veque} of $\mathfrak{D}$,
\begin{equation}
  \mathfrak{D}=\displaystyle\bigcup\limits_{i=1}^N \omega_i,
\end{equation}
with $\sigma^i$ being the area of the cell $\omega_i$.  One
builds a spatial discrete approximation of the model by
integrating the continuous equations on each finite volume
$\omega_i$
%\begin{widetext}
  \begin{equation}
    \begin{split}
      \displaystyle\partial_t\int\limits_{\omega_i}\theta h{\rm d}x+ \int\limits_{\partial\omega_i}\theta h \boldsymbol{v}\cdot\boldsymbol{n}{\rm d}s
      &=\displaystyle\int\limits_{\omega_i}\mathfrak{M}{\rm d}x,\\
      \displaystyle\partial_t\int\limits_{\omega_i}\theta h v^a{\rm d}x+ \int\limits_{\partial\omega_i}\theta h v^a\boldsymbol{v}\cdot\boldsymbol{n}{\rm d}s+ \int\limits_{\omega_i}\theta h\partial_a w{\rm d}x
      &=\displaystyle-\int\limits_{\omega_i}\mathcal{K}\|\boldsymbol{v}\|v^a{\rm d}x, \quad a=1,2,
    \end{split}
    \label{eq_swe_fvm}
  \end{equation}
  \begin{equation}
    \begin{split}
      \displaystyle\partial_t\int\limits_{\omega_i}\theta h\rho_\alpha{\rm d}x+ \int\limits_{\partial\omega_i}\theta \rho_\alpha h \boldsymbol{v}\cdot\boldsymbol{n}{\rm d}s
      &=\displaystyle\int\limits_{\omega_i}\theta (e_\alpha+e^r_\alpha-d_\alpha){\rm d}x,\\
      \displaystyle\partial_t\int\limits_{\omega_i}m_\alpha{\rm d}x
      &=\displaystyle\int\limits_{\omega_i}\theta (d_\alpha-e^r_\alpha){\rm d}x, \quad \alpha=\overline{1,M},
    \end{split}
    \label{eq_eros_fvm}
  \end{equation}
%\end{widetext}
and then by defining an approximations of the integrals.
Here, $\boldsymbol{n}=(n_1,n_2)^T$ stands for the unit
normal vector pointing towards the outside of the boundary
$\partial\omega_i$ of $\omega_i$.  

Denote by $h^i$, $v_a^i$, $\rho_{\alpha}^i$, $m_{\alpha}^i$,
with $\alpha=\overline{1,M}$, $a=1,2$ the average of the
unknown variables $h$, $v^a$, $\rho_{\alpha}$, $m_{\alpha}$
on the cell $\omega_i$, respectively, for
$i=\overline{1,N}$.  Let also $\theta^i$ be the average
value of the parameter $\theta$ on $\omega_i$.  In order to
later write a compact form of the system
(\ref{eq_swe_fvm}-\ref{eq_eros_fvm}), we introduce the
following arrays:
\begin{equation}
  \begin{split}
    {\boldsymbol h} &:= \left[ h^1, \ldots, h^N \right];\\
    {\boldsymbol v} &:= \left[ v^i_a \right]_{\overset{i=\overline{1,N}}{a=1,2}}, \quad 
    {\boldsymbol v}^i = (v^i_1,v^i_2);\\
    {\boldsymbol m} &:= \left[ m^i_\alpha \right]_{\overset{i=\overline{1,N}}{\alpha=\overline{1,M}}},
    \quad {\boldsymbol m}^i := (m_1^i, \ldots, m_M^i); \\
    {\boldsymbol \rho} &:= \left[ \rho^i_\alpha \right]_{\overset{i=\overline{1,N}}{\alpha=\overline{1,M}}}, 
    \quad {\boldsymbol \rho}_\alpha := (\rho_\alpha^1, \ldots, \rho_\alpha^N).
  \end{split}
  \label{eq_h_v_m_rho_arrays}
\end{equation}

Similar as in \cite{sds_apnum}, introducing the
approximations
\begin{equation}
  \int\limits_{\omega_i}\theta h{\rm d}x \approx \sigma^i\theta^i h^i, \quad
  \int\limits_{\partial\omega_i}\theta h \boldsymbol{v}\cdot\boldsymbol{n}{\rm d}s 
  \approx \sigma^i \mathcal{A}^i(\boldsymbol{h},\boldsymbol{v}), \quad
  \int\limits_{\omega_i}\mathfrak{M}{\rm d}x \approx \sigma^i \mathcal{M}^i,
  \label{eq_approx_swe1}
\end{equation}
\begin{equation}
  \int\limits_{\omega_i}\theta h v_a {\rm d}x \approx \sigma^i\theta^i h^i v_a^i, \quad
  \int\limits_{\partial\omega_i}\theta h v_a \boldsymbol{v}\cdot\boldsymbol{n}{\rm d}s +
  \int\limits_{\omega_i}\theta h\partial_a w{\rm d}x
  \approx \sigma^i \mathcal{B}_a^i(\boldsymbol{h},\boldsymbol{v}), \quad
  \int\limits_{\omega_i}\mathcal{K}|\boldsymbol{v}|v_a{\rm d}x 
  \approx -\sigma^i \mathcal{C}_a^i(\boldsymbol{h},\boldsymbol{v}),
  \label{eq_approx_swe2}
\end{equation}
the ODE model (\ref{eq_swe_fvm}) can be compactly written as
\begin{equation}
  \left \{
    \begin{array}{cccccl}
      \partial_t (\theta^i h^i) & + & \mathcal{A}^i & = & \mathcal{M}^i, & \\
      \partial_t (\theta^i h^i v_a^i) & + & \mathcal{B}_a^i & = & \mathcal{C}_a^i, & \quad a={1,2},
    \end{array}
  \right .
  \label{eq_swe_compact}
\end{equation}
where
\begin{equation}
  \begin{split}
    \mathcal{A}^i &:= \frac{1}{\sigma ^i} \sum_{j\in N(i)}l^{ij} \Psi^{ij} v_{\rm norm}^{ij},\\
    \mathcal{B}_a^i &:= \frac{1}{\sigma ^i}\sum_{j\in N(i)}l^{ij} \Gamma_a^{ij}
    +\frac{1}{\sigma ^i} \sum_{j\in N(i)}l^{ij} (w^{ij}-w^i) \Lambda^{ij} n_a^{ij}, \quad 
    a={1,2},\\
    \mathcal{C}_a^i &:= -\mathcal{K}^i |\boldsymbol{v}^i|v_a^i, \quad a={1,2}.
  \end{split}
  \label{eq_AiBaiCai}
\end{equation}
Here, $N(i)$ denotes the set of all cell-indices $j$ for
which the cell $\omega_j$ has a common side $(i,j)$ with
$\omega_i$, $l^{ij}$ is the length of this common side, and
$\boldsymbol{n}^{ij}$ is the unitary normal to this
interface pointing towards $\omega_j$.  The quantities found
in (\ref{eq_AiBaiCai}) are defined as
\begin{equation*}
  \begin{array}{l}
    \displaystyle 
    \Psi^{ij} = \Psi^{ij}(\boldsymbol{h},\boldsymbol{v}) :=
    \left\{
    \begin{array}{l}
      \theta^ih^i,
      \quad {\rm if}\quad
      v_{\rm norm}^{ij}(\boldsymbol{v})\geq
      0,\\
      \theta^jh^j,
      \quad {\rm if}\quad
      v_{\rm norm}^{ij}(\boldsymbol{v}) < 0,
    \end{array}
    \right.\\
    \displaystyle 
    \boldsymbol{v}^{ij} = (v_1^{ij},v_2^{ij}) := 
    \frac{\boldsymbol{v}^i + \boldsymbol{v}^j}{2},\\
    \displaystyle 
    v_{\rm norm}^{ij} = \boldsymbol{v}^{ij}\cdot 
    \boldsymbol{n}^{ij},\\
    \Gamma^{ij}=\Gamma_a^{ij}(\boldsymbol{h},\boldsymbol{v}):=
    \Psi^{ij}(\boldsymbol{h},\boldsymbol{v})\,
    v_a^{ij}v_{\rm norm}^{ij}-
    \nu^{ij} \triangle v_a^{ij},\qquad a={1,2},\\
    (\triangle \boldsymbol{v})^{ij} = 
    ((\triangle v_1)^{ij}, (\triangle v_2)^{ij}):=
    \boldsymbol{v}^j- \boldsymbol{v}^i,\\
    \displaystyle 
    \nu^{ij} := \max\{c^i, c^j\}\frac{2\theta^i h^i \theta^j h^j}{\theta^i h^i + \theta^j h^j}, 
    \qquad {\rm where} \qquad 
    c^i :=|\boldsymbol{v}^i|+\sqrt{gh^i},\\
    w^i = w^i(\boldsymbol{h}):=h^i+z^i,\\
    \displaystyle \Lambda ^{ij} = \Lambda ^{ij}(\boldsymbol{h},\boldsymbol{v}) :=
    \left\{
    \begin{array}{l}
      \Psi ^{ij}(\boldsymbol{h},\boldsymbol{v}),
      \quad {\rm if}\quad
      v_{\rm norm}^{ij}(\boldsymbol{v}) \neq 0,\\
      \theta^i \,
      h^i,
      \quad {\rm if}\quad
      v_{\rm norm}^{ij}(\boldsymbol{v}) = 0,
      \quad w^i \geq w^j,\\
      \theta^j \, 
      h^j,
      \quad {\rm if}\quad
      v_{\rm norm}^{ij}(\boldsymbol{v}) = 0,
      \quad w^i<w^j,
    \end{array}
    \right.\\
    w^{ij} = w^{ij}(\boldsymbol{h}):=\displaystyle\frac{w^i+w^j}{2}.
  \end{array}
\end{equation*}

Similarly, the ODE model (\ref{eq_eros_fvm}) for the erosion
process can be compactly written as
\begin{equation}
  \left\{
    \begin{array}{cccccccl}
      \partial_t (\theta^i h^i \rho_\alpha^i) & + & \mathcal{P}_\alpha^i
      & = & \mathcal{Q}_\alpha^i & + & \mathcal{R}_\alpha^i, & \\
      \partial_t m_\alpha^i & + & 0
      & = & -\mathcal{Q}_\alpha^i & + & 0, & \quad \alpha =\overline{1,M},\\
    \end{array}
  \right.
  \label{eq_eros_compact}
\end{equation}
where
\begin{equation}
  \mathcal{P}_\alpha^i := \frac{1}{\sigma ^i}
  \sum_{j\in N(i)}l^{ij} \Phi_{\alpha}^{ij} v_{\rm norm}^{ij}, \quad
  \mathcal{Q}_\alpha^i := \theta^i \left[ (e_\alpha^r)^i-d_\alpha^i \right], \quad
  \mathcal{R}_\alpha^i := \theta^i e_\alpha^i,
  \label{eq_PaliQaliRali}
\end{equation}
for all $\alpha = \overline{1,M}$ and with
\begin{equation*}
  \begin{array}{l}
    \Phi_{\alpha}^{ij} =
    \Phi_{\alpha}^{ij}({\boldsymbol \rho}; \boldsymbol{h}, \boldsymbol{v}) :=
    \left\{
    \begin{array}{l}
      \rho_{\alpha}^i \theta^i h^i,\quad {\rm if} \quad v_{\,{\rm norm}}^{ij} \geq 0,\\
      \rho_{\alpha}^j \theta^j h^j,\quad {\rm if} \quad v_{\,{\rm norm}}^{ij} < 0,
    \end{array}
    \right. \\
    \displaystyle e_\alpha^i =
    e_\alpha^i({\boldsymbol m}; \boldsymbol{v}) := 
    p_\alpha[1-H^i({\boldsymbol m})]\frac{F(\Omega^i(\boldsymbol{v})-\Omega_{cr})_{+}}{J},\\
    \displaystyle (e_\alpha^r)^i =
    (e_\alpha^r)^i({\boldsymbol m}; \boldsymbol{h}, \boldsymbol{v}) :=
    H^i({\boldsymbol m}) \frac{m_\alpha^i}{m_t^i}
    \frac{\gamma_s}{\gamma_s-1} \frac{F(\Omega^i(\boldsymbol{v})-\Omega_{cr})_{+}}{g \cdot h^i},\\
    \displaystyle d_\alpha^i = 
    d_\alpha^i({\boldsymbol \rho}) := \nu_{s, \alpha} \rho_\alpha^i, \\
    \displaystyle H^i = H^i({\boldsymbol m}) :=
    \min\left\{\frac{m_t^i}{m_t^{\star}},1\right\} =
    \min\left\{\frac{\sum_{\alpha=1}^Nm_\alpha^i}{m_t^{\star}},1\right\},\\
    \displaystyle \Omega^i = \Omega^i(\boldsymbol{v}) := 
    \theta^i\rho_{w}|\boldsymbol{\mathfrak{\tau}}_s^i||{\boldsymbol{v}}^i| = 
    (\theta^i)^2\rho_{w}\alpha_s\left|{\boldsymbol{v}}^i\right|^3.
  \end{array}
\end{equation*}

Therefore, the spatial discrete approximation system
(\ref{eq_swe_fvm}-\ref{eq_eros_fvm}) on a finite volume
$\omega_i$ reads now as
\begin{equation}
  \frac {{\rm d}\mathcal{U}^i}{{\rm d}t} + \mathcal{F}^i= \mathcal{S}^i + \mathcal{T}^i,
  \label{eq_swe_eros_omega_i}
\end{equation}
where
\begin{equation*}
  \mathcal{U}^i :=
  \left(
    \begin{array}{c}
      \theta^i h^i\\
      \theta^i h^i v_a^i\\
      \theta^i h^i\rho_\alpha ^i\\
      m_\alpha^i
    \end{array}
  \right), \quad
  \mathcal{F}^i :=
  \left(
    \begin{array}{c}   
      \mathcal{A}^i\\
      \mathcal{B}_a^i\\
      \mathcal{P}_\alpha ^i\\
      \boldsymbol{0}_M
    \end{array}
  \right), \quad
  \mathcal{S}^i :=
  \left(
    \begin{array}{c}   
      \mathcal{M}^i\\
      \mathcal{C}_a^i\\
      \mathcal{Q}_\alpha ^i\\
      -\mathcal{Q}_\alpha ^i
    \end{array}
  \right), \quad
  \mathcal{T}^i :=
  \left(
    \begin{array}{c}   
      0\\
      \boldsymbol{0}_2\\
      {\cal R}_\alpha ^i\\
      \boldsymbol{0}_M
    \end{array}
  \right),
\end{equation*}
with $\boldsymbol{0}_M$ representing the column zero vector
of order $M$.  Using (\ref{eq_swe_eros_omega_i}), the
resulting semi-discrete scheme takes the form of a system of
ordinary differential equations (ODEs) written as
\begin{equation}
  \frac{{\rm d}\mathcal{U}}{{\rm d}t} + 
  \mathcal{F}(\mathcal{U}) = \mathcal{S}(\mathcal{U}) + \mathcal{T}(\mathcal{U}),
  \label{eq_ode_compact}
\end{equation}
where $\mathcal{U}$, $\mathcal{F}$, $\mathcal{S}$,
$\mathcal{T}$ are vectors of order $N(3+2M)$ with components
$\mathcal{U}^i$, $\mathcal{F}^i$, $\mathcal{S}^i$,
$\mathcal{T}^i$, respectively.  In order to find the
solution of (\ref{eq_ode_compact}), we should observe that
we can firstly solve the water dynamics and then deal with
the sediment one.

For water, one has to solve
\begin{equation}
  \frac{{\rm d}\mathcal{U}_w^i}{{\rm d}t} + \mathcal{F}_w^i = \mathcal{S}_w^i,
  \label{eq_ode_water_omegai}
\end{equation}
where
\begin{equation*}
  \mathcal{U}_w^i :=
  \left(
    \begin{array}{c}
      \theta^i h^i\\
      \theta^i h^i v_a^i
    \end{array}
  \right), \quad
  \mathcal{F}_w^i :=
  \left(
    \begin{array}{c}   
      \mathcal{A}^i\\
      \mathcal{B}_a^i
    \end{array}
  \right), \quad
  \mathcal{S}_w^i :=
  \left(
    \begin{array}{c}   
      \mathcal{M}^i\\
      \mathcal{C}_a^i
    \end{array}
  \right),
\end{equation*}
for all $i=\overline{1,N}$, a system which can be compactly
written as
\begin{equation}
  \frac {{\rm d}\mathcal{U}_w}{{\rm d}t} + \mathcal{F}_w(\mathcal{U}_w)= \mathcal{S}_w(\mathcal{U}_w).
  \label{eq_ode_water_compact}
\end{equation}

For sediment, one has to solve
\begin{equation}
  \frac{{\rm d}\mathcal{U}_s^i}{{\rm d}t} + \mathcal{F}_s^i= \mathcal{S}_s^i + \mathcal{T}_s^i,
  \label{eq_ode_sediment_omegai}
\end{equation}
where
\begin{equation*}
  \mathcal{U}_s^i :=
  \left(
    \begin{array}{c}
      \theta^i h^i\rho_\alpha ^i\\
      m_\alpha^i
    \end{array}
  \right), \quad
  \mathcal{F}_s^i :=
  \left(
    \begin{array}{c}
      \mathcal{P}_\alpha ^i\\
      \boldsymbol{0}_M
    \end{array}
  \right), \quad
  \mathcal{S}_s^i :=
  \left(
    \begin{array}{c}
      \mathcal{Q}_\alpha ^i\\
      -\mathcal{Q}_\alpha ^i
    \end{array}
  \right), \quad
  \mathcal{T}_s^i :=
  \left(
    \begin{array}{c}   
      \mathcal{R}_\alpha ^i\\
      \boldsymbol{0}_M
    \end{array}
  \right),
\end{equation*}
for all $i=\overline{1,N}$, a system which can be compactly
written as
\begin{equation}
  \frac{{\rm d}\mathcal{U}_s}{{\rm d}t} + \mathcal{F}_s(\mathcal{U}_s) =
  \mathcal{S}_s(\mathcal{U}_s) + \mathcal{T}_s(\mathcal{U}_s).
  \label{eq_ode_sediment_compact}
\end{equation}
%----------------------------------------------------------
%----------------------------------------------------------

%----------------------------------------------------------
%----------------------------------------------------------
\subsection{Time integration scheme. Fractional step method}
\label{subsection_time_integration_scheme}
To obtain a numerical solution, we use a fractional time
step method to integrate the ODEs.  Basically, this means
that we split initial ODE system into two subsystems,
integrate each of them separately and then combine the two
solutions to obtain the solution of the original system
\cite{veque-phd, strang}.  For our model, if we denote by
$\mathcal{E}_w^1(t)$ and $\mathcal{E}_w^2(t)$ the evolution
operators for
\begin{equation}
  \frac{{\rm d}\mathcal{U}_w}{{\rm d}t} = \mathcal{S}_w(\mathcal{U}_w) \quad
  {\rm and} \quad
  \frac{{\rm d}\mathcal{U}_w}{{\rm d}t} + \mathcal{F}_w(\mathcal{U}_w) = \boldsymbol{0}_3,
  \label{eq_ode_water-e1_si_e2}
\end{equation}
respectively, and by $\mathcal{E}_s^1(t)$ and $\mathcal{E}_s^2(t)$ the evolution
operators for
\begin{equation}
  \frac{{\rm d}\mathcal{U}_s}{{\rm d}t} = \mathcal{S}_s(\mathcal{U}_s) \quad
  {\rm and} \quad
  \frac{{\rm d}\mathcal{U}_s}{{\rm d}t} + \mathcal{F}_s(\mathcal{U}_s) = \mathcal{T}_s(\mathcal{U}_s),
  \label{eq_ode_sediment-e1_si_e2}
\end{equation}
respectively, then an evolution operator of
(\ref{eq_ode_water_compact}) is
\begin{equation}
  \mathcal{U}_w(t+\triangle t) := 
  \mathcal{E}_w^1(\triangle t/2)\mathcal{E}_w^2(\triangle t) 
  \mathcal{E}_w^1(\triangle t/2)\mathcal{U}_w(t),
\end{equation}
and of (\ref{eq_ode_sediment_compact}) is
\begin{equation}
  \mathcal{U}_s(t+\triangle t) := 
  \mathcal{E}_s^1(\triangle t/2)\mathcal{E}_s^2(\triangle t) 
  \mathcal{E}_s^1(\triangle t/2)\mathcal{U}_s(t).
\end{equation}
Compactly, an approximate evolution operator of
(\ref{eq_ode_compact}) can be written as
\begin{equation}
  \mathcal{U}(t+\triangle t) := 
  \mathcal{E}^1(\triangle t/2)\mathcal{E}^2(\triangle t) 
  \mathcal{E}^1(\triangle t/2)\mathcal{U}(t),
\end{equation}
where
\begin{equation*}
  \mathcal{E}^1 :=
  \left(
    \begin{array}{l}
      \mathcal{E}_w^1\\
      \mathcal{E}_s^1
    \end{array}
  \right), \quad
  \mathcal{E}^2 :=
  \left(
    \begin{array}{l}
      \mathcal{E}_w^2\\
      \mathcal{E}_s^2
    \end{array}
  \right).
\end{equation*}

%----------------------------------------------------------
\bigskip\noindent {\bf The construction of $\mathcal{E}_w^1$
  and $\mathcal{E}_w^2$}

A complete description for the construction these two
operators can be found in \cite{sds_apnum}, and therefore we
will not insist on this step.  Still, it is important to
mention that time step $\triangle t_n :=t^{n+1}-t^n$ must be
bounded by
\begin{equation}
  \tau_n = CFL \displaystyle \frac{\phi_{\rm min}}{c^n_{\rm max}},
  \label{eq_tau_n}
\end{equation}
due to the hyperbolic character of the shallow-water
equations and to the positivity requirement of the water
depth, where $CFL$ is a number between $0$ and $1$ (the
Courant-Friedrichs-Lewy condition) and
\begin{equation}
  c_i = \|\boldsymbol{v}_i\|+\sqrt{gh_i}, \quad 
  c_{\rm max} = \max\limits_i \{c_i\}, \quad 
  \phi_{\rm min} = \min\limits_i
  \left\{
    \displaystyle\frac{\sigma_i}{\sum\limits_{j\in N(i)} l_{(i,j)}}
  \right\} .
  \label{eq_ci_cmax_phimin}
\end{equation}
For the hexagonal network we work with, the positivity
condition of $h^i$ reads as
\begin{equation}
  {\triangle t} \leq \frac{\sigma}{6l\cdot \underset{ij}{\max}\{|v_{\rm norm}^{ij}|\}},
  \label{eq_h_positivity_condition}
\end{equation}
where $l$ and $\sigma$ are the radius and the area of any
hexagonal cell $\omega_i$, respectively.  This previous
condition is satisfied when $\tau_n$ is chosen using the
Courant-Friedrichs-Lewy condition:
\begin{equation}
  \tau_n = CFL \cdot \frac{\theta_{min}}{\theta_{max}} \cdot
  \frac{\sigma}{6l\cdot (||\boldsymbol{v}||_{max}+\sqrt{gh_{max}})}.
  \label{eq_tau_n_hexag}
\end{equation}
%----------------------------------------------------------

%----------------------------------------------------------
\bigskip\noindent {\bf The construction of
  ${\mathcal E}_s^1$ and ${\mathcal E}_s^2$}

For the construction of ${\mathcal E}_s^1$, one must solve
\begin{equation}
  \left\{
    \begin{aligned}
      \partial_t(\theta^i h^i \rho_\alpha^i) & = \mathcal{Q}_\alpha^i\\
      \partial_tm_\alpha^i & = -\mathcal{Q}_\alpha^i
    \end{aligned}
  \right. ,
  \label{ec_sediment_E1s_forma1}
\end{equation}
for all $\alpha =\overline{1,M}$, where
$\mathcal{Q}_\alpha^i$ is given by
\begin{equation*}
  \mathcal{Q}_\alpha^i := \theta^i [ (e_\alpha^r)^i - d_\alpha^i ],
\end{equation*}
with
\begin{equation*}
  \displaystyle (e_\alpha^r)^i := 
  H^i\frac{m_\alpha^i}{m_t^i} \frac{\gamma_s}{\gamma_s-1} \frac{F(\Omega^i(\boldsymbol{v})-\Omega_{cr})_{+}}{gh^i}, \quad 
  H^i := \min\left \{\frac{m_t^i}{m_t^{\star}},1\right \}, \quad
  d_\alpha^i := \nu_{s, \alpha} \rho_\alpha^i .
\end{equation*}
Let us denote
\begin{equation}
  \mathcal{H}^i(\boldsymbol{m}; \boldsymbol{h}, \boldsymbol{v}) :=
  \frac{1}{\chi^i(\boldsymbol{m})}
  \frac{\gamma_s}{\gamma_s-1}
  \frac{F(\Omega^i(\boldsymbol{v})-\Omega_{cr})_{+}}{gh^i},
  \label{eq_H_i}
\end{equation}
with
\begin{equation}
  \chi^i(\boldsymbol{m}) := \frac{m_t^i(\boldsymbol{m})}{H^i(\boldsymbol{m})} = 
  \max \{m_t^i(\boldsymbol{m}), m_t^{\star}\}, \quad
  \Omega^i(\boldsymbol{v}) := \alpha_s\rho_{w}(\theta^i)^2|\boldsymbol{v}^i|^3 .
  \label{eq_chi_i_Omega_i}
\end{equation}

\begin{proposition}
  If $\underline {\mathcal{H}}^i$ is an approximation of
  $\mathcal{H}^i(\boldsymbol{m}; \boldsymbol{h}, \boldsymbol{v})$
  on the time interval $(t,t+\triangle t)$, then the system
  (\ref{ec_sediment_E1s_forma1}) can be analytically solved and its
  solution has an exponential form.
  \label{prop_E1s}
\end{proposition}

\noindent
\begin{proof}
  Let $\underline {\mathcal{H}}^i$ be an approximation of
  $\mathcal{H}^i(\boldsymbol{m}; \boldsymbol{h}, \boldsymbol{v})$
  on the time interval $(t,t+\triangle t)$, e.g.
  \begin{equation}
    \underline{\mathcal{H}}^i \approx
    \mathcal{H}^i (\underline{\boldsymbol{m}}; \underline{\boldsymbol{h}}, \underline{\boldsymbol{v}}),
    \label{eq_H_underline}
  \end{equation}
  with $\underline{\boldsymbol{m}}$,
  $\underline{\boldsymbol{h}}$, $\underline{\boldsymbol{v}}$
  being the values of $\boldsymbol{m}$, $\boldsymbol{h}$,
  $\boldsymbol{v}$ at the moment $t$, respectively.

  Observe that $\mathcal{Q}_\alpha^i$ can be written as
  \begin{equation}
    \mathcal{Q}_\alpha^i \approx \theta^i 
    [\underline {\mathcal{H}}^i m_\alpha ^i - \nu_{s, \alpha} \rho_\alpha ^i]
    \label{eq_Q_alfa_i}
  \end{equation}
  and the system (\ref{ec_sediment_E1s_forma1}) becomes
  \begin{equation}
    \left\{
      \begin{array}{ccr}
        \partial_t(\theta^i h^i \rho_\alpha^i) & = 
        &-\theta^i \nu_{s, \alpha} \rho_\alpha^i + \theta^i \underline{\mathcal{H}}^i m_\alpha^i\\
        \partial_t m_\alpha^i & = 
        & \theta^i \nu_{s, \alpha} \rho_\alpha^i - \theta^i \underline{\mathcal{H}}^i m_\alpha^i
      \end{array}
    \right. .
    \label{ec_sediment_E1s_forma2}
  \end{equation}
  If for any arbitrarily fixed $i$ and $\alpha$ one uses the
  notations
  \begin{equation}
    x := \rho_\alpha^i, \quad 
    y := m_\alpha^i, \quad
    a := \theta^i \nu_{s, \alpha}, \quad
    b := \theta^i \underline{\mathcal{H}}^i, \quad
    c := \theta^i \underline{h^i},
  \end{equation}
  then the system (\ref{ec_sediment_E1s_forma2}) takes the
  form
  \begin{equation}
    \left\{
      \begin{array}{ccrcc}
        \dot{x} & = & \displaystyle -\frac{a}{c}x & + & \displaystyle\frac{b}{c}y\\
        \dot{y} & = & ax & - & by
      \end{array}
    \right. ,
  \end{equation}
  whose solution is
  \begin{equation}
    \left\{
      \begin{array}{l}
        \displaystyle x(s)=\frac{cx_0+y_0}{a+bc}b+\frac{ax_0-by_0}{a+bc}e^{-(a/c+b)\cdot s}\\
        \displaystyle y(s)=\frac{cx_0+y_0}{a+bc}a-\frac{ax_0-by_0}{a+bc}c \cdot e^{-(a/c+b)\cdot s}
      \end{array}
    \right. .
    \label{eq_sol-sol}
  \end{equation}
  Therefore, the solution of (\ref{ec_sediment_E1s_forma1})
  can be written as
  \begin{equation}
    \left\{
      \begin{array}{l}
        \rho_\alpha^i (t+s) = x(s)\\
        m_\alpha^i (t+s) =y(s)
      \end{array}
    \right. ,
    \label{eq_ro_alfa_m_alfa_E1s}
  \end{equation}
  for any $s \in [0, \triangle t]$, where
  \begin{equation}
    \left\{
      \begin{array}{l}
        x_0 = \rho_\alpha^i (t)\\
        y_0 = m_\alpha^i (t)
      \end{array}
    \right. .
    \label{eq_ro_alfa0_m_alfa0_E1s}
  \end{equation}
\end{proof}

Now, for the construction of $\mathcal{E}_s^2$, we will
integrate the system
\begin{equation}
  \left\{
    \begin{array}{ccccc}
      \partial_t \theta^i h^i \rho_\alpha^i&+&\mathcal{P}_\alpha^i&=&\mathcal{R}_\alpha^i\\
      \partial_t m_\alpha^i&+&0&=&0\\
    \end{array}
    \right. , \quad \alpha = \overline{1,M},
    \label{ec_sediment_E2s_forma1}
\end{equation}
using an explicit Euler time step method on the interval
$(t, t + \triangle t)$ to get
\begin{equation}
  \left\{
    \begin{array}{ccccc}
      \theta^i h^i(t+\triangle t) \rho_\alpha^i(t+\triangle t) & = 
      & \theta^i h^i(t) \rho_\alpha^i(t) & + & \triangle t [{\cal R}_\alpha^i(t)-{\cal P}_\alpha^i(t)]\\
      m_\alpha^i(t+\triangle t) & = & m_\alpha^i(t) & &
    \end{array}
    \right. .
    \label{eq_ec_sediment_E2s_explicit_euler}
\end{equation}

\begin{proposition}[$\rho_\alpha$-positivity]
  The positivity condition for the water depth $h_i$ ensures
  also the positivity of $\rho_\alpha^i$.
  \label{prop_E2s}
\end{proposition}

\noindent
\begin{proof}
  The equation for $\rho_\alpha^i$ in
  (\ref{eq_ec_sediment_E2s_explicit_euler}) reads as
  \begin{equation}
    \begin{split}
      \theta^i h^i(t+\triangle t) \rho_\alpha^i(t+\triangle t) =
      & \theta^i h^i(t) \rho_\alpha ^i(t) - \displaystyle \triangle t\frac{1}{\sigma ^i} \sum_{j\in N(i)}l^{ij} \Phi_{\alpha}^{ij} v_{\rm norm}^{ij}+\\
      &\displaystyle + \triangle t \theta^i p_\alpha[1-H^i({\boldsymbol m})]\frac{F(\Omega^i-\Omega_{cr})_{+}}{J}.
    \end{split}
    \label{eq_ro_alfa_E2s_forma1}
  \end{equation}
  Let us denote the positive and the negative parts of any
  $v\in\mathbb{R}$ by
  \begin{equation*}
    [v]^{+} := \max\{v,0\}, \quad [v]^{-}:= -\min\{v,0\},
  \end{equation*}
  respectively.  Then, decomposing $v_{\rm norm}^{ij}$ into
  \begin{equation*}
    v_{\rm norm}^{ij} = \left[ v_{\rm norm}^{ij} \right]^{+} - \left[ v_{\rm norm}^{ij} \right]^{-}
  \end{equation*}
  gives
  \begin{equation}
    \begin{split}
      \theta^i h^i(t+\triangle t) \rho_\alpha^i(t+\triangle t) = 
      & \theta^i h^i(t) \rho_\alpha ^i(t) 
      \left[
        1 - {\triangle t} \frac{1}{\sigma^i} \sum_{j\in N(i)} l^{ij} \left[ v_{\rm norm}^{ij} \right]^{+}
      \right] +\\
      & + {\triangle t}
      \left\{
        \frac{1}{\sigma^i}
        \sum_{j\in N(i)} l^{ij} \theta^j h^j(t) \rho_\alpha^j(t) \left[ v_{\rm norm}^{ij} \right]^{-}+
      \right.\\
      &\left.
        \displaystyle \theta^i p_\alpha [1-H^i({\boldsymbol m})] \frac{F(\Omega^i-\Omega_{cr})_{+}}{J}
      \right\} ,
    \end{split}
    \label{eq_ro_alfa_E2s_forma2}
  \end{equation}
  from where one can easily observe that $\rho_\alpha^i$
  remains positive.
\end{proof}

We can now formulate the following extension (including the
sediment) of the result from \cite{sds_apnum}:
\begin{theorem}
  For a proper definition of:\\
  - the flux interface term $(\theta h)^{ij} v_{\rm norm}^{ij}$ \\
  - the discrete gradient of the free surface $w^{ij}$\\
  - and the time step bound of $\triangle t$\\
  the numerical scheme is:\\
  - {\bf well-balanced}
  \begin{center}
    $\left. \boldsymbol{v}=0,\; h+z=ct. \right|_{t=0} \, 
    \longrightarrow 
    \left. \boldsymbol{v}=0,\; h+z=ct. \right|_{t>0}$ ,
  \end{center}
  - $h$, $\rho_\alpha$, $m_\alpha$ {\bf positive scheme}
  \begin{center}
    $\left. h>0,\; \rho_\alpha>0,\; m_\alpha>0\right|_{t=0}\, 
    \longrightarrow
    \left. h>0,\; \rho_\alpha>0,\; m_\alpha>0\right|_{t>0}$ .
  \end{center}
\end{theorem}
%----------------------------------------------------------

%----------------------------------------------------------
%----------------------------------------------------------
%==========================================================
%==========================================================

%==========================================================
%==========================================================
\section{A stationary solution for sediment}
\label{section_stationary_sol_sed}
Despite the complexity of the model equations
(\ref{eq_swe1}-\ref{eq_eros2}), there are some
configurations of the soil surface and vegetation density
distribution that allow us to obtain analytical solutions.
Such a solution can be helpful when trying to validate the
model.

%----------------------------------------------------------
%----------------------------------------------------------
\subsection{Exact solution}
\label{subsec_AnalyticalSolution}
Let us consider the case of the plain soil surface with
constant vegetation density.  For such case, the problem
reduces to a 1D model equation.  Let $\partial_x z = -s_0$
be the constant gradient of the soil surface and
$\theta(x)=\theta_0$ be the porosity of the cover plant.  If
$h_0$ and $v_0$ satisfy
\begin{equation}
  \label{rep-ns.01}
  v_0^2 =
  \displaystyle \frac {\theta_0gh_0s_0}
  {\alpha_vh_0(1-\theta_0)+\theta_0\alpha_s},
\end{equation}
then $h(t,x)=h_0$, $v(t,x)=v_0$ is a solution of the
shallow water equation (\ref{eq_swe1}-\ref{eq_swe2}).

As in \cite{sander}, one can now use a uniform flow
$h(t,x)=h_0$, $v(t,x)=v_0$ to find certain exact solutions
for the sediment variables.  Let us first introduce the
notations
\begin{equation}
  \Gamma := \frac{\gamma_s}{\gamma_s-1}
  \frac{F(\Omega(v_0)-\Omega_{crt})_{+}}{g h_0},\quad
  \Lambda:=\frac{F(\Omega(v_0)-\Omega_{crt})_+}{J},\quad 
  q := h_0 v_0,
  \label{eq_rep-ns.04}
\end{equation}
where $(x)_+:=\max\{x,0\}$.

Water flow generates a net erosion of the soil if the total
mass $m_t$ of the deposited sediment is smaller than
$m_t^{\star}$, i.e. $H = m_t/m^{\star}_t < 1$.  In order to have a
steady state of $m_\alpha$, the following must hold
\begin{equation}
  \label{rep-ns.06}
  d_\alpha = e_\alpha^r,
  \qquad {\rm i.e.} \qquad
  \nu_{s, \alpha} \cdot\rho_\alpha = \Gamma \displaystyle\frac{m_\alpha}{m_t^{\star}},
\end{equation}
which gives
\begin{equation}
  H = \left(\sum\limits_\beta m_\beta\right)/{m_t^{\star}} = 
  \frac{1}{\Gamma}\sum\limits_{\beta =1}^N \nu_{s, \beta} \cdot\rho_\beta.
  \label{rep-ns.06_2}
\end{equation}
The steady state of the suspended sediment solves the equations
\begin{equation}
  q\frac{{\rm d} \rho_\alpha}{{\rm d} x} = e_\alpha + e^r_\alpha - d_\alpha,
  \qquad {\rm i.e.} \qquad
  q\frac{{\rm d} \rho_\alpha}{{\rm d} x} = p_\alpha \Lambda
  \left(
    1-\displaystyle\frac{1}{\Gamma}\sum\limits_{\beta =1}^N \nu_{s, \beta} \cdot\rho_\beta 
  \right),
  \label{rep-ns.07}
\end{equation}
since $e_{\alpha} = p_{\alpha} \Lambda (1-H)$.  Multiplying
the derivative of (\ref{rep-ns.06_2}) by $q$ and using
(\ref{rep-ns.07}) gives an equation for the ratio
$m_t/m_t^{\star}$
\begin{equation}
  q\frac{{\rm d}}{{\rm d} x}
  \frac{m_t}{m_t^{\star}}=\frac{\Lambda}{\Gamma}
  \left(
    \sum\limits_\alpha 
    \nu_{s, \alpha} \cdot p_\alpha
  \right) 
  \left( 1- \frac{m_t}{m_t^{\star}} \right)
  \label{rep-ns.09}
\end{equation}
which has the solution
\begin{equation}
  \frac{m_t}{m_t^{\star}}(x) = 1+
  \left.\left(\frac{m_t}{m_t^{\star}}-1\right)\right|_{x=0}
  \cdot\exp
  \left[
    -\frac{\Lambda}{q\Gamma}
    \left(
      \sum\limits_\alpha
      \nu_{s, \alpha} \cdot p_\alpha
    \right)
    x
  \right]
  \label{eq_sol_stationara_cont_mt}
\end{equation}
Therefore, the condition $H<1$ is satisfied for all $x>0$ if
and only if it is satisfied at $x=0$, i.e. if
\begin{equation}
  \frac{m_t}{m_t^{\star}}(0)=\frac{1}{\Gamma}\sum\limits_\alpha \nu_{s, \alpha}\cdot\rho_\alpha(0)<1.
  \label{eq_cond_H_1_cont}
\end{equation}
Using the solution (\ref{eq_sol_stationara_cont_mt}), one obtains
\begin{equation}
  \begin{array}{l}
    \rho_\alpha(x) = \rho_\alpha(0) + \\
    \displaystyle +\frac{p_\alpha \Gamma}{\sum\limits_\beta \nu_{s, \beta} \cdot p_\beta}
    \left[
    \frac{1}{\Gamma}
    \left(\sum\limits_\beta \nu_{s, \beta} \cdot\rho_\beta (0)\right) -1
    \right]
    \left\{
    \exp
    \left[
    -\frac{\Lambda}{q\Gamma}
    \left(\sum\limits_\beta \nu_{s, \beta} \cdot p_\beta \right) x
    \right]
    -1
    \right\},
  \end{array}
  \label{eq_sol_stationara_cont_rhoalfa}
\end{equation}
and then
\begin{equation}
  \mu_\alpha(x) = \mu_\alpha(0) + 
  \frac{\nu_{s, \alpha} \cdot p_\alpha}{\sum\limits_\beta \nu_{s, \beta} \cdot p_\beta}
  \left( \mu_\alpha(0) - 1 \right)
  \left\{
    \exp
    \left[
      -\frac{\Lambda}{q\Gamma}
      \left(\sum\limits_\beta \nu_{s, \beta} \cdot p_\beta \right) x
    \right]
    -1
  \right\},
  \label{eq_sol_stationara_cont_malfa_pe_mt}
\end{equation}
where we used the notation
\begin{equation}
  \mu_\alpha := \displaystyle\frac{m_\alpha}{m_t^{\star}}.
\end{equation}
%----------------------------------------------------------
%----------------------------------------------------------

%----------------------------------------------------------
%----------------------------------------------------------
\subsection{Numerical solution}
\label{subsec_num_sol_sedim}
For water flowing on a channel with constant slope $s_0$, we
consider
\begin{equation}
  h(x,y)=h_0,\quad v_y(x,y)=v_0,\quad v_0^2=\frac{\theta_0gh_0s_0}{\alpha _vh_0(1-\theta_0)+\theta_0\alpha s},\quad v_x(x,y)=0.
  \label{eq_hv_steady}
\end{equation}
Denote $\boldsymbol{v}_0(x,y):=(0,v_0)^T$.  From
(\ref{eq_ode_sediment_omegai}), the discrete stationary
solution of the sediment is given by solving
\begin{equation}
  \begin{split}
    {\cal P}_\alpha ^i &= {\cal Q}_\alpha ^i+{\cal R}_\alpha ^i\\
    {\cal Q}_\alpha ^i &= 0
  \end{split},
  \label{eq_stationary_discr_sedim-sys-form1}
\end{equation}
for all $i=\overline{1,N}$ and $\alpha=\overline{1,M}$.

As in the continuous case, the condition $H<1$ is satisfied
along the channel if and only if it is satisfied at the
upper bound of the channel and in this case, using the
notations from (\ref{eq_rep-ns.04}), we ca rewrite the
system as
\begin{equation}
  \begin{split}
    \sum_{j\in N(i)} l^{ij} \Phi_{\alpha}^{ij} v_{\rm norm}^{ij} 
    & = \sigma^i \theta^i p_\alpha \Lambda
    \left(
      1-\frac{1}{\Gamma}\sum\limits_\beta \nu_{s, \beta} \rho_\beta ^i
    \right)\\
    \mu_\alpha^i\Gamma & = \nu_{s, \alpha} \rho_\alpha^i
  \end{split},
  \quad \alpha=\overline{1,M}, \quad i=\overline{1,N},
  \label{eq_stationary_discr_sedim-sys-form2}
\end{equation}
where
\begin{equation}
  \mu_\alpha^i := \displaystyle\frac{m_\alpha^i}{m_t^{\star}}.
\end{equation}

{ASTERIX} is a software developed by the authors in
\cite{sds_ASTERIX} for modeling the water flow on vegetated
hillslopes.  The discretization scheme behind it was thought
of in \cite{sds_apnum} and built on hexagonal networks for
various reasons, e.g. as the ones described in
\cite{sds-ADataPortingTool}.  To incorporate a ``sediment
module'' into ASTERIX and to validate the proposed model, we
focus our attention on the numerical solution over a
hexagonal network.

\begin{figure}[htbp]
  \centering
  \includegraphics[width=0.4\textwidth]{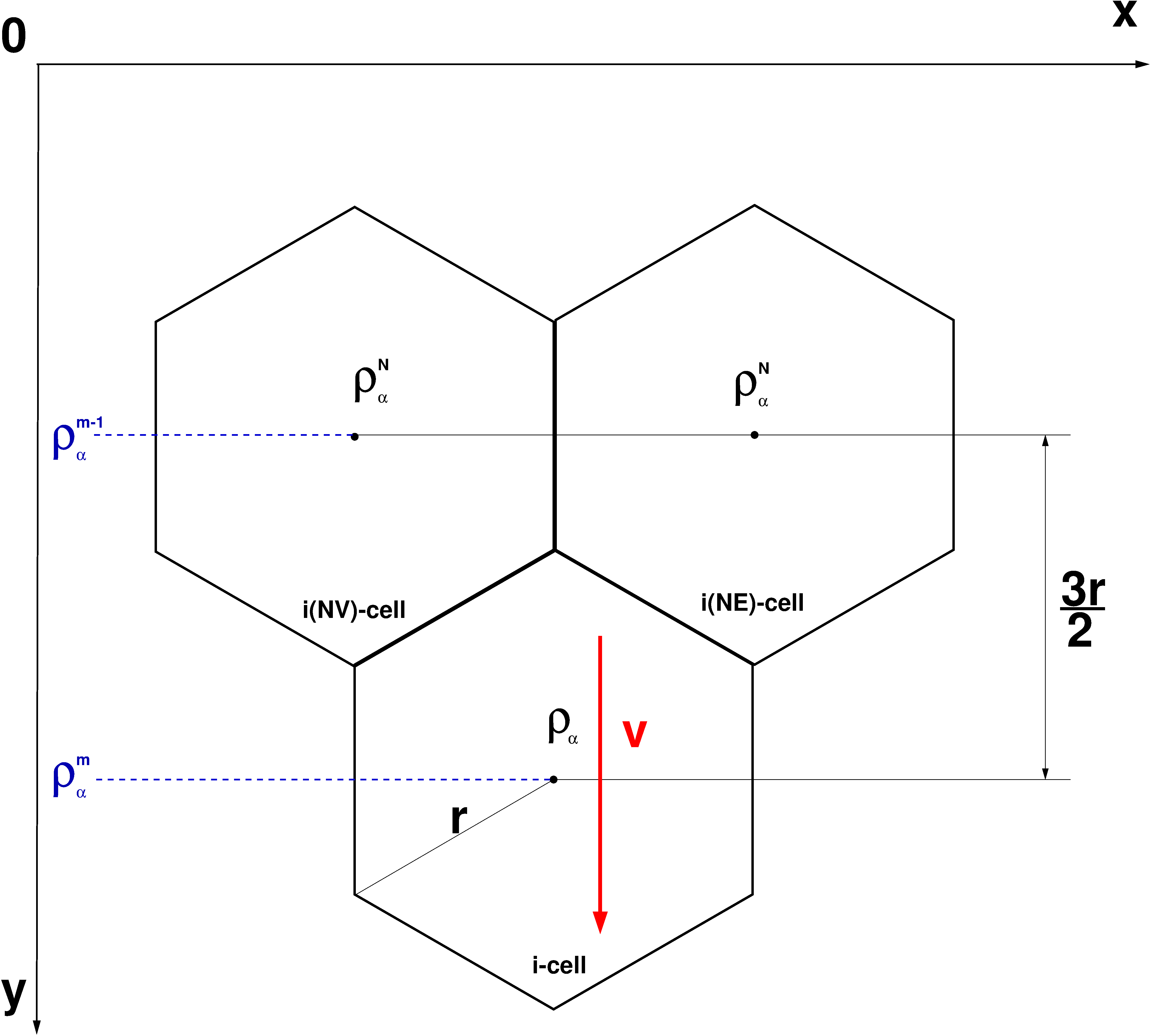}
  \caption{Indexing the rows of hexagonal cells along a
    channel by $m$.  The distance between two adjacent rows
    is ${3r}/{2}$.}
  \label{fig_hexagoane}
\end{figure}

Thus, let us consider the case of a hexagonal network with
cells of radius $r$, i.e.
\begin{equation}
  l^{ij} = r, \quad
  \sigma^i = \sigma = \frac{3\sqrt{3}}{2} r^2, \quad
  \frac{l}{\sigma} = \frac{2\sqrt{3}}{9r}, \quad \forall i,j,
  \label{eq_hexag_network_notations}
\end{equation}
and for constant vegetation
(i.e. $\theta^i=\theta, \,\forall i$).  Using the notations
\begin{equation}
  A(r) := \frac{3}{2}\frac{\Lambda}{\Gamma} 
  \left( \sum\limits_\alpha p_\alpha \nu_{s, \alpha} \right) \cdot r, \quad
  \gamma=\gamma(r) := \frac{q}{q+A(r)}, \quad
  \mu^i := \frac{m_t^i}{m_t^{\star}},
  \label{eq_AAA}
\end{equation}
and indexing the rows of hexagonal cells along the channel
by $m$ from $0$ to $M$, it can be shown that the solution of
the system (\ref{eq_stationary_discr_sedim-sys-form2}) can
be compactly written as
\begin{equation}
  \mu^m = \gamma^m \mu^0 + 1 - \gamma^m \quad {\rm and} \quad
  \rho_\alpha^m = \rho_\alpha^0 + \frac{3r}{2q} p_\alpha \Lambda\left(1-\mu^0 \right)
  \gamma \frac{1-\gamma^m}{1-\gamma}.
  \label{eq_stationary_discr_sedim-sol-mt_rhoalfa}
\end{equation}
Note that although we now use the same capital letter as for
the number of sediment classes, $M$ here represents the
total number of rows of hexagonal cells along the channel.
Fig.~\ref{fig_hexagoane} pictures an example of indexing the
rows of hexagonal cells along a channel from top to bottom
where water flows along the $y$-direction.
%----------------------------------------------------------
%----------------------------------------------------------

%----------------------------------------------------------
%----------------------------------------------------------
\subsection{Convergence}
\label{subsec_convergence}
In order to show the convergence of the discrete numerical
solution (\ref{eq_stationary_discr_sedim-sol-mt_rhoalfa}) to
the analytical stationary solution of the sediment described
by (\ref{eq_sol_stationara_cont_mt}) and
(\ref{eq_sol_stationara_cont_rhoalfa}), let us firstly
observe that if we denote by $L$ the channel length, and for
any $y\in [0,L]$ define
\begin{equation}
  p=p(y,r):= \left[ \frac{y}{ar} \right],
  \label{eq_definition_of_p}
\end{equation}
where $a:=3/2$ and $\left[ \cdot \right]$ denotes the floor
function, then
\begin{equation}
  \lim_{r \to 0} p(y,r) = \infty \quad {\rm and} \quad
  \lim_{r \to 0} p(y,r) \cdot a r = y .
\end{equation}

\begin{theorem}
  The numerical stationary solutions for the total deposited
  sediment $\mu^m$, for the suspended sediment classes
  $\rho_\alpha^m$, and for the deposited sediment classes
  $\mu_\alpha^m$ converge to the analytical solutions from
  (\ref{eq_sol_stationara_cont_mt}),
  (\ref{eq_sol_stationara_cont_rhoalfa}), and
  (\ref{eq_sol_stationara_cont_malfa_pe_mt}), respectively.
\end{theorem}
\begin{proof}
  Using the previously introduced notation, the expression
  for the numerical stationary solution of the total
  deposited sediment can now be rewritten as
  \begin{equation}
    \mu(y,r) = \gamma(r)^{p(y,r)} \mu^0 + 1 - \gamma(r)^{p(y,r)}.
    \label{eq_stationary_discr_sedim-sol-mt_theorem}
  \end{equation}
  Observing that
  \begin{equation*}
    \lim\limits_{r\to 0} \gamma(r)^{p(y,r)} = 
    \exp{ \left[ -\frac{\Lambda}{q\Gamma} \left( \sum\limits_\alpha \nu_{s, \alpha} p_\alpha \right) y \right] }
  \end{equation*}
  will immediately give
  \begin{equation*}
    \lim\limits_{r\to 0} \mu(y,r) = 1 + (\mu^0 - 1)
    \exp{ \left[ -\frac{\Lambda}{q\Gamma} \left( \sum\limits_\alpha \nu_{s, \alpha} p_\alpha \right) y \right] }, \quad
    \forall y\in [0,L].
  \end{equation*}
  The convergence for the numerical stationary solution of
  the suspended sediment $\rho_\alpha^m$ can be also
  obtained after rewriting its expression as
  \begin{equation}
    \rho_\alpha(y,r) = \rho_\alpha^0 + 
    \frac{3r}{2q} p_\alpha \Lambda \left( 1-\mu^0 \right)
    \gamma(r) \frac{\gamma(r)^{p(y,r)}-1}{\gamma(r) -1}.
    \label{eq_stationary_discr_sedim-sol-rhoalfa_theorem}
  \end{equation}
  and observing that
  \begin{equation*}
    \frac{3r}{2q} p_\alpha \Lambda \left( 1-\mu^0 \right) \frac{\gamma(r)}{\gamma(r)-1} =
    p_\alpha\left (\mu^0 -1\right )\frac{\Gamma}{\sum\limits_\beta \nu_{s, \beta} p_\beta}.
  \end{equation*}
  The convergence for $\mu_\alpha$ follows immediately
  from
  \begin{equation*}
    \mu_\alpha(y,r) = \frac{\nu_{s, \alpha}}{\Gamma}\rho_\alpha(y,r).
  \end{equation*}
\end{proof}
%----------------------------------------------------------
%----------------------------------------------------------

%----------------------------------------------------------
%----------------------------------------------------------
\subsection{A stability result on the channel}
\label{subsec_stability}
Recall that the evolution of the sediment variables was
given by (\ref{eq_ode_sediment_omegai})
\begin{equation*}
  \frac{{\rm d}\mathcal{U}_s^i}{{\rm d}t} + \mathcal{F}_s^i= \mathcal{S}_s^i + \mathcal{T}_s^i,
\end{equation*}
where
\begin{equation*}
  \mathcal{U}_s^i :=
  \left(
    \begin{array}{c}
      \theta^i h^i\rho_\alpha ^i\\
      m_\alpha^i
    \end{array}
  \right), \quad
  \mathcal{F}_s^i :=
  \left(
    \begin{array}{c}
      \mathcal{P}_\alpha ^i\\
      \boldsymbol{0}_M
    \end{array}
  \right), \quad
  \mathcal{S}_s^i :=
  \left(
    \begin{array}{c}
      \mathcal{Q}_\alpha ^i\\
      -\mathcal{Q}_\alpha ^i
    \end{array}
  \right), \quad
  \mathcal{T}_s^i :=
  \left(
    \begin{array}{c}   
      \mathcal{R}_\alpha ^i\\
      \boldsymbol{0}_M
    \end{array}
  \right),
\end{equation*}
for all $i=\overline{1,N}$.

Let us analyze the case $\theta^i=1$ and one class of
sediment (thus $\alpha =1$ also, so we drop this index).
Assume the water on the channel reached the steady state
$h^i=h$, $\mathbf{v}^i=(0,v)$.  We seek to see what happens
with the sediment variables $\rho^i$ and $m^i$ on a
hexagonal network as in (\ref{eq_hexag_network_notations})
for the case $H < 1$.  Without losing the generality, if we
index the rows of hexagonal cells along the channel by $n$,
the previous system can be carefully rewritten as
\begin{equation}
  \left\{
    \begin{aligned}
      \frac{d\rho^n}{dt} + k (\rho^n - \rho^{n-1}) 
      &= - a \rho^n + b \mu^n + c (1 - \mu^n)\\
      \frac{m_t^{\star}}{h} \frac{d\mu^n}{dt} &= a \rho^n - b \mu^n
    \end{aligned}
  \right. , \quad
  n = \overline{1,M},
  \label{eq_stability_sys_forma1}
\end{equation}
where
\begin{equation}
  k := \frac{2q}{3rh}, \quad
  a := \frac{\nu_s}{h}, \quad
  b := \frac{\gamma_s}{\gamma_s-1}\frac{F(\Omega-\Omega_{cr})_{+}}{gh^2}, \quad 
  c := \frac{F(\Omega-\Omega_{cr})_{+}}{Jh},
  \label{eq_stability_k_notations}
\end{equation}
and
\begin{equation}
  \mu^n := \frac{m^n}{m_t^{\star}}.
\end{equation}

In order to prove a stability result for the equilibrium
point of (\ref{eq_stability_sys_forma1}), we need the
following classical result \cite{Silvester01112000} known
from linear algebra:
\begin{proposition}
  Let
  \begin{equation*}
    A = \left( \begin{array}{cc} P&Q\\ R&S \end{array} \right)
  \end{equation*}
  be $2 \times 2$ block matrix with $P,\,Q,\,R,\,S$ being
  square matrices of order $n$.  If $P$ is invertible and
  $PR=RP$, then $\det(A)=\det(PS-RQ)$.
  \label{proposition_lin_alg}
\end{proposition}
\noindent
Obs.: This formula need not hold if $PR \neq RP$.

\begin{theorem}
  The system (\ref{eq_stability_sys_forma1}) invariates the
  domain $D := [ 0, b/a ]^M \times [0,1]^M$ for $b > c$ and
  its equilibrium solution is asymptotically stable.
\end{theorem}
\begin{proof}
  (1) {\bf Invariance}\\
  Due to continuity of the solution, a trajectory of
  (\ref{eq_stability_sys_forma1}) must cross the boundary
  $\partial D$ in order to leave the domain $D$.  We will
  show that if a trajectory reaches this boundary, then it
  does not cross it.  For this, let $\tilde{t}$ be a moment
  of time at which the boundary $\partial D$ is touched,
  i.e. there is $n_0 \in \{1,2,\ldots,M\}$ such that
  $\rho^{n_0}(\tilde{t}) \in \{0,b/a\}$ or/and
  $\mu^{n_0}(\tilde{t}) \in \{0,1\}$, and the trajectory is
  inside $D$ for $t \leq \tilde{t}$.
  
  Using (\ref{eq_stability_sys_forma1}), one can easily
  verify that
  \begin{equation*}
    \rho^{n_0}(\tilde{t}) = 0 \Longrightarrow \frac{d\rho^{n_0}}{dt}(\tilde{t}) \geq 0,
    \quad
    \rho^{n_0}(\tilde{t}) = \frac{b}{a} \Longrightarrow \frac{d\rho^{n_0}}{dt}(\tilde{t}) \leq 0,
  \end{equation*}
  and
  \begin{equation*}
    \mu^{n_0}(\tilde{t}) = 0 \Longrightarrow \frac{d\mu^{n_0}}{dt}(\tilde{t}) \geq 0,
    \quad
    \mu^{n_0}(\tilde{t}) = 1 \Longrightarrow \frac{d\mu^{n_0}}{dt}(\tilde{t}) \leq 0,
  \end{equation*}
  These relations ensure that the trajectory of
  (\ref{eq_stability_sys_forma1}) remains inside $D$.

  \medskip
  (2) {\bf Stability}\\
  We can rewrite the system (\ref{eq_stability_sys_forma1})
  as
  \begin{equation}
    \left\{
      \begin{aligned}
        \frac{d\rho^n}{dt} &= -(a+k) \rho^n + k \rho^{n-1} + (b-c) \mu^n + c\\
        \frac{d\mu^n}{dt}  &= -\frac{b h}{m_t^{\star}} \mu^n + \frac{a h}{m_t^{\star}} \rho^n
      \end{aligned}
    \right. , \quad
    n = \overline{1,M}.
    \label{eq_stability_sys_forma2}
  \end{equation}
  The matrix form of (\ref{eq_stability_sys_forma2}) reads
  as
  \begin{equation}
    \frac{dX}{dt} = AX + B,
    \label{eq_odematrice}
  \end{equation}
  where
  $X = \left( \rho^1, \rho^2, \ldots, \rho^M, \mu^1, \mu^2, \ldots, \mu^M \right)^T$
  and
  $B = \left( c + k\rho^0, c, \ldots, c, 0, 0, \ldots, 0 \right)^T$
  are column vectors of sizes $2M$, while
  \begin{equation}
    A = \left( \begin{array}{cc} P&Q\\ R&S \end{array} \right)
    \label{eq_matrice_A}
  \end{equation}
  is a $2 \times 2$ block matrix of order $2M$ with
  $$Q = (b-c) I_M, \quad R = \frac{a h}{m_t^{\star}} I_M, \quad S = -\frac{b h}{m_t^{\star}} I_M,$$
  and $P$ being the lower bidiagonal matrix having all the
  diagonal elements equal to $-(a+k)$ and all the
  subdiagonal elements equal to $k$.  Here, $I_M$ denotes
  the identity matrix of order $M$.

  Let $p_A(\lambda)$ be the characteristic polynomial of the
  matrix $A$ defined in (\ref{eq_matrice_A}),
  \begin{equation*}
    p_A(\lambda) = \det \left( A-\lambda\mathbf{I}_{2M} \right).
  \end{equation*}
  Using Proposition~\ref{proposition_lin_alg}, one can
  easily deduce that
  \begin{equation*}
    p_A(\lambda) = \det\left( \left(\frac{b h}{m_t^{\star}}+\lambda\right) (P-\lambda I_M) + \frac{a h}{m_t^{\star}}(b-c) I_M \right),
  \end{equation*}
  and then observe that 
  \begin{equation}
    p_A(\lambda) = 0 \Longleftrightarrow d(\lambda) = 0,
  \end{equation}
  where $d$ is the polynomial
  \begin{equation*}
    d(\lambda) = (a+k+\lambda) \left(\frac{b h}{m_t^{\star}}+\lambda\right) - \frac{a h}{m_t^{\star}}(b-c) .
  \end{equation*}
  Now, since the discriminant of $d$ is always positive and
  $kb + ac > 0$, then the eigenvalues of $A$ are all real
  and negative, which concludes the proof.
\end{proof}
%----------------------------------------------------------
%----------------------------------------------------------
%==========================================================
%==========================================================

%==========================================================
%==========================================================
\section{Analytic Solutions}
\label{section_analytic_sols}
It is difficult to find analytic solutions of the PDE system
(\ref{eq_swe1}-\ref{eq_eros2}) for the general case, but
they can sometimes be found for particular cases.  In this
sense, as in the previous subsection, we will consider the
steady state flow on a constant slope inclined plane where
$h$ and $\boldsymbol{v}$ are constant and related by
(\ref{eq_hv_steady}).  In addition, we consider the case of
a single-class sediment ($M=1$).  The erosion equations
(\ref{eq_eros1}-\ref{eq_eros2}) reduce to
\begin{equation}
  \left\{
    \begin{aligned}
      \partial_t \rho + v \partial_x \rho & = -a\rho   + b H(m) + c (1-H(m))\\
      \frac{1}{h}\partial_t m & = a \rho - b H(m)
    \end{aligned}
  \right. ,
  \label{eq_HR1_eqs}
\end{equation}
where $\rho=\rho_1$, $m=m_1$,
$H(m) = \min\left\{ {m}/{m_t^{\star}}, 1 \right\}$, and
\begin{equation}
  a := \frac{\nu_s}{h}, \quad
  b := \frac{\gamma_s}{\gamma_s-1}\frac{F(\Omega-\Omega_{cr})_{+}}{gh^2}, \quad 
  c := \frac{F(\Omega-\Omega_{cr})_{+}}{Jh}.
  \label{eq_HR1_abc}
\end{equation}

\noindent
{\bf Case $H=1$}

In this case, the system (\ref{eq_HR1_eqs}) reads as
\begin{equation}
  \left\{
    \begin{aligned}
      \partial_t \rho + v \partial_x \rho & = -a\rho   + b\\
      \frac{1}{h}\partial_t m & = a \rho - b
    \end{aligned}
  \right. ,
  \label{eq_HR1_Hequals1}
\end{equation}
whose general solution has the form
\begin{equation}
  \rho (t,x) =
  \left\{
    \begin{array}{ll}
      \displaystyle\frac{b}{a}-{\rm e}^{-a t}
      \left(\displaystyle\frac{b}{a}-\rho _{0}(x - v t)\right), & x - v t \geq 0\\
      \displaystyle\frac{b}{a}-{\rm e}^{-a x/v} 
      \left(\displaystyle\frac{b}{a}-\rho _{b}(t-x/v)\right), & x -v t < 0
    \end{array}
  \right. ,
  \label{eq_HR1_Hequals1_sol_ro}
\end{equation}
\begin{equation}
  m (t,x) = m_0 (x) + h \cdot
  \left\{
    \begin{array}{ll}
      g(t,x), & x -v t \geq 0\\
      \displaystyle g(x/v,x) - 
      a {\rm e}^{-a x/v} \int_{x/v}^t \left(\frac{b}{a}-\rho_{ b}(\tau-x/v) \right) d\tau, & x - v t < 0
    \end{array}
  \right. ,
  \label{eq_HR1_Hequals1_sol_m}
\end{equation}
where $\rho_{0}(x) := \rho(0,x)$ and $m_0(x) := m(0,x)$ are
the initial conditions, $\rho_{b}(t) := \rho(t,0)$ is the
boundary condition at $x=0$, and
\begin{equation}
  g(s,x) := - \frac{b}{a}(1 - e^{-as}) + a\int_0^s \rho _0 (x-v\tau) e^{-a\tau}d\tau.
  \label{eq_HR1_Hequals1_fcn_g}
\end{equation}

\begin{proposition}(Full deposition)

  Assume that $m_0(x) \geq m_t^{\star}$.  The solution $(\rho, m)$ of
  the system (\ref{eq_HR1_Hequals1}) given by
  (\ref{eq_HR1_Hequals1_sol_ro}) and
  (\ref{eq_HR1_Hequals1_sol_m}) has the properties:
  \begin{enumerate}
  \item If
    \begin{equation}
      \rho_b \geq \frac{b}{a} \quad \text{and} \quad \rho_0 \geq \frac{b}{a},
      \label{ineg-1}
    \end{equation}
    then 
    \begin{equation}
      \frac{b}{a} \leq \rho(t,x) \leq \max\left\{\sup_x\rho_0,\, \sup_t\rho_b\right\}
      \quad \text{and} \quad m(t,x) \geq m_t^{\star} ,
    \end{equation}
    for all $t,x \geq 0$, and therefore
    $(\rho(t,x), m(t,x))$ is global solution of
    (\ref{eq_HR1_eqs}).
  \item If
    \begin{equation}
      \rho_b \geq \frac{b}{a}, \quad \rho_0 \leq \frac{b}{a}, \quad \text{and} \quad m_0 \geq h\frac{b}{a} + m_t^{\star},
      \label{ineg-2}
    \end{equation}
    then
    \begin{equation}
      \begin{array}{ll}
        \displaystyle \inf_x\rho_0 \leq \rho(t,x) \leq \frac{b}{a}, & \text{for} \quad x-v t \geq 0,\\[7pt]
        \displaystyle \frac{b}{a} \leq \rho(t,x) \leq \sup_t\rho_b, & \text{for} \quad x-v t < 0,\\[7pt]
        m(t,x) \geq m_t^{\star} , & {}
      \end{array}
    \end{equation}
    for all $t,x \geq 0$, and therefore
    $(\rho(t,x), m(t,x))$ is global solution of
    (\ref{eq_HR1_eqs}).  Moreover, under either hypothesis
    (\ref{ineg-1}) or (\ref{ineg-2}), if the boundary
    condition $\rho_b$ is constant, then the function
    $\rho(t, x)$ is spatially decreasing for large values of
    time, i.e.
    \begin{equation}
      \partial_x \rho \leq 0, \quad \text{for} \quad t > x/v
      \label{eq_init_data_no_longer_felt}
    \end{equation}
    (that is, the initial datum is no longer felt).
  \item If
    \begin{equation}
      \rho_b \leq \frac{b}{a}, \quad \rho_0 \geq \frac{b}{a}, \quad \text{and} \quad m_0 > m_t^{\star},
      \label{ineg-3}
    \end{equation}
    then 
    \begin{equation}
      \begin{array}{ll}
        \displaystyle \frac{b}{a} \leq \rho(t,x) \leq \sup_x \rho_0, & \text{for} \quad x-v t \geq 0,\\
        \displaystyle \inf_t \rho_b \leq \rho(t,x) \leq \frac{b}{a}, & \text{for} \quad x-v t < 0,
      \end{array}
    \end{equation}
    and $m(t,x) \geq m_t^{\star}$ on some time interval
    $[0,T]$, and therefore $(\rho(t,x), m(t,x))$ is a local
    solution of (\ref{eq_HR1_eqs}).  Moreover, for a
    constant boundary condition $\rho_b$ in (\ref{ineg-3}),
    the function $\rho(t,x)$ is spatially increasing for
    large values of time, i.e.
    \begin{equation}
      \partial_x \rho \geq 0, \quad \text{for} \quad t > x/v.
    \end{equation}
  \end{enumerate}
  \label{proposition_Hmaimic1}
\end{proposition}

\begin{remark}
  The previous result provides conditions for which the
  solutions of (\ref{eq_HR1_Hequals1}) are also solutions
  for (\ref{eq_HR1_eqs}).  We also note that, from a
  physical point of view, $\rho$ must be bounded from above
  by the density of the sediment,
  i.e. $\rho(t,x) < \gamma_s \rho_w$.  This condition is
  obviously fulfilled in the first case of
  Proposition~\ref{proposition_Hmaimic1} if it is satisfied
  by $\rho_b$ and $\rho_0$.  In the other two cases, it is
  fulfilled provided that, in addition, the inequality
  $b/a < \gamma_s \rho_w$ holds.
\end{remark}

\bigskip\noindent
{\bf Case $H<1$}

In this case, the system (\ref{eq_HR1_eqs}) reads as
\begin{equation}
  \left\{
    \begin{split}
      \partial_t \rho + v \partial_x \rho & = -a\rho   + b \mu + c (1-\mu)\\
      \frac{m_t^{\star}}{h}\partial_t \mu & = a \rho - b \mu
    \end{split}
  \right. ,
  \label{eq_HR1_Hmaimic1}
\end{equation}
where $\mu = m/{m_t^{\star}}$.  We do not have a general
solution for (\ref{eq_HR1_Hmaimic1}), but for the case when
\begin{equation}
  ({m_t^{\star}}/{h}) \alpha^2 + (b + a ({m_t^{\star}}/{h})) \alpha + ac = 0
  \label{eq_HR1_Hmaimic1_eq_alfa}
\end{equation}
has a positive discriminant, one can write a particular
solution of the form
\begin{equation}
  \mu (t,x) = 1 + A e^{\alpha t} + B e^{\beta x},
  \label{eq_HR1_Hmaimic1_sol_mu}
\end{equation}
\begin{equation}
  \rho (t,x) = \frac{b}{a} + 
  \frac{b+\alpha ({m_t^{\star}}/{h})}{a} A e^{\alpha t} + \frac{b}{a} B e^{\beta x},
  \label{eq_HR1_Hmaimic1_sol_ro}
\end{equation}
where $\alpha$ is one of the negative solutions of
(\ref{eq_HR1_Hmaimic1_eq_alfa}) and
\begin{equation}
  \beta =- \frac{ac}{bv}.
  \label{eq_HR1_Hmaimic1_beta}
\end{equation}
This particular solution is not compatible with any initial
and boundary data.  It holds for initial data of the form
\begin{equation*}
  \mu_0(x) := \mu (0,x) = 1 + A + B e^{\beta x},
\end{equation*}
\begin{equation*}
  \rho_0(x) := \rho (0,x) = \frac{b}{a} + 
  \frac{b+\alpha ({m_t^{\star}}/{h})}{a} A + \frac{b}{a} B e^{\beta x},
\end{equation*}
and boundary data of the form
\begin{equation*}
  \rho_b(t) := \rho (t,0) = \frac{b}{a} (1+B) +
  \frac{b+\alpha ({m_t^{\star}}/{h})}{a} A e^{\alpha t}.
\end{equation*}
Furthermore, for
\begin{equation*}
  b \leq a(\gamma_s \rho_w),
\end{equation*}
the solution
(\ref{eq_HR1_Hmaimic1_sol_mu}-\ref{eq_HR1_Hmaimic1_sol_ro})
of (\ref{eq_HR1_Hmaimic1}) retains a physical meaning
($0 \leq \mu \leq 1$ and $0 \leq \rho \leq \gamma_s \rho_w$)
if the constants $A$ and $B$ are chosen such that
\begin{equation*}
  -1 \leq B \leq 0 
  \qquad {\rm and} \qquad
  %\max\left\{-(1+B),\frac{a(\gamma_s \rho_w) - b}{\alpha ({m_t^{\star}}/{h})}\right\} \leq A \leq 0.
  -\min\left\{(1+B),\frac{b - a(\gamma_s \rho_w)}{\alpha ({m_t^{\star}}/{h})}\right\} \leq A \leq 0.
\end{equation*}
In this case, $\rho(t,x)$ and
$m(t,x) := m_t^{\star} \cdot \mu(t,x)$ given by
(\ref{eq_HR1_Hmaimic1_sol_mu}-\ref{eq_HR1_Hmaimic1_sol_ro})
is a solution for (\ref{eq_HR1_eqs}).
%==========================================================
%==========================================================

%==========================================================
%==========================================================
\section{Numerical Applications}
\label{section_num_app}

To assess the confidence of both the model and the numerical
scheme, we will use internal and external validation
approaches.  For internal validation, the numerical results
are usually analyzed in a theoretical framework (e.g.,
comparison with analytic solutions, stability).  Although
such a method validates the results against the mathematical
model and not against the physical processes, this type of
validation is absolutely necessary to ensure the
mathematical consistency of the method.  For the external
validation, one can perform a quantitative analysis (e.g.,
comparison of numerical data with measured real data when
they are available) and/or a qualitative one (e.g.,
comparison of the evolution given by the numerical model
with the observed/expected one).  The main advantage of
these methods is that a good consistency of data validates
both the numerical data and the mathematical model.

%----------------------------------------------------------
\subsection{Comparison between the exact and numerical
  solutions}
\label{sect_comp_exact_num_sols}
We consider the flow on a channel with constant slope $s_0$,
similar as in \cite{sds_apnum, sds_ASTERIX}.  Briefly, the
experimental installation consists of an $18\ {\rm m}$ long
and $1\ {\rm m}$ width laboratory flume with a longitudinal
bottom slope $s_0 = 1.05\ {\rm mm}/{\rm m}$.  Starting with
\begin{equation*}
  h(0,\boldsymbol{x}) = h_0, \quad \boldsymbol{v}(0,\boldsymbol{x}) = (0,v_0)
\end{equation*}
along the bare ($\theta = 1$) soil surface with
$\alpha_s=0.00709$, we considered the experiment of a
constant upstream flow rate $q_L = h_0 v_0$ and free
downstream discharge.  For
\begin{equation}
  h_0=0.05\ {\rm m}, \quad v_0 = 0.26952\ {\rm m/s},
  \label{eq_exact_num_sols_h0v0}
\end{equation}
we have simulated the flow and erosion processes using the
numerical scheme described in this article.  In what
follows, we present the results of two different tests.
\medskip

\noindent {\bf Test 1: } Stationary solution for
$m_t/m_t^{\star}<1$.

The properties of the soil and of the three sediment classes
considered in this test are given in
Table~\ref{table_exact_num_sols}.
\begin{table}[h!]
  \caption{Errors between the exact and numerical solutions
    of the erosion variables $\rho_{\alpha}$ and
    $m_{\alpha}$ for the uniform flow with
    $h_0=0.05\ {\rm m}$ and $v_0 = 0.26952\ {\rm m/s}$ on
    a constant slope bare channel with
    $s_0 = 1.05\ {\rm mm}/{\rm m}$ and $\alpha_s=0.00709$.}
  \centering
  \begin{tabular}{ccccccccc}
    \hline
    $\gamma_s$ & $F$ & $J$      & $\Omega_{\rm cr}$ & $\alpha$ & $p_\alpha$ & $\nu_{s,\alpha}$
    & \multicolumn{2}{c}{relative errors in $l^1$} \\
    -          & -   & {[J/kg]} & {[${\rm W/m}^2$]} & -        & -          & {[${\rm m/s}$]}
    & Erosion & Susp. \\
    \hline
               &      &     &       & 1 & 0.3 & 0.001 & 0.015734 & 0.018958\\
    2.6 & 0.01 & 0.2 & 0.007 & 2 & 0.5 & 0.002 & 0.026172 & 0.010333\\
               &      &     &       & 3 & 0.2 & 0.003 & 0.037517 & 0.019639\\
    \hline
  \end{tabular}
  \label{table_exact_num_sols}
\end{table}
The values for $h_0$ and $v_0$ were chosen as in
(\ref{eq_exact_num_sols_h0v0}) in order to have an exact
solution for the erosion variables with which the numerical
one can be compared.

\begin{figure}[htbp]
  \centering
  \begin{tabular}{cc}
    \includegraphics[width=0.45\linewidth]{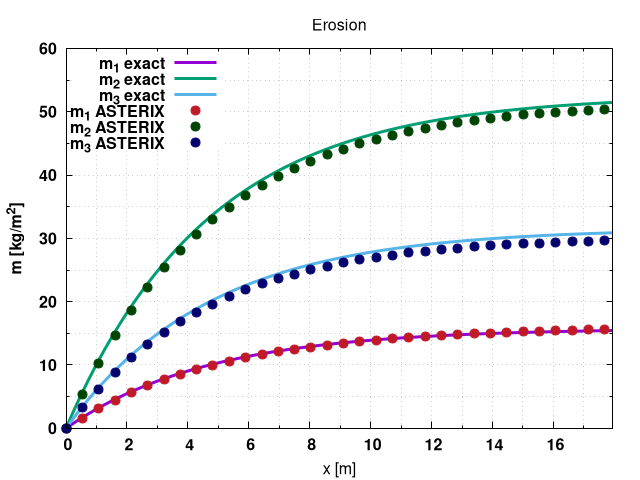}
    &\includegraphics[width=0.45\linewidth]{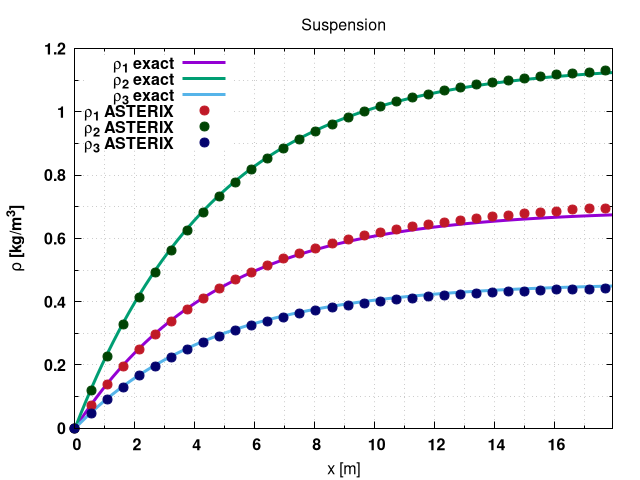}
  \end{tabular}
  \caption{The numerical and the exact solutions for the
    mass densities $m_{\alpha}$ of the deposited sediment
    (left picture) and for the mass densities
    $\rho_{\alpha}$ of the suspended sediment (right
    picture)}
  \label{fig_exact_num_sols_ro_m_case1}
\end{figure}

For this test, using the numerical scheme constructed in
Section~\ref{section_num_approx}, we obtain a uniform steady
water flow
\begin{equation}
  h(t,\boldsymbol{x}) = h_0, \quad \boldsymbol{v}(t,\boldsymbol{x}) = (0,v_0),
  \label{eq_unif_flow_h0v0}
\end{equation}
with the erosion variables approaching the steady state
solution described in Subsection~\ref{subsec_num_sol_sedim}.
For each $\alpha=\overline{1,3}$, a comparison between the
numerical solution for the mass density $\rho_{\alpha}$ of
the suspended sediment of size class $\alpha$ and the
exact solution given by
(\ref{eq_sol_stationara_cont_rhoalfa}) is pictured in
Fig.~\ref{fig_exact_num_sols_ro_m_case1} (the right
picture).  The value of the relative error between these two
solutions (last column in Table~\ref{table_exact_num_sols})
confirms the small distance between them.
\medskip

\noindent {\bf Test 2: } Time evolution for erosion process
when $m_t/m_t^{\star} \geq 1$.

Since we have an analytic solution
(\ref{eq_HR1_Hequals1_sol_ro}-\ref{eq_HR1_Hequals1_sol_m})
only for the case of a single-class sediment, we consider
for this test the uniform steady water flow
(\ref{eq_unif_flow_h0v0}) on a soil surface with properties
given in Table~\ref{table_exact_num_sols}, but with $M=1$,
i.e. $p_1=1$, $\nu_s=\nu_{s,1}$.

\begin{figure}[!htbp]
  \centering
  \begin{tabular}{ccc}
    {} & {$m$ [kg/m$^2$]} & {$\rho$ [kg/m$^2$]}\\
    \begin{turn}{90} \hspace{15mm} {t = 0 s} \end{turn}
    &\includegraphics[width=0.40\linewidth]{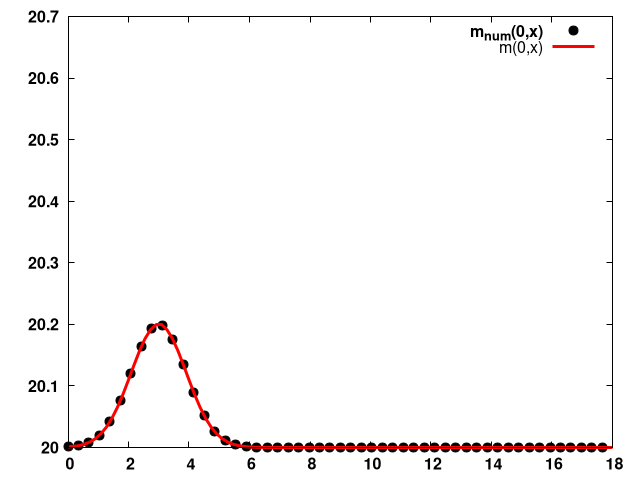}
    &\includegraphics[width=0.40\linewidth]{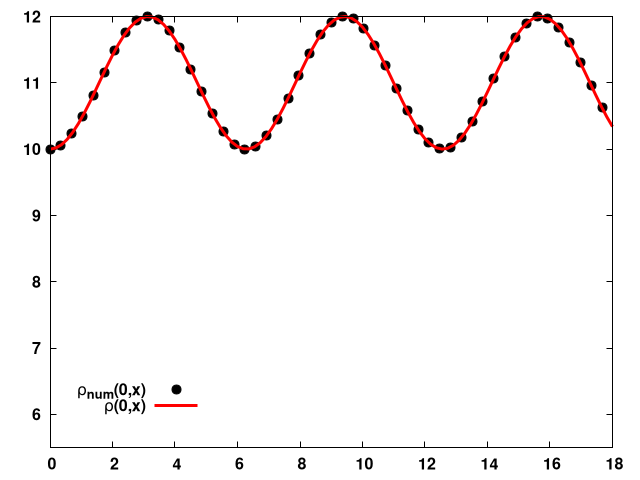}\\
    \begin{turn}{90} \hspace{13mm} {t = 10 s} \end{turn}
    &\includegraphics[width=0.40\linewidth]{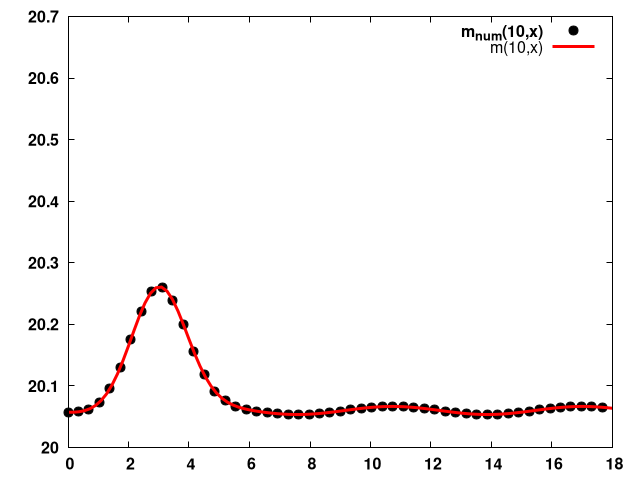}
    &\includegraphics[width=0.40\linewidth]{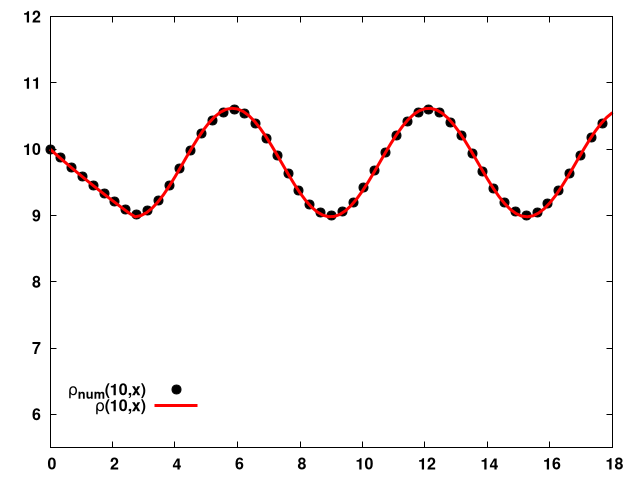}\\
    \begin{turn}{90} \hspace{13mm} {t = 30 s} \end{turn}
    &\includegraphics[width=0.40\linewidth]{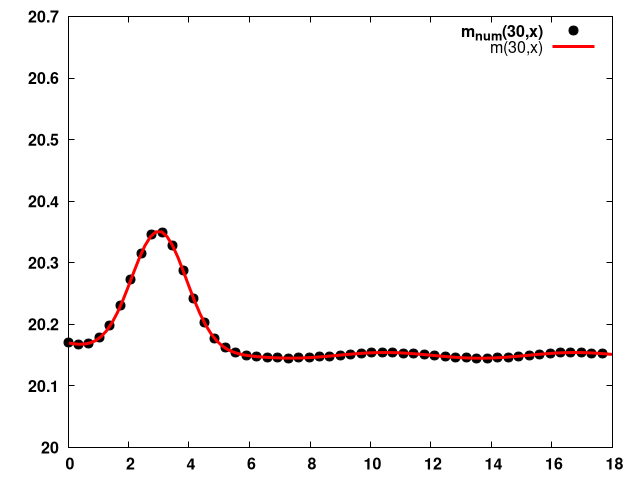}
    &\includegraphics[width=0.40\linewidth]{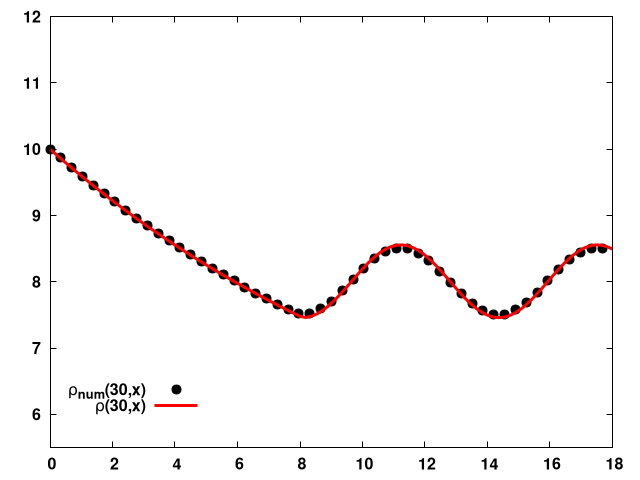}\\
    \begin{turn}{90} \hspace{12mm} {t = 100 s} \end{turn}
    &\includegraphics[width=0.40\linewidth]{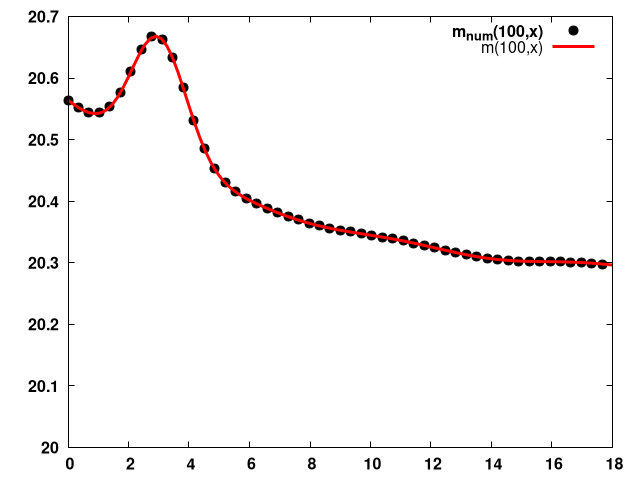}
    &\includegraphics[width=0.40\linewidth]{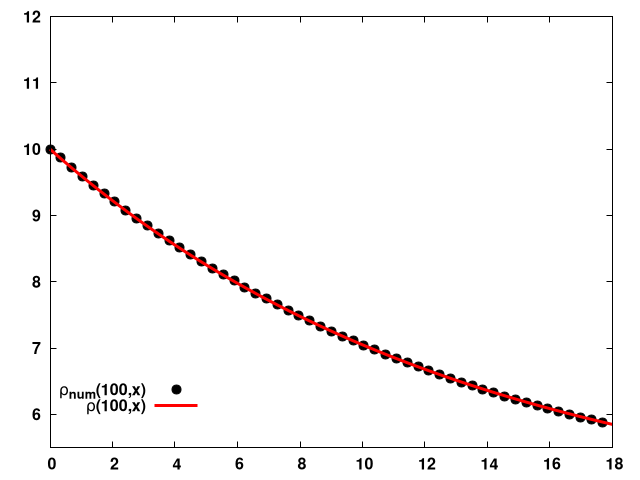}\\
    {} & {$x$ [m]} & {$x$ [m]}
  \end{tabular}
  \caption{Snapshots from the evolution of the numerical and
    the exact solutions for the mass density $m$ of the
    deposited sediment (left picture) and for the mass
    density $\rho$ of the suspended sediment (right picture)
    for Test~2 at four moments of time.}
  \label{fig_exact_num_sols_ro_m_case2}
\end{figure}

For the mass density $\rho$ of the suspended sediment and
for the mass density $m$ of the deposited sediment, we
consider a constant boundary (at $x=0$) value
\begin{equation*}
  \rho_b(t) = 10
\end{equation*}
and
\begin{equation*}
  \rho_0(x) = \rho_b(t) + 1 + \sin \left(x-\frac{\pi}{2}\right), \quad
  m_0(x) = 20 + 0.2\exp\left(-0.6(x-3)^2\right)
\end{equation*}
as initial data.  The values of these parameters and of the
above chosen functions place us in the first case of
Proposition~\ref{proposition_Hmaimic1} for any
$m_t^{\star} < 20$.  We note the reader that, having such
expressions for $\rho_b$, $\rho_0$, and $m_0$, one can
easily calculate the analytic solutions for the sediment
variables $\rho(t,x)$ and $m(t,x)$ using
(\ref{eq_HR1_Hequals1_sol_ro}) and
(\ref{eq_HR1_Hequals1_sol_m}), respectively, and therefore,
a comparison between these solutions and the numerical ones
built with the algorithm presented in this paper can be
provided.  In this regard,
Fig.~\ref{fig_exact_num_sols_ro_m_case2} presents four
snapshots (at 0, 10, 30, and 100 s) of the evolution of the
analytical and numerical solutions for $\rho$ and $m$.
Also, since $\rho_b$ is constant, one can observe that the
property (\ref{eq_init_data_no_longer_felt}) is verified,
i.e. the initial datum is no longer felt for large values of
time.
The small maximal values of the relative error between the
numerical and exact solution for $m$ and $\rho$ over the
entire time interval $t \in [0,100]$, $1.5\cdot 10^{-5}$ and
$1.7\cdot 10^{-3}$, respectively, confirms the small
distance between these two solutions.
%----------------------------------------------------------

%----------------------------------------------------------
\subsection{Wave propagation through heterogeneous media}
\label{sect_wave_propag}
In order to better understand the Dam Break phenomena, some
laboratory experiments have been designed and accomplished
around the world.  The experimental data were compared to
theoretical results obtained using Saint-Venant equations
and the reader is referred to CADAM (the European Concerted
Action Project on Dambreak Modelling) project for condense
knowledge and best practice on dambreak modelling.  The
theoretical results reported by \cite{sds_apnum, delestre,
  noelle, frazao} show a good agreement with laboratory
data.  This fact increases the confidence in the
Saint-Venant system as a mathematical model of water
dynamics in the dam break phenomena.  The flush flood is
another important problem that deserves the attention of
hydrologists.  Generally, the flash food propagates in very
heterogeneous environmental media (different soil surface,
plant cover or building structures).  We appreciate that the
vegetated Saint-Venant system (\ref{eq_swe1}-\ref{eq_swe2})
is an adequate mathematical model to study the flash food
problem.  Unfortunately, there are very few experimental
results concerning this problem.  To help in this regard, we
consider a numerical experiment where the soil surface
heterogeneity is of main concern.  We got inspired from the
laboratory experiment in \cite{Dupuis2016} consisting of a
rectangular long flume partially covered by vegetation.  The
plant cover is uniformly distributed along the width of the
flume, and therefore one deals with a 1D problem.

\begin{figure}[htbp]
  \centering
  \includegraphics[width=0.45\textwidth,,height=4cm]{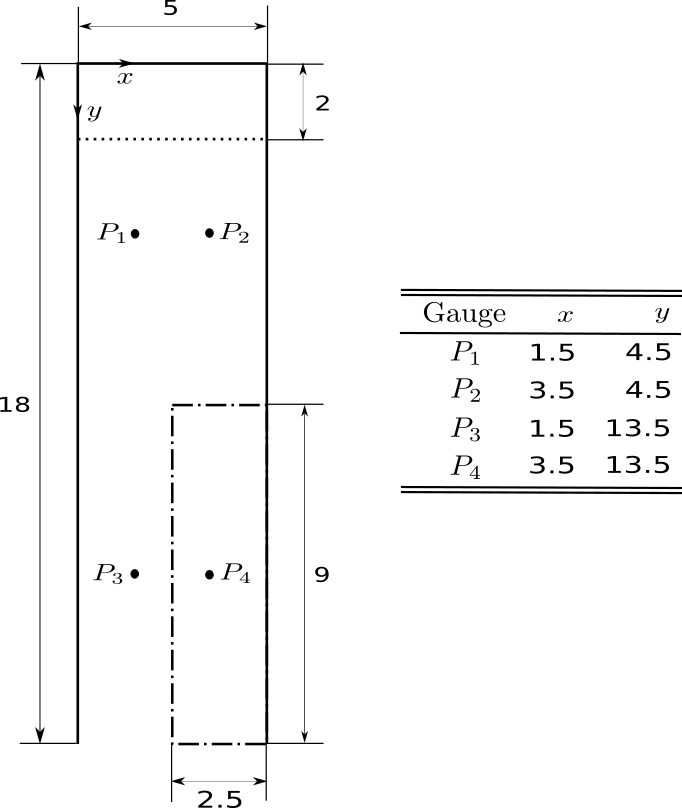}
  \caption{Scheme of the experimental installation for the
    Dam Break flow in a rectangular domain}
  \label{fig_sch_dam_break_DomainOmega}
\end{figure}

For our experiment, we consider a flume given by the
rectangular domain
\begin{equation*}
  \mathfrak{D} = \left\{ (x,y)| \, 0 \leq x\leq 5, \, 0 \leq y \leq 18 \right\}
\end{equation*}
who has two distinct complementary subdomains
\begin{equation*}
  \mathfrak{D}_1 = \left\{ (x,y)| \, 2.5 \leq x \leq 5, \, 9 \leq y \leq 18 \right\}
  \quad {\rm and } \quad \mathfrak{D}_0 = \mathfrak{D} \setminus \Omega_1
\end{equation*}
as pictured in Fig.~\ref{fig_sch_dam_break_DomainOmega}.

\noindent
We assume than only $\mathfrak{D}_0$ is covered by vegetation
\begin{equation*}
  \theta = \left\{
    \begin{array}{cc}
      \theta_1, & (x,y) \in \mathfrak{D}_1\\
      1, & (x,y) \in \mathfrak{D}_0
    \end{array}
  \right.
\end{equation*}
and/or that it has a soil surface different from the rest of
the domain:
\begin{equation*}
  \alpha_s = \left\{
    \begin{array}{cc}
      \tau_1, & (x,y) \in \mathfrak{D}_1\\
      \tau_0, & (x,y) \in \mathfrak{D}_0
    \end{array}
  \right.
\end{equation*}
The boundaries of flume are considered to be impermeable
walls, except the right side where we assume to have free
discharge.  Initially, the flume is only partially covered
by water at rest
\begin{equation*}
  h(x,y) = \left\{
    \begin{array}{cc}
      1, & y \in [0,2]\\
      0, & y \in (2,18]
    \end{array}
  \right.
  \quad {\rm and } \quad \boldsymbol{v} = 0.
\end{equation*}
This is now a 2D Riemann Problem that cannot be reduced to
1D.  In order to have physical relevant parameters, the
values of $\alpha_s$ and $\alpha_p$ are chosen by fitting
\cite{sds_apnum} the experimental data from
\cite{Dupuis2016}.  For our experiment, we slightly modify
the density of cover plants and use $\theta=0.99$ instead of
$\theta=0.99366$.

\begin{figure}[!h]
  \centering
  \begin{tabular}{ ccc }
    \hspace{-5mm}\begin{turn}{90}\hspace{2mm}\small{Water depth [m]}\end{turn}\hspace{-5mm}
    &\includegraphics[width=0.47\linewidth,height=3.0cm]{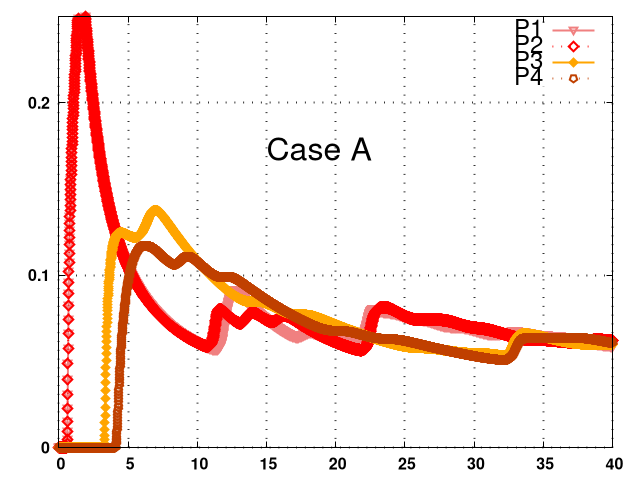}
    &\includegraphics[width=0.47\linewidth,height=3.0cm]{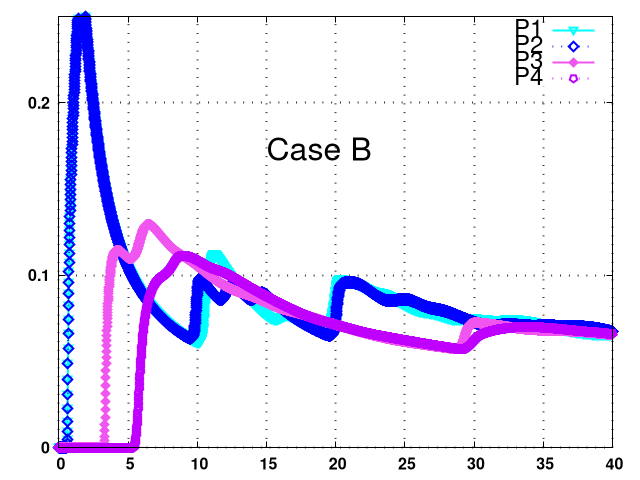}\\
    \hspace{-5mm}\begin{turn}{90}\hspace{2mm}\small{Water depth [m]}\end{turn}\hspace{-5mm}
    &\includegraphics[width=0.47\linewidth,height=3.0cm]{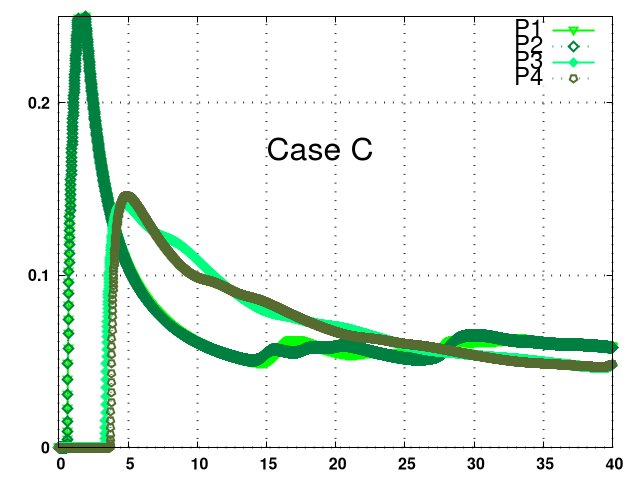}
    &\includegraphics[width=0.47\linewidth,height=3.0cm]{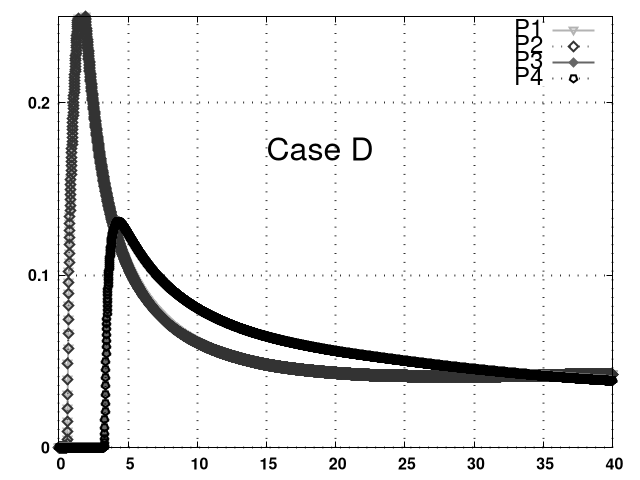}\\
    {} & \small{Time [s]} & \small{Time [s]}
  \end{tabular}
  \caption{Dynamics of the water depth from the Dam Break
    flow at gauges P1 - P4.  Each figure reveals the water
    depth evolution at all four gauges for the case of a
    soil surface and plant cover configuration given in
    Table~\ref{table_4scenarios}.}
  \label{fig_water_dynamics1_P1_P4_RectangularDomain}
\end{figure}

\begin{figure}[!h]
  \vspace{-5mm}
  \centering
  \begin{tabular}{ ccc }
    \hspace{-5mm}\begin{turn}{90}\hspace{2mm}\small{Water depth [m]}\end{turn}\hspace{-5mm}
    &\includegraphics[width=0.47\linewidth,height=3.0cm]{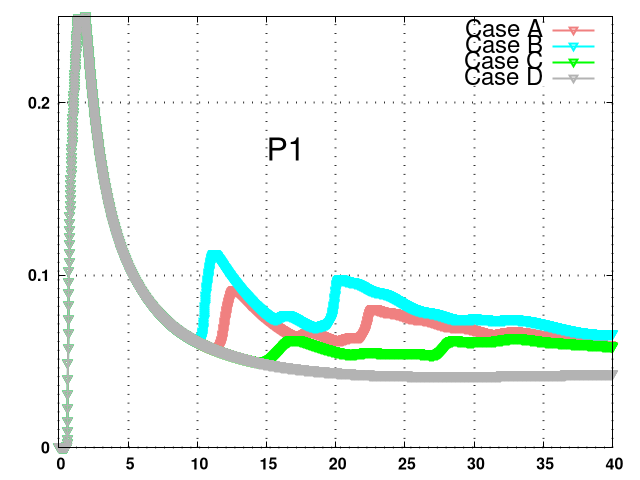}
    &\includegraphics[width=0.47\linewidth,height=3.0cm]{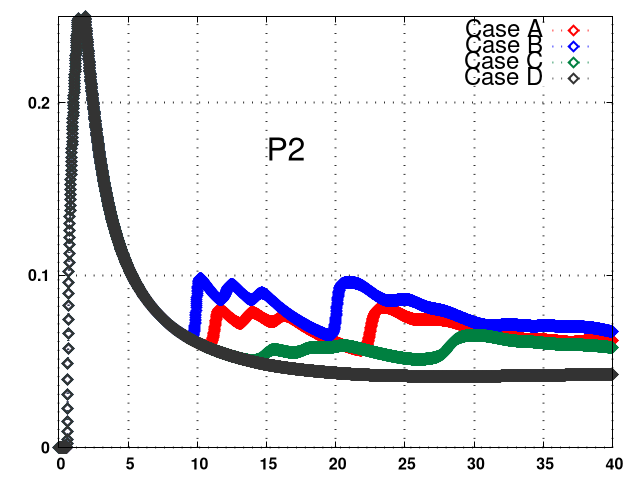}\\
    \hspace{-5mm}\begin{turn}{90}\hspace{2mm}\small{Water depth [m]}\end{turn}\hspace{-5mm}
    &\includegraphics[width=0.47\linewidth,height=3.0cm]{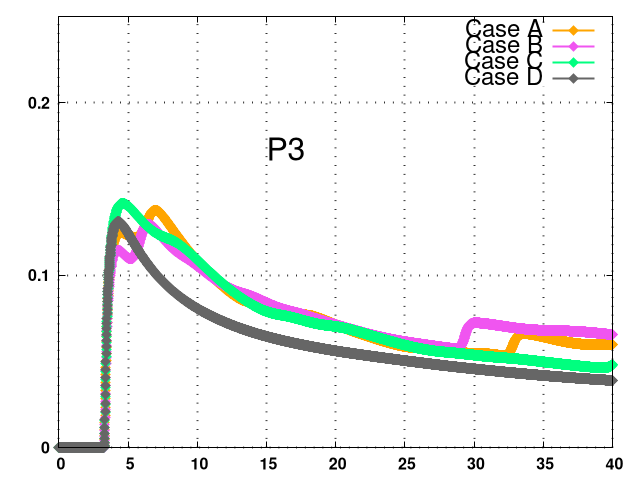}
    &\includegraphics[width=0.47\linewidth,height=3.0cm]{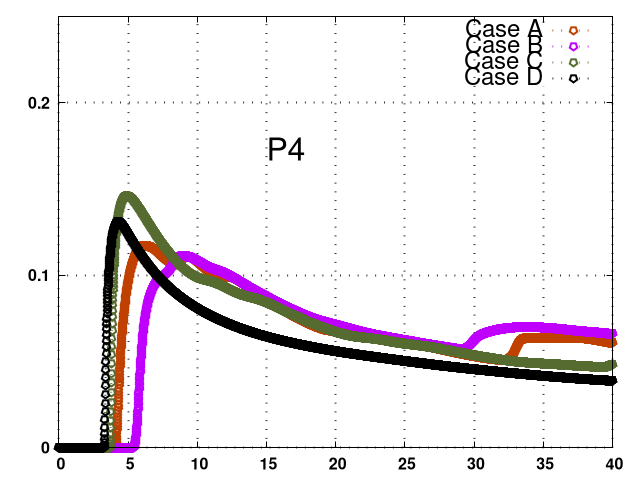}\\
    {} & \small{Time [s]} & \small{Time [s]}
  \end{tabular}
  \caption{Dynamics of the water depth from the Dam Break
    flow at gauges P1 - P4.  Each figure reveals the water
    depth evolution at a gauge for all four different soil
    surface and plant cover configurations given in
    Table~\ref{table_4scenarios}.}
  \label{fig_water_dynamics2_P1_P4_RectangularDomain}
\end{figure}

\begin{figure}[htbp]
    \begin{tabular}{ccc}
      \includegraphics[width=0.32\linewidth]{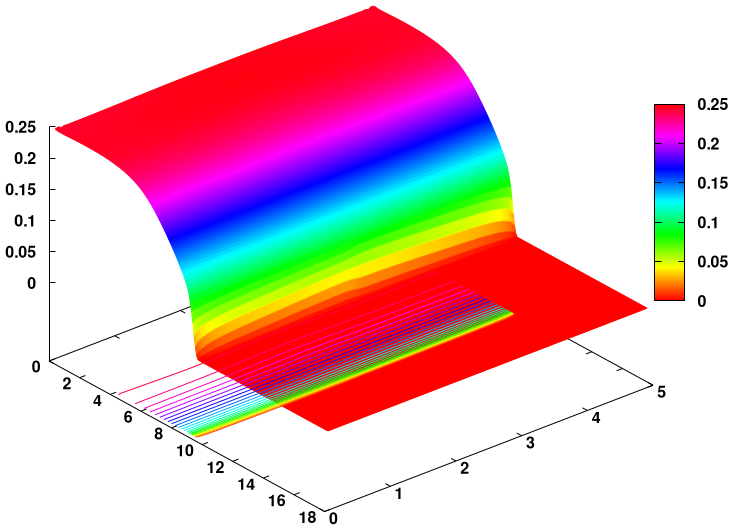}
      &\includegraphics[width=0.32\linewidth]{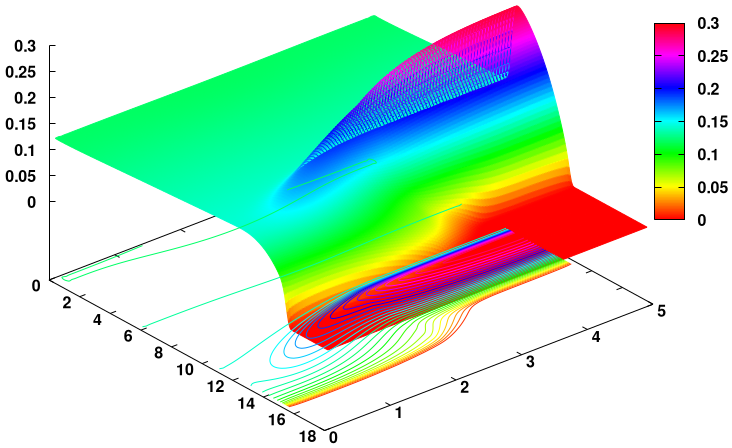}
      &\includegraphics[width=0.32\linewidth]{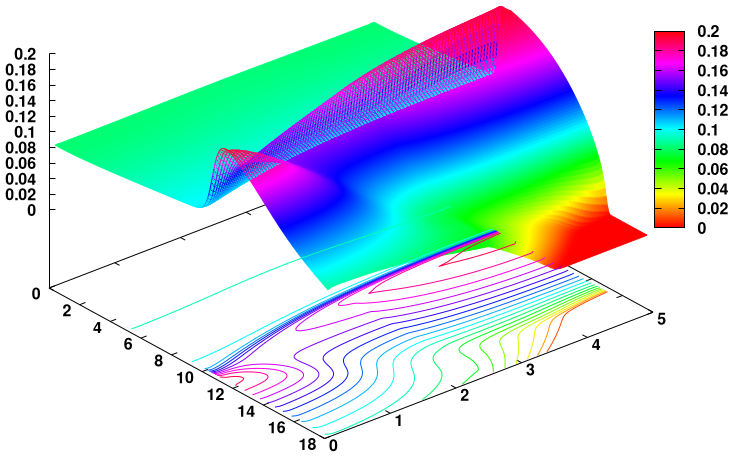}\\
      (a) $t=2\,{\rm s}$
      &(b) $t=4\,{\rm s}$
      &(c) $t=6\,{\rm s}$\\
      &&\\
      \includegraphics[width=0.32\linewidth]{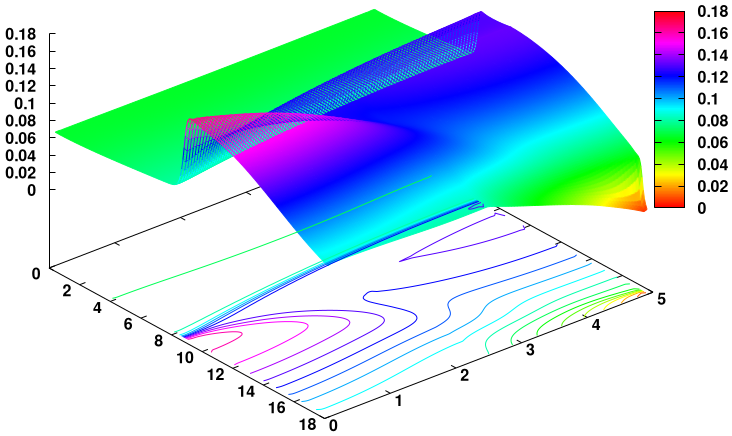}
      &\includegraphics[width=0.32\linewidth]{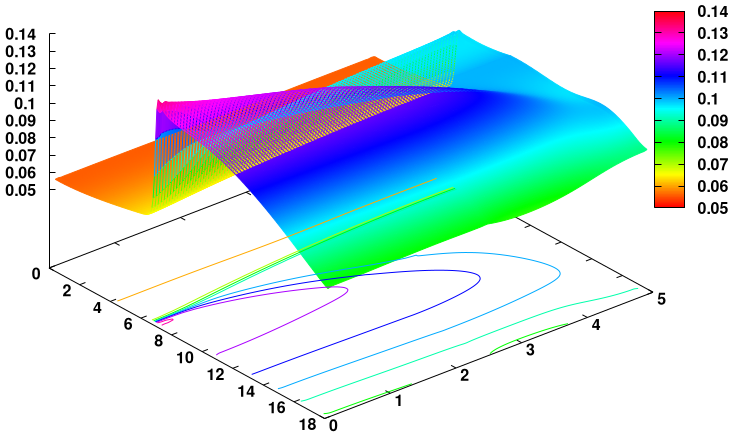}
      &\includegraphics[width=0.32\linewidth]{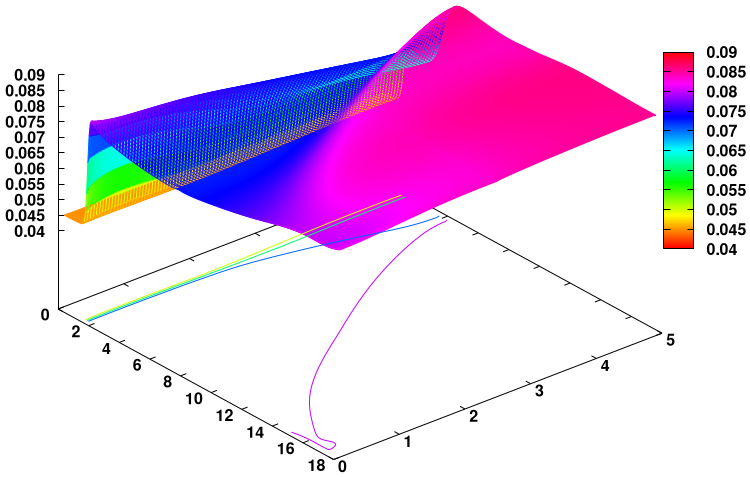}\\
      (d) $t=8\,{\rm s}$
      &(e) $t=10\,{\rm s}$
      &(f) $t=15\,{\rm s}$
    \end{tabular}
    \caption{Snapshots from the water dynamics at different
      moments of time for the case of terrain partial
      covered by plants (Case A from
      Table~\ref{table_4scenarios}).}
  \label{fig_water_distribution_snapshots_RectangularDomain}
\end{figure}

The numerical results for the water depth of the flow
resulting from the sudden removal of the door from $y=2$ in
four different scenarios described in
Table~\ref{table_4scenarios} are pictured in
Fig.~\ref{fig_water_dynamics1_P1_P4_RectangularDomain},
Fig.~\ref{fig_water_dynamics2_P1_P4_RectangularDomain}, and
Fig.~\ref{fig_water_distribution_snapshots_RectangularDomain}.
\begin{table}[h!]
  \centering
  \caption{The four configurations of plant cover and
    soil surface in the Dam Break Problem on the flume
    pictured in Fig.~\ref{fig_sch_dam_break_DomainOmega}}
  \begin{tabular}{|c|c|c|c|c|}
    \hline
    &{Partially vegetated flume}&\multicolumn{3}{c|}{Non-vegetated flume}\\
    &{Uniform soil surface}&\multicolumn{2}{c|}{Non-uniform soil surface}&{Uniform soil surface}\\\hline
    &{Case A}&{Case B}&{Case C}&{Case D}\\\hline
    $\theta$&$\theta_1=0.99$&$\theta_1=1$&$\theta_1=1$&$\theta_1=1$\\\hline
    $\alpha_s$&$\tau_1=\tau_0$&$\tau_1=0.4$&$\tau_1=0.04$&$\tau_1=\tau_0$\\\hline
  \end{tabular}
  \label{table_4scenarios}
\end{table}

The first picture from
Fig.~\ref{fig_water_dynamics1_P1_P4_RectangularDomain}
presents the evolution of the water depth in four shades of
red at all four gauges P1 - P4 for the soil surface and
plant cover configuration given by the Case A in
Table~\ref{table_4scenarios}.  Similarly, the water depth
evolution for the other three cases B, C, and D are drawn in
shades of blue, green, and black in
Fig.~\ref{fig_water_dynamics1_P1_P4_RectangularDomain}(b),
(c), and (d), respectively.  All these $16$ graphs are
rearranged in
Fig.~\ref{fig_water_dynamics2_P1_P4_RectangularDomain} in
order to better observe the influence of vegetation and soil
surface friction parameter $\alpha_s$ on the flow dynamics.
In this sense,
Fig.~\ref{fig_water_dynamics2_P1_P4_RectangularDomain}(a)
presents the evolution of the water depth at gauge P1 for
all four cases described in Table~\ref{table_4scenarios}.

For an overview of the evolution of the water surface on the
entire domain,
Fig.~\ref{fig_water_distribution_snapshots_RectangularDomain}
provides some snapshots at different moments of time for the
case of a terrain partial covered by plants (Case A in
Table~\ref{table_4scenarios}).

\begin{figure}[htbp]
  \centering
  \begin{tabular}{ lccc }
    \hspace{-3mm}\begin{turn}{90}\hspace{17mm}\small{Case A}\end{turn}
    &\includegraphics[width=0.3\linewidth,height=4.3cm]{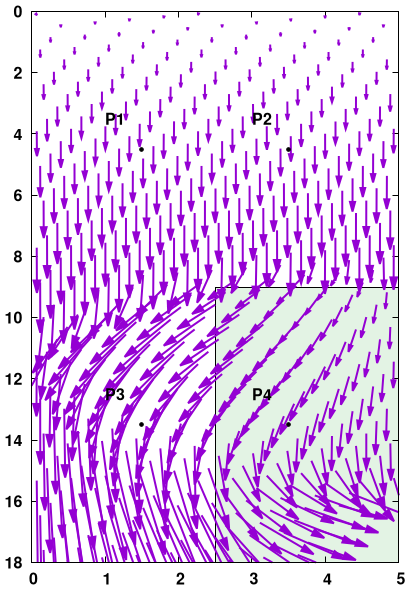}
    &\includegraphics[width=0.3\linewidth,height=4.3cm]{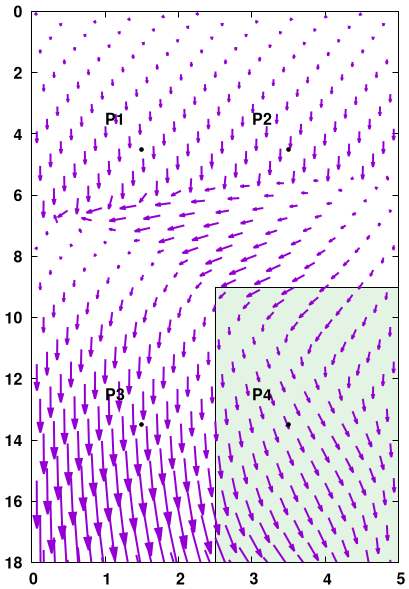}
    &\includegraphics[width=0.3\linewidth,height=4.3cm]{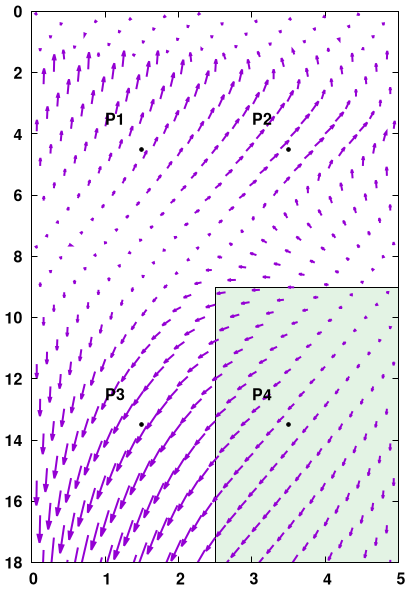}\\
    \hspace{-3mm}\begin{turn}{90}\hspace{17mm}\small{Case B}\end{turn}
    &\includegraphics[width=0.3\linewidth,height=4.3cm]{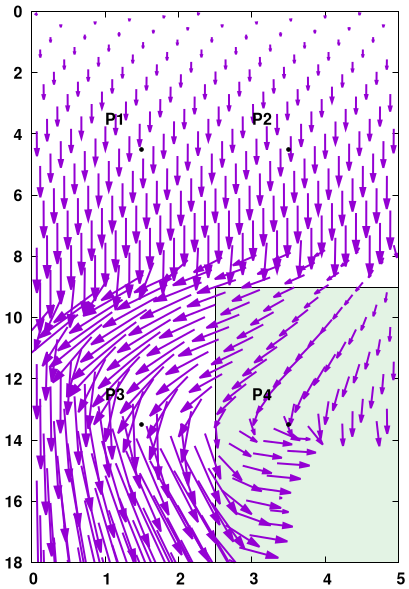}
    &\includegraphics[width=0.3\linewidth,height=4.3cm]{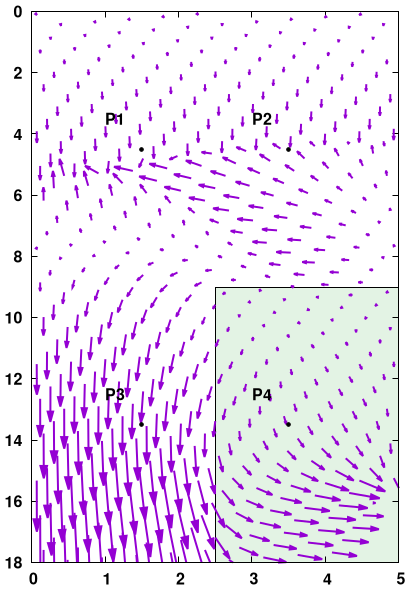}
    &\includegraphics[width=0.3\linewidth,height=4.3cm]{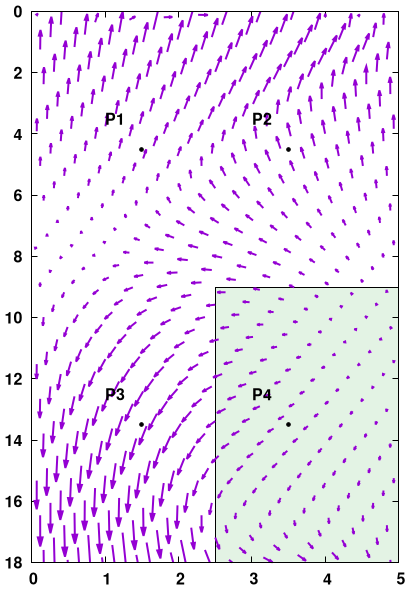}\\
    \hspace{-3mm}\begin{turn}{90}\hspace{17mm}\small{Case C}\end{turn}
    &\includegraphics[width=0.3\linewidth,height=4.3cm]{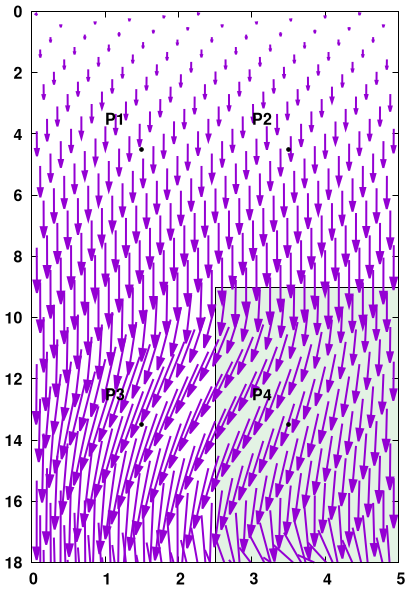}
    &\includegraphics[width=0.3\linewidth,height=4.3cm]{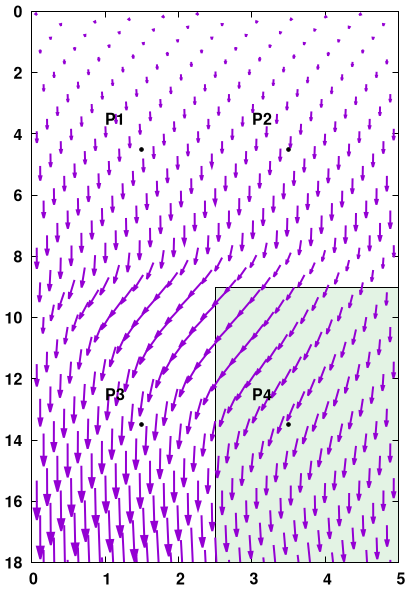}
    &\includegraphics[width=0.3\linewidth,height=4.3cm]{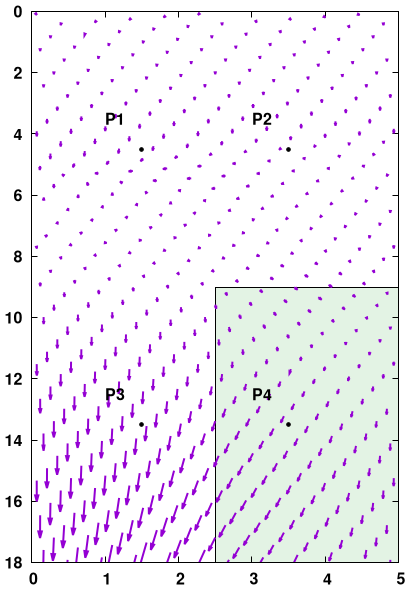}\\
    \hspace{-3mm}\begin{turn}{90}\hspace{17mm}\small{Case D}\end{turn}
    &\includegraphics[width=0.3\linewidth,height=4.3cm]{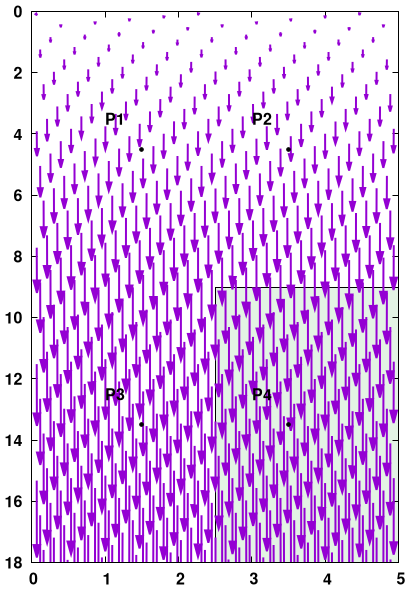}
    &\includegraphics[width=0.3\linewidth,height=4.3cm]{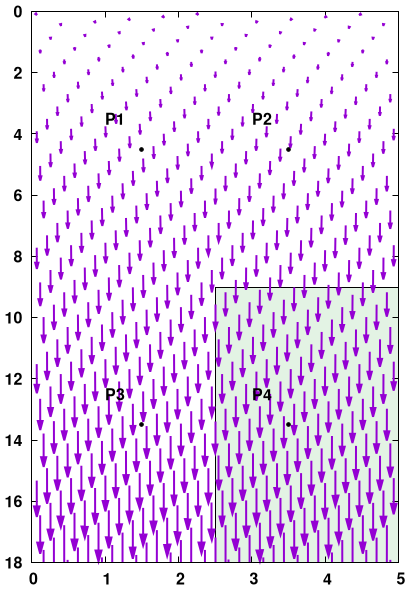}
    &\includegraphics[width=0.3\linewidth,height=4.3cm]{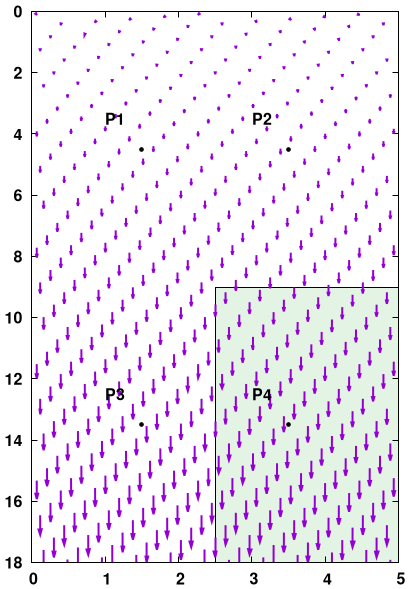}\\
    {} & \small{$t=6\,{\rm s}$} & \small{$t=10\,{\rm s}$} & \small{$t=15\,{\rm s}$}
  \end{tabular}
  \caption{Velocity profile of the flow inside
    $\mathfrak{D}$ at different moments of time.  The value
    of $\tau_0$ from the ``ambient environment'' $\Omega_0$
    was kept constant for all four simulations:
    $\tau_0 = 0.01$.}
  \label{fig_wave_propag_velo_profile}
\end{figure}

Snapshots of the velocity profile at three moments of time
for each case considered in Table~\ref{table_4scenarios} are
also presented in Fig.~\ref{fig_wave_propag_velo_profile}.

The results show that, despite the simplicity of the
configuration, the presence of heterogeneity gives rise to
very complicated dynamics: water accumulation, water
redistribution, backward waves (generated by the cover plant
and the walls).

In the case of homogeneous soil surface (Case D in
Table~\ref{table_4scenarios}), one deals with a 1D Riemann
Problem which has a very simple rarefaction wave solution,
see
Fig.~\ref{fig_water_dynamics1_P1_P4_RectangularDomain}(d)
(identical solution at P1 and P2, identical solution at P3
and P4) and
Fig.~\ref{fig_water_distribution_snapshots_RectangularDomain}(a)
(the wave solution did not reach the vegetated zone
$\mathfrak{D}_1$).

The presence of the cover plant induces:

1) a complicated topology of the water velocity, see
Fig.~\ref{fig_wave_propag_velo_profile}(first row);

2) water accumulation, see
Fig.~\ref{fig_wave_propag_velo_profile}(first row) and
Fig.~\ref{fig_water_dynamics1_P1_P4_RectangularDomain}(a);

3) water redistribution, see
Fig.~\ref{fig_water_distribution_snapshots_RectangularDomain}(b,c)
and Fig.~\ref{fig_wave_propag_velo_profile}(first row);

4) bakward wave, see
Fig.~\ref{fig_water_distribution_snapshots_RectangularDomain}(d,e,f)
and
Fig.~\ref{fig_water_dynamics1_P1_P4_RectangularDomain}(a,b);

5) water slowing down,
Fig.~\ref{fig_wave_propag_velo_profile}(first row).

Also, when comparing the three simulations on the
non-vegetated flume ($\theta_1=1$), one can observe that,
from a qualitative point of view, significant higher
water-soil frictional forces in $\mathfrak{D}_1$
(i.e. significant higher values of $\tau_1$) tend to give
phase portraits closer to the one given by the simulation on
the partially vegetated flume ($\theta_1=0.99$).

We do not have any experimental data for the water flow with
sediment  inside such  a  flume, but  since  the water  flow
module has been  validated in many ways and  seems to behave
well,  we  decided  to accomplish  a  different  theoretical
experiment which also includes  the erosion process in order
to see if  the extended model (\ref{eq_swe1}-\ref{eq_eros2})
can capture its qualitative behavior.

The soil properties and the three sediment classes
considered for theses new simulations of the {\bf erosion}
are given in Table~\ref{table_soil_and_sedim_flume}.
\begin{table}[h!]
  \caption{Values of the soil and sediment parameters used
    for the simulations on the flume.}
  \centering
  %\begin{ruledtabular}
    \begin{tabular}{ccccccc}
      \hline
      $\gamma_s$ & $F$ & $J$      & $\Omega_{\rm cr}$ & $\alpha$ & $p_\alpha$ & $\nu_{s,\alpha}$ \\
      -          & -   & {[J/kg]} & {[${\rm W/m}^2$]} & -        & -          & {[${\rm m/s}$]} \\
      \hline
          &      &            &       & 1 & 0.65 & 0.0038\\
      2.6 & 0.06 & 5.0 \& 0.1 & 0.007 & 2 & 0.30 & 0.0827\\
          &      &            &       & 3 & 0.05 & 0.2317\\
      \hline
    \end{tabular}
  %\end{ruledtabular}
  \label{table_soil_and_sedim_flume}
\end{table}

\begin{figure}[!htbp]
  \begin{tabular}{ccccc}
    {} & $\rho_1$ & $\rho_2$ & $m_1$ & $m_2$\\
    %&&&&\\
    \begin{turn}{90}\hspace{00mm}{$\theta=0.97, \quad J=1\,{\rm J}\cdot{\rm kg}^{-1}$}\end{turn}
    &\includegraphics[width=2.7cm,height=4.5cm]{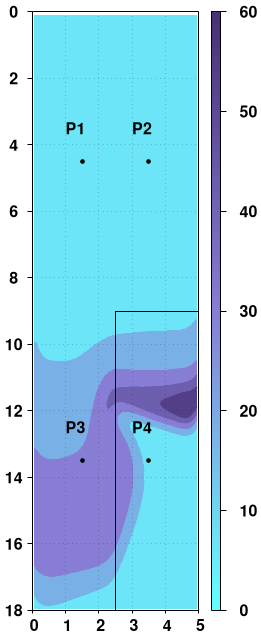}
    &\includegraphics[width=2.7cm,height=4.5cm]{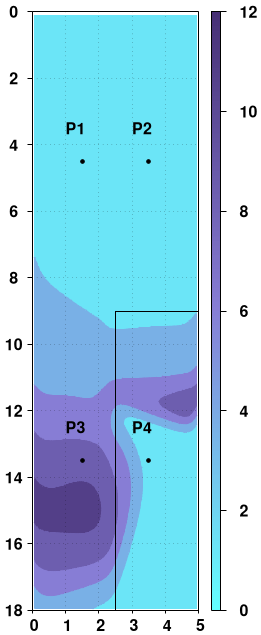}
    &\includegraphics[width=2.7cm,height=4.5cm]{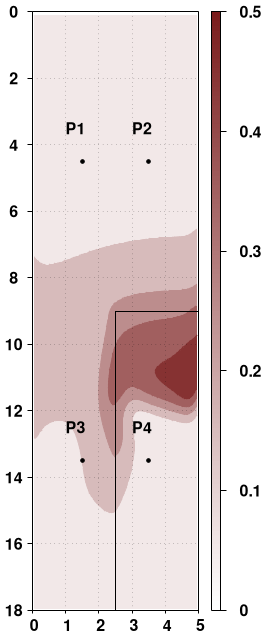}
    &\includegraphics[width=2.7cm,height=4.5cm]{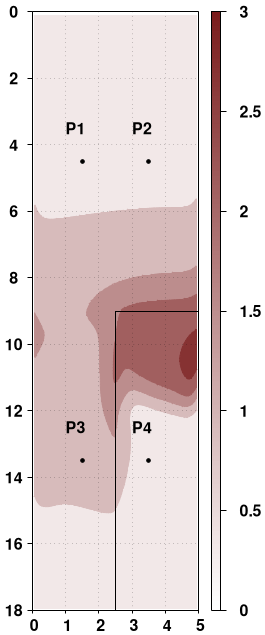}\\
    %&&&&\\
    \begin{turn}{90}\hspace{00mm}{$\theta=0.99, \quad J=1\,{\rm J}\cdot{\rm kg}^{-1}$}\end{turn}
    &\includegraphics[width=2.7cm,height=4.5cm]{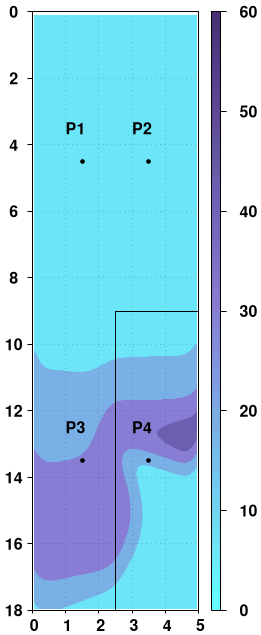}
    &\includegraphics[width=2.7cm,height=4.5cm]{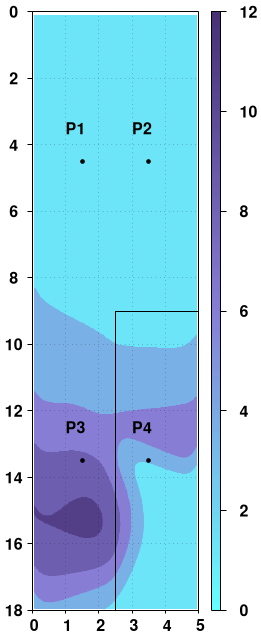}
    &\includegraphics[width=2.7cm,height=4.5cm]{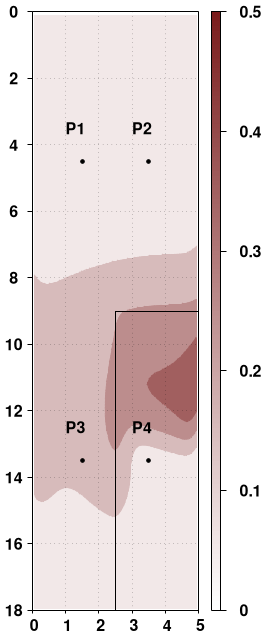}
    &\includegraphics[width=2.7cm,height=4.5cm]{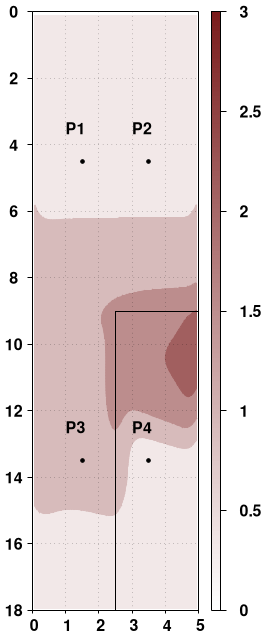}\\
    %&&&&\\
    \begin{turn}{90}\hspace{00mm}{$\theta=0.99, \quad J=50\,{\rm J}\cdot{\rm kg}^{-1}$}\end{turn}
    &\includegraphics[width=2.8cm,height=4.5cm]{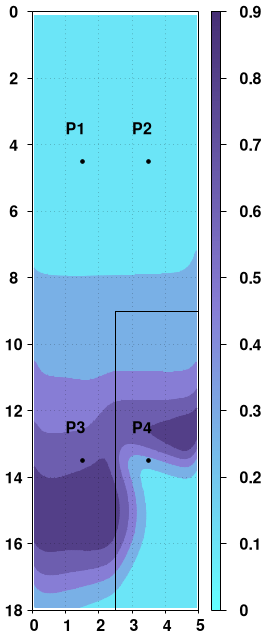}\hspace{-01mm}
    &\includegraphics[width=2.8cm,height=4.5cm]{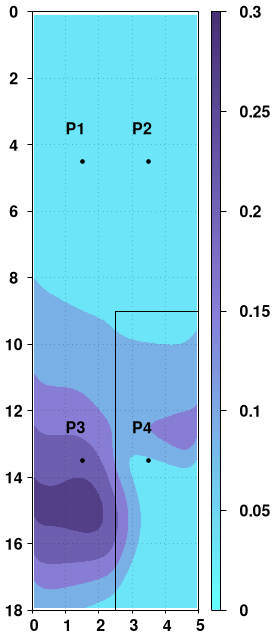}\hspace{-01mm}
    &\includegraphics[width=3.2cm,height=4.5cm]{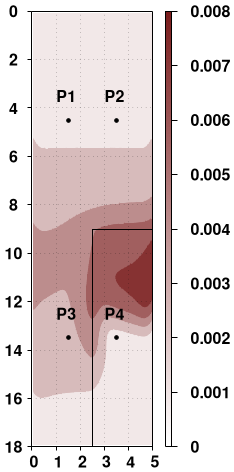}\hspace{-03mm}
    &\includegraphics[width=3.2cm,height=4.5cm]{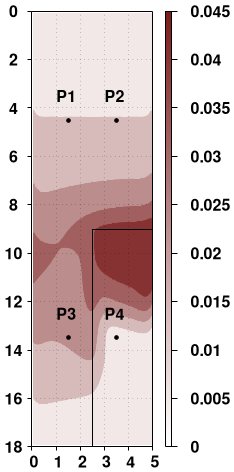}\hspace{-03mm}
  \end{tabular}
  \caption{Mass densities (after $t=6\,{\rm s}$) of two size
    classes of suspended sediment and of deposited sediment
    found on flow simulations on two types of vegetated soil
    when $J$ varies.}
  \label{fig_flume_sediment}
\end{figure}

The initial water level at rest is $h=1.0\, {\rm m}$ for
$y \in [0,2]$ and $h=0.1\, {\rm m}$ for $y \in (2,18]$.  The
initial values of the mass densities of the suspended and
deposited sediment are considered to be 0:
\begin{equation*}
  \rho_{\alpha}(0) = 0, \quad m_{\alpha}(0) = 0, \quad \alpha=\overline{1,3}.
\end{equation*}
We ran ASTERIX for four simulations corresponding to two
different types of vegetated soil
\begin{equation*}
  \theta_1=0.99 \quad {\rm and} \quad \theta_1=0.97
\end{equation*}
and to two different values of the energy $J$ of soil
particles detachment from (\ref{eq_da_ea_era})
\begin{equation*}
  J=50.0 \quad {\rm and} \quad J=1.0.
\end{equation*}
The snapshots of the mass densities $\rho_{\alpha}$ and
$m_{\alpha}$ of the sediment in the size classes $1$ and $2$
after $t=6\,{\rm s}$ from the removal of the door between
the thank and the channel are presented in
Fig.~\ref{fig_flume_sediment}.
The suspended sediment is represented in shades of purple,
while the mass density of the sediment deposited on the soil
is represented in shades of brown.

\begin{figure}[!htbp]
  \begin{tabular}{ccc}
    {}
    &{$\rho_1$}
    &{$m_1$}\\
    %&&&&\\
    \begin{turn}{90}\hspace{2mm}\small{$\theta=0.97, \quad J=1\,{\rm J}\cdot{\rm kg}^{-1}$}\end{turn}
    &\includegraphics[width=0.45\linewidth]{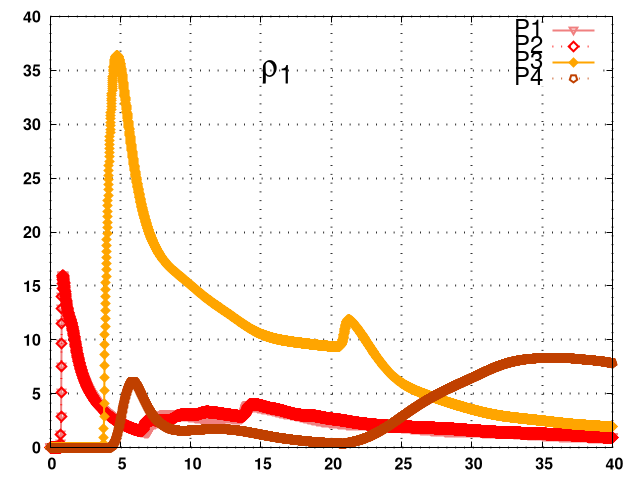}
    &\includegraphics[width=0.45\linewidth]{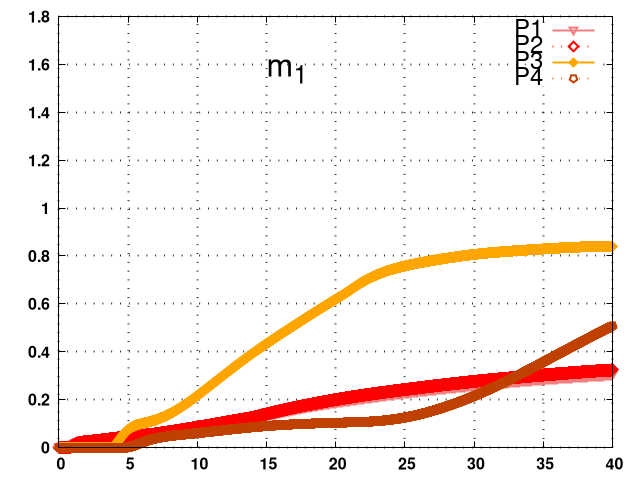}\\
    %&&&&\\
    \begin{turn}{90}\hspace{2mm}\small{$\theta=0.99, \quad J=1\,{\rm J}\cdot{\rm kg}^{-1}$}\end{turn}
    &\includegraphics[width=0.45\linewidth]{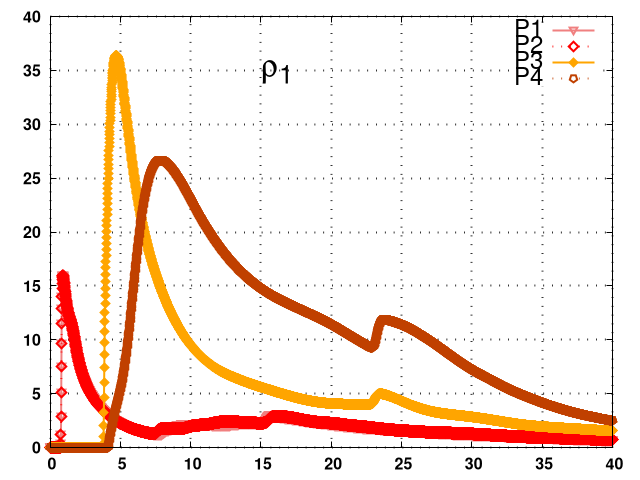}
    &\includegraphics[width=0.45\linewidth]{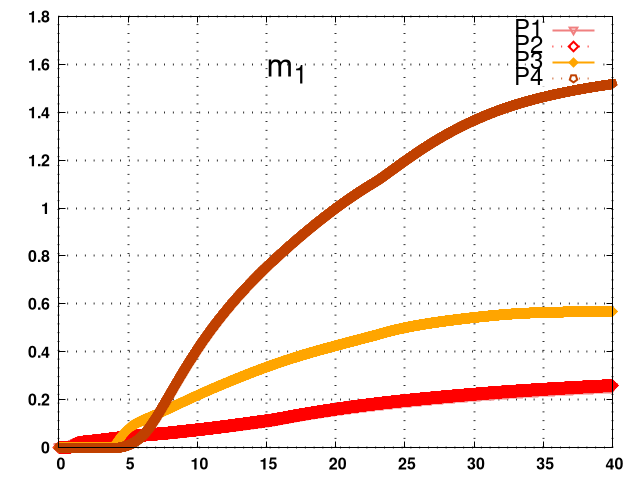}\\
    %&&&&\\
    \begin{turn}{90}\hspace{2mm}\small{$\theta=0.99, \quad J=50\,{\rm J}\cdot{\rm kg}^{-1}$}\end{turn}
    &\includegraphics[width=0.45\linewidth]{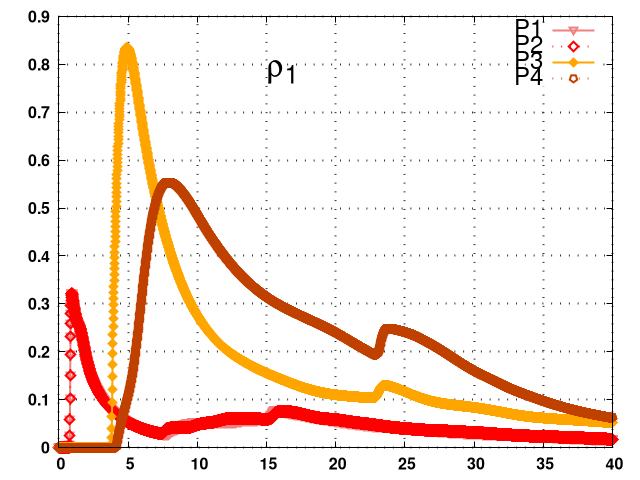}
    &\includegraphics[width=0.45\linewidth]{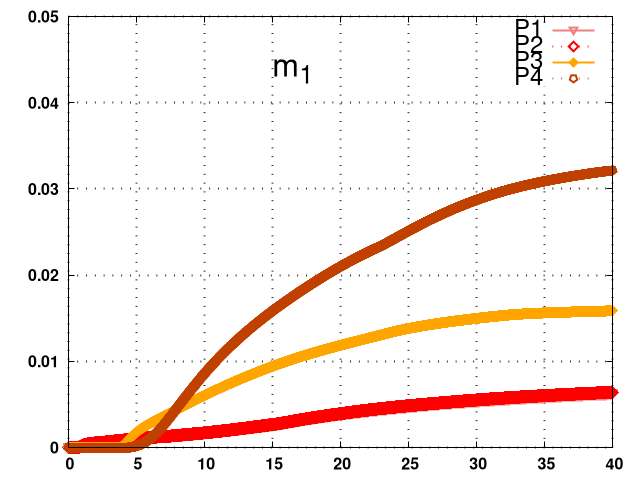}\\
    %&&&&\\
    {}
    &{t [s]}
    &{t [s]}
  \end{tabular}
  \caption{Evolution in time of the mass densities $\rho_1$
    and $m_1$ of the suspended sediment and deposited
    sediment found on flow simulations on two types of
    vegetated soil when $J$ varies.}
  \label{fig_flume_sediment_time_evolution}
\end{figure}

The time evolution of the mass densities $\rho_{\alpha}$ and
$m_{\alpha}$ of the sediment class with the smallest falling
velocity ($\alpha = 1$) at the four gauges for the three
cases considered in Fig.~\ref{fig_flume_sediment} is
pictured in Fig.~\ref{fig_flume_sediment_time_evolution}.

A first observation one can draw when comparing the pictures
from the first to the second row in
Figs.~\ref{fig_flume_sediment} and
\ref{fig_flume_sediment_time_evolution} is that both the
presence of the suspended sediment and its deposition are
delayed when vegetation is denser (smaller values of
$\theta$).  When comparing the columns (but same two rows)
in Fig.~\ref{fig_flume_sediment}, one can easily notice that
the particles with higher falling velocity are transported
more slowly and settle faster than the ones with smaller
falling velocity.  We note the reader that the same color
code legend was used for all four pictures of the suspended
sediment $\rho_{\alpha}$, but it represents a scale of
smaller values on the second column as compared to the ones
from the first column.

When the energy $J$ of soil particles detachment is small
({\bf weak erosion}, see second row in
Fig.~\ref{fig_flume_sediment}), the maximum concentration of
the suspended sediment is reached closer to the tank gate as
it gets heavier (as the falling velocity of the sediment
class increases) and there is only a little amount of
suspended sediment that is deposited faster than
transported.

For higher values of the energy $J$ of soil particles
detachment ({\bf strong erosion}, see third row in
Fig.~\ref{fig_flume_sediment}), the maximum concentration of
suspended sediment, at high values of its falling velocity,
is reached further away from the tank gate.  Also, the
quantity of the suspended sediment in the water is higher
and a part of it is deposited, but a significant part of it
is also transported.
%-----------------------------------------------------------

%-----------------------------------------------------------
\subsection{Simulation on Lipaia's area}
\label{sect_SimulationOnLipaia}
\begin{figure}[h!]
  \centering
  \includegraphics[width=0.44\linewidth]{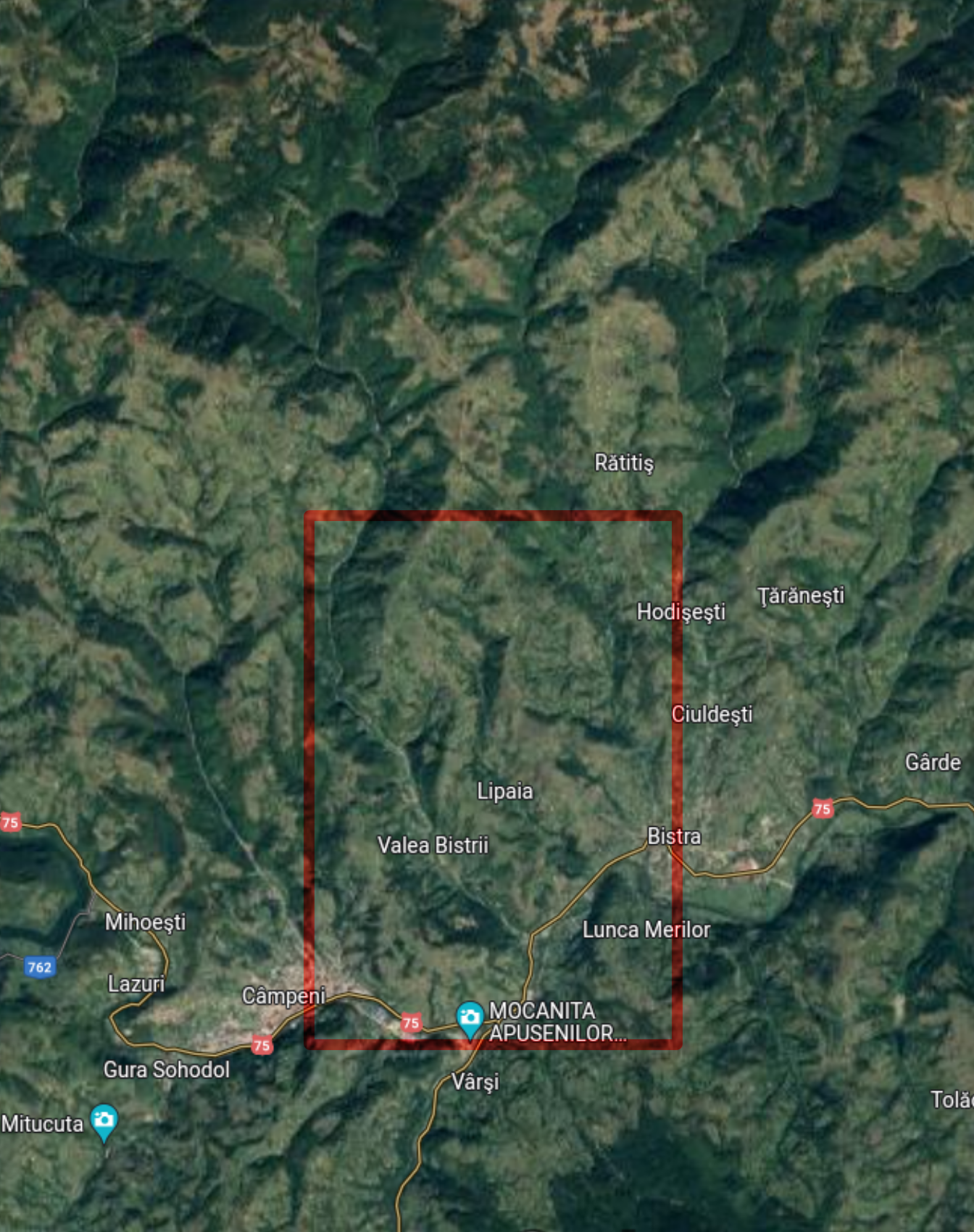}
  \hspace{0.5em}
  \includegraphics[width=0.44\linewidth]{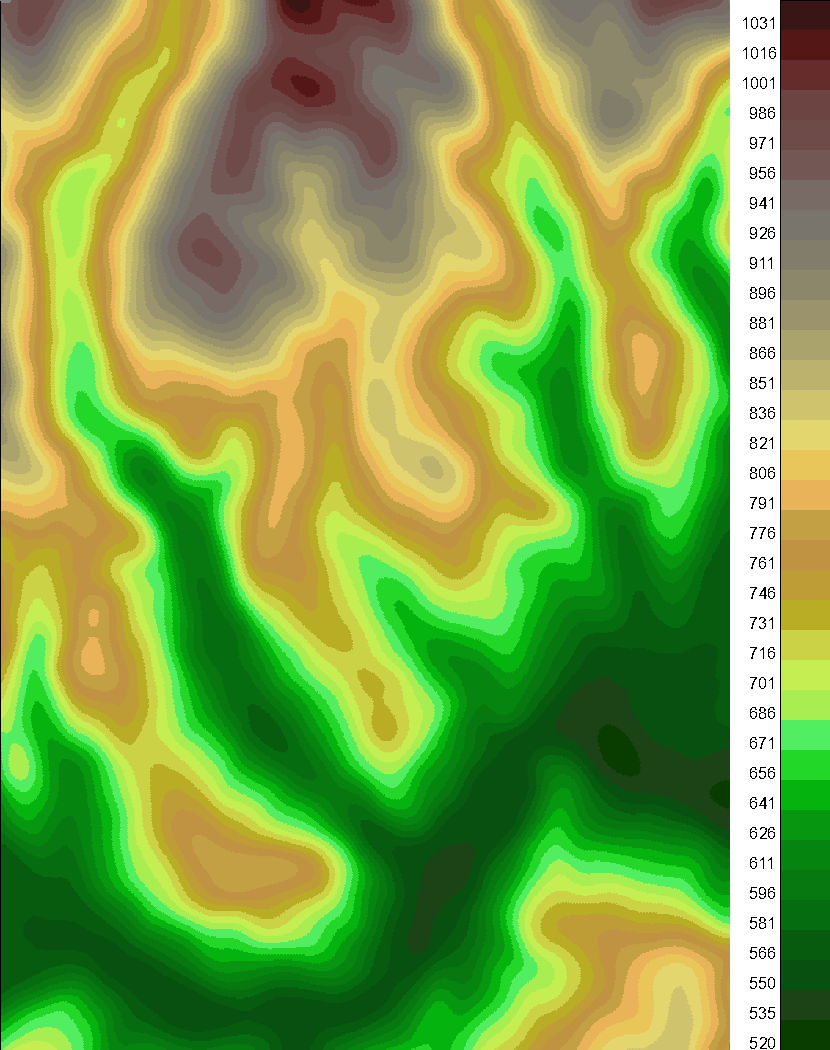}
  \caption{Lipaia's area.  Satellite image
    \cite{google_earth_lipaia} and the terrain reconstructed
    (from GIS data) on a hexagonal network.}
  \label{fig_lipaia_foto}
\end{figure}

Lipaia is a valley in Alba County, Romania located north of
Arie\c{s} River and surrounding the village (with the same
name) whose geographical coordinates are $46^{\circ}$
$23^{\prime}$ $23^{\prime\prime}$ North, $23^{\circ}$
$5^{\prime}$ $19^{\prime\prime}$ East.  To be closer to
reality, we use the Geographic Information System (GIS) data
for an approximately $29\, {\rm km}^2$ soil surface included
in a $4.50 \times 6.46\, {\rm km}^2$ surrounding rectangular
area, see Fig.~\ref{fig_lipaia_foto}.
Since we do not have experimental data for plant cover
density, water and velocity distributions, erosion process,
we decided to accomplish the following theoretical
experiment: starting with a uniform shallow water layer of
$4.5\, {\rm cm}$ in depth on the entire landscape, and using
two different uniform vegetation densities
($\theta_1 = 0.999$ and $\theta_2 = 0.969$), we ran our
software ASTERIX on a network with $264451$ hexagonal cells
of radius $6.495\, {\rm m}$ to simulate the water flow
across the cells for a set time interval.  The hexagonal
network was generated from the Lipaia's GIS raster data
using the porting method introduced in
\cite{sds-ADataPortingTool}.  For the water-plant and
water-soil frictional terms (\ref{eq_tau_p_s}), we
considered
\begin{equation*}
  \alpha_p = 73.39\, {\rm m}^{-1}, \quad \quad \alpha_s = 0.007.
\end{equation*}

\begin{figure}[!htbp]
  \centering
  \includegraphics[width=0.9\textwidth]{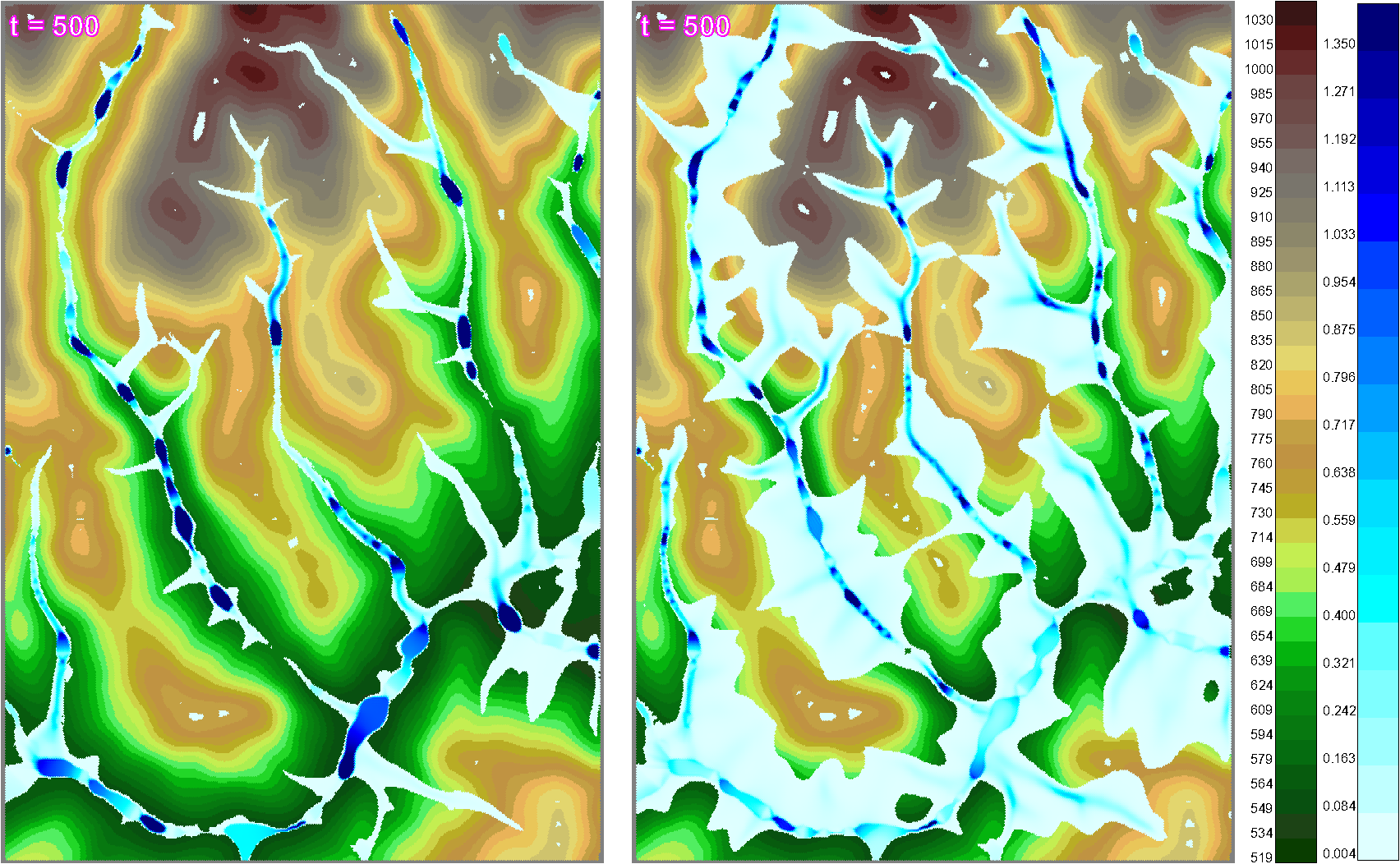}
  \caption{Snapshots of water distribution on Lipaia's Valley
    for two different uniform vegetation densities:
    $\theta=0.999$ and $\theta=0.969$ on the left and right
    picture, respectively.  As expected, our numerical data
    are consistent with terrain observations: the amount of
    water leaving the basin is greater in the case of lower
    vegetation density.}
    \label{fig_2Dcomparison_water}
\end{figure}

\begin{figure}[!htbp]
  \centering
  \includegraphics[width=0.9\textwidth]{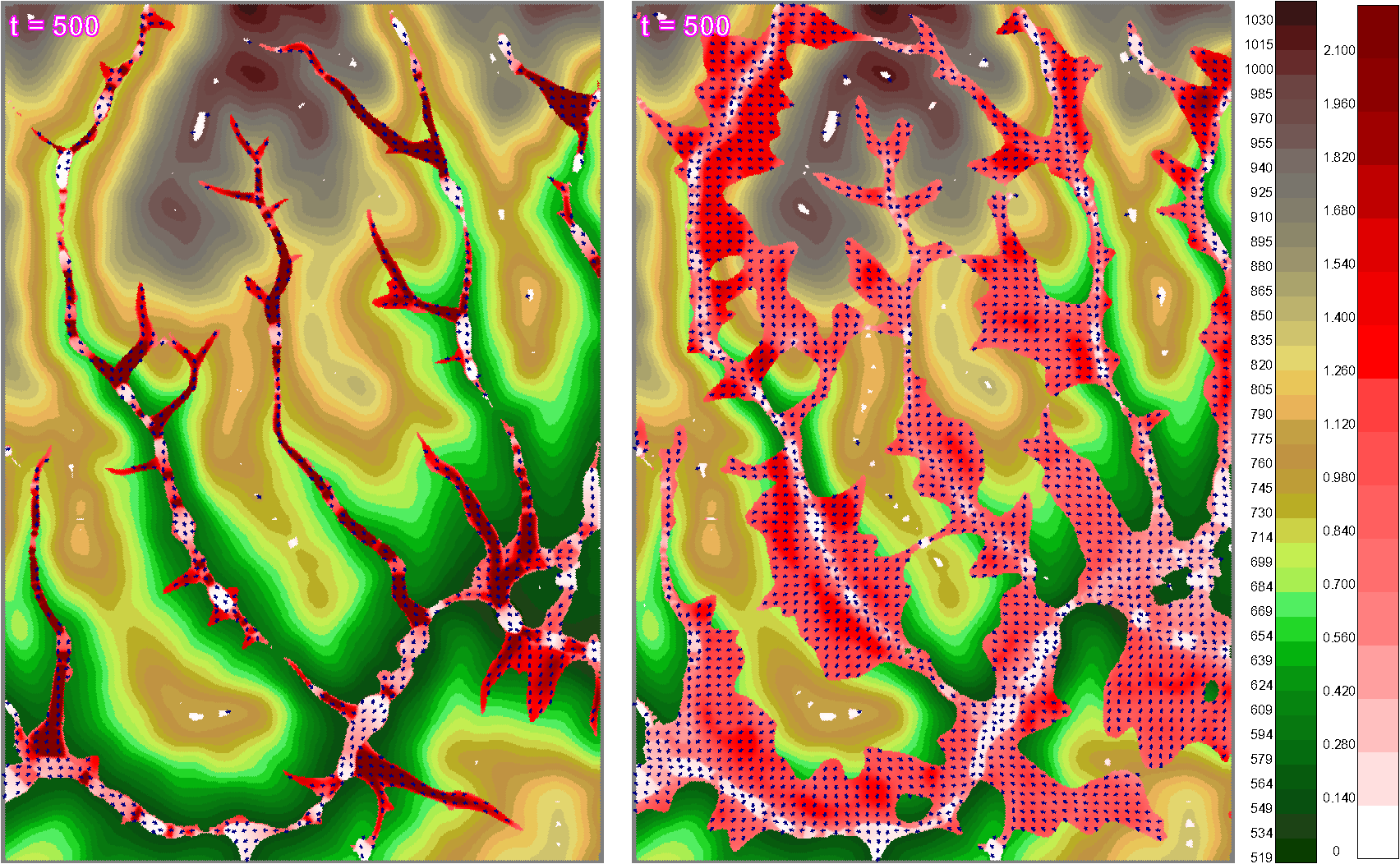}
  \caption{Snapshots of water velocity distribution on
    Lipaia's Valley for two different uniform vegetation
    densities: $\theta=0.999$ and $\theta=0.969$ on the left
    and right picture, respectively.  As expected, the
    velocities are smaller when vegetation is denser.}
    \label{fig_2Dcomparison_velocity}
\end{figure}

\begin{figure}[!htbp]
  \centering
  \includegraphics[width=0.9\textwidth]{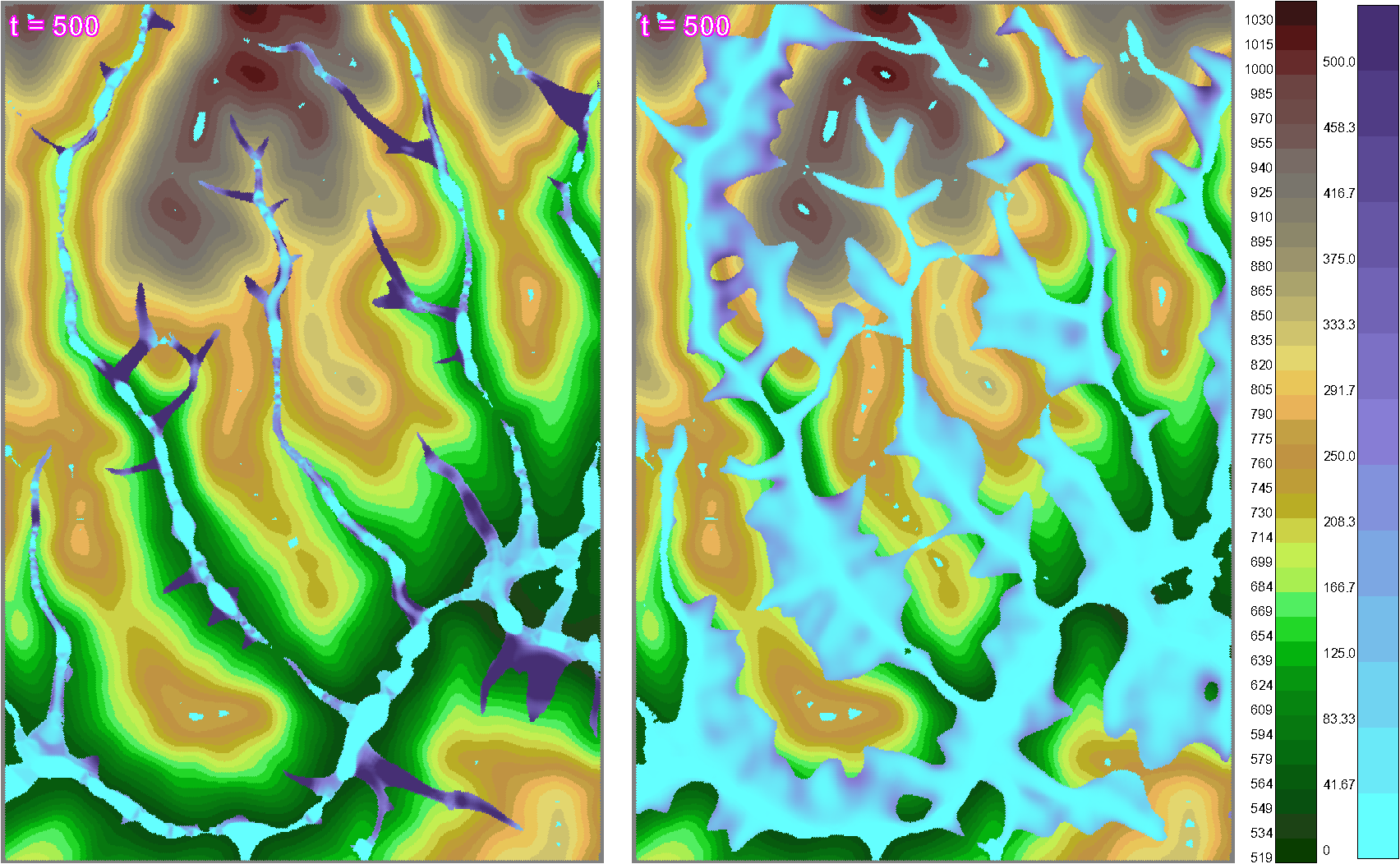}
  \caption{Snapshots of the concentration of the suspended
    sediment on Lipaia's Valley for two different uniform
    vegetation densities: $\theta=0.999$ and $\theta=0.969$
    on the left and right picture, respectively.  As
    expected, water is muddier when the soil is covered with
    less vegetation.}
    \label{fig_2Dcomparison_suspension}
\end{figure}

\begin{figure}[!htbp]
  \centering
  \includegraphics[width=0.9\textwidth]{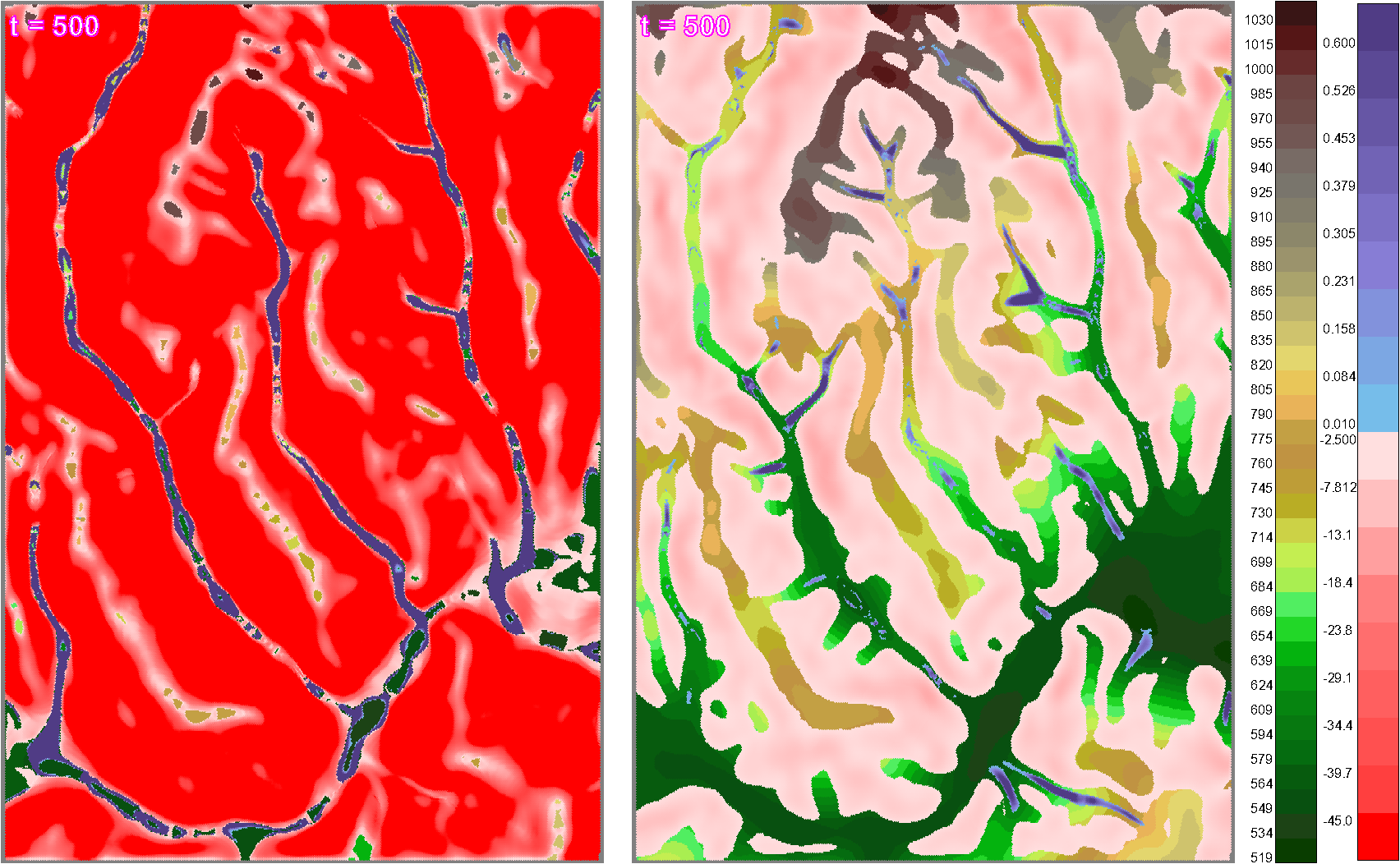}
  \caption{Snapshots of the sediment deposited on the ground
    (shades of purple) and of the eroded sediment (shades of
    red) on Lipaia's Valley for two different uniform
    vegetation densities: $\theta=0.999$ and $\theta=0.969$
    on the left and right picture, respectively.  It can be
    seen, as expected, that the erosion process is more
    intense when the plants are absent.}
    \label{fig_2Dcomparison_erosion}
\end{figure}

The water distribution at $t = 500\, {\rm s}$ is
comparatively presented in Fig.~\ref{fig_2Dcomparison_water}
using a blue color gradient (the darkest blue corresponds to
the cells with $\theta h \geq 30 h_0$, where
$h_0 = 0.045\,{\rm m}$ is the initial water depth).  The
water velocity distribution at the same moment in time is
pictured in Fig.~\ref{fig_2Dcomparison_velocity} using a red
color gradient for the modulus of the velocity and arrows
for the flow direction.  The concentration of the suspended
sediment is given in Fig.~\ref{fig_2Dcomparison_suspension}
in shades of purple and is obviously higher for the case
with less vegetation.  The convention of coloring the relief
where $\theta h \leq 0.1 h_0$ is applied for all three
figures.  The sediment deposited on the ground and the
eroded sediment are drawn in
Fig.~\ref{fig_2Dcomparison_erosion} with shades of purple
and red, respectively.  The data in
Table~\ref{table_BasinLipaia} confirms that more water is
retained in the basin if the vegetation is denser.
\begin{table}
  \caption{Numerical output from the theoretical experiment
    on Lipaia's Valley.  ASTERIX was run for
    $t=500\,{\rm s}$.} 
  \centering
  %\begin{ruledtabular}
    \begin{tabular}{cccc}
      \hline
      $\theta$ & {Water in} & {Water out} & {Water left} \\
      -        & {[m$^3$]}  & {[m$^3$]}   & {[m$^3$]} \\
      \hline
      $0.999$ & $1293921.12$ & $87124.23$ & $1206796.89$ \\
      $0.969$ & $1293921.12$ & $62112.39$ & $1231808.73$ \\ \hline
    \end{tabular}
  %\end{ruledtabular}
  \label{table_BasinLipaia}
\end{table}


\begin{figure}[!htbp]
  \centering
  \begin{tabular}{ cc }
    \begin{turn}{90}\hspace{15mm}{Discharge rate [m$^3$/s]}\end{turn}
       &\includegraphics[width=0.68\textwidth]{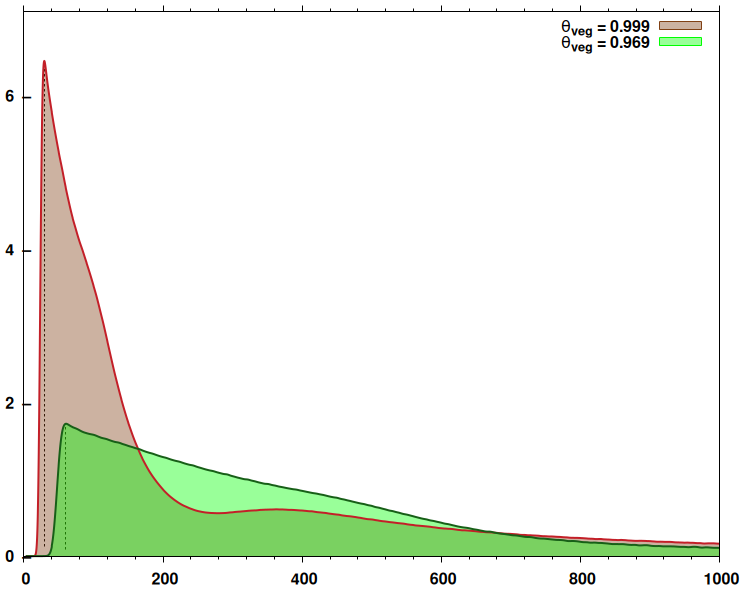}\\
    {} &Time [s]
  \end{tabular}
  \caption{Discharge rates on Lipaia's Valley for two
    different uniform vegetation densities ($\theta=0.969$
    and $\theta=0.999$).}
  \label{fig_discharge_on_lipaia}
\end{figure}

Furthermore, Fig.~\ref{fig_discharge_on_lipaia} emphasizes
that the maximal discharge rate from the entire valley is
higher and occurs earlier when vegetation is sparser.

\begin{figure}[!htbp]
  \centering
  \begin{tabular}{ cc }
    \begin{turn}{90}\hspace{10mm}{$\theta=0.999$}\end{turn} 
    & \includegraphics[width=0.91\textwidth]{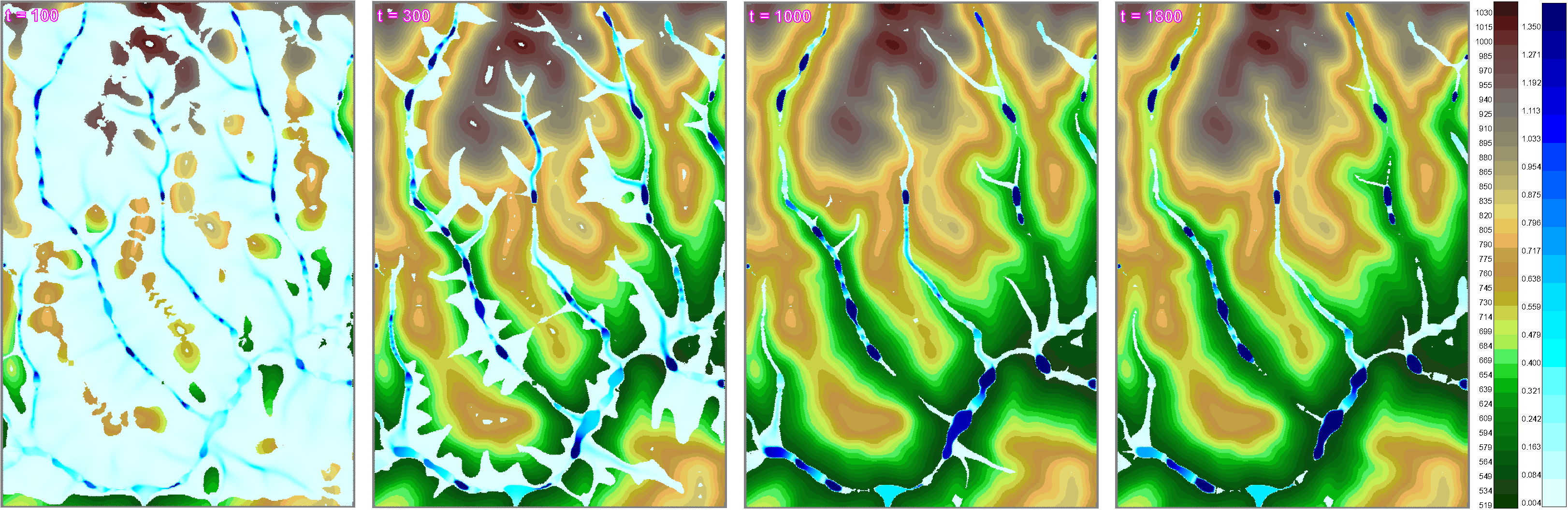}\\
    \begin{turn}{90}\hspace{10mm}{$\theta=0.969$}\end{turn} 
    & \includegraphics[width=0.91\textwidth]{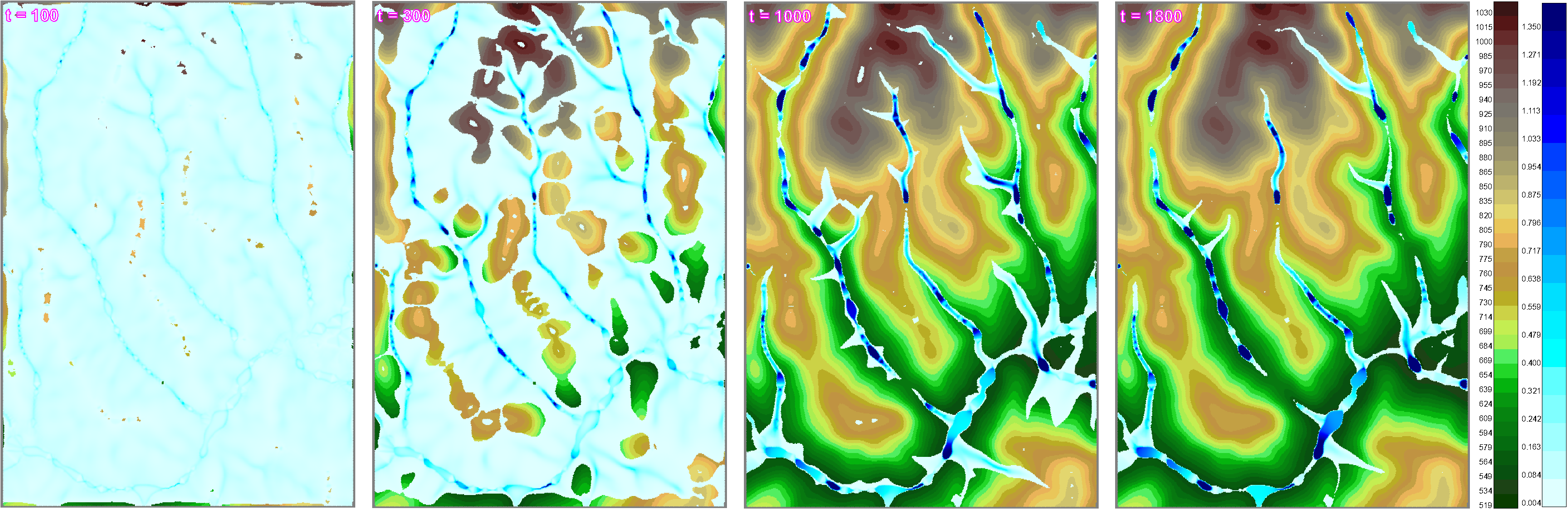}
  \end{tabular}
  \caption{Snapshots from the water dynamics on Lipaia's
    basin at four moments of time: 100, 300, 1000, and 1800
    s.}
  \label{fig_water_dynamics_lipaia}
\end{figure}

We note that the properties of the soil and sediment we used
for the simulation on Lipaia are given in
Table~\ref{table_soil_and_sedim_lipaia} and some snapshots
from the water dynamics at four moments of time are revealed
in Fig.~\ref{fig_water_dynamics_lipaia}.
\begin{table}[h!]
  \caption{Values of the soil and sediment parameters we considered
    for the simulation on Lipaia's area.}
  \centering
  %\begin{ruledtabular}
    \begin{tabular}{ccccccc}
      \hline
      $\gamma_s$ & $F$ & $J$      & $\Omega_{\rm cr}$ & $\alpha$ & $p_\alpha$ & $\nu_{s,\alpha}$ \\
      -          & -   & {[J/kg]} & {[${\rm W/m}^2$]} & -        & -          & {[${\rm m/s}$]} \\
      \hline
          &     &     &      & 1 & 0.1 & 0.05 \\
      2.6 & 0.2 & 9.0 & 0.09 & 2 & 0.3 & 0.08 \\
          &     &     &      & 3 & 0.6 & 0.25 \\ \hline
    \end{tabular}
  %\end{ruledtabular}
  \label{table_soil_and_sedim_lipaia}
\end{table}
%----------------------------------------------------------
%==========================================================
%==========================================================

%==========================================================
%==========================================================
\section{Conclusions and Remarks}
\label{section_conclusions_and_remarks}
The paper focuses on a mathematical model of water and
sediment dynamics on hillslopes in the presence of
vegetation.  The flow and erosion processes are described by
balance equations and some closure empirical relations.  The
numerical scheme to approximate the solution of the model
(\ref{eq_swe1}-\ref{eq_eros2}) is built as simple as
possible in order to avoid high computational complexity,
but to still be suitable to be used from a theoretical or
didactic level up to practical applications on wide flow
surfaces from various environmental study areas.

The paper presents three numerical experiments as prototypes
of three real hydrological processes: the propagation of a
flash flood in a real environment, the dam break, and the
water flow and soil erosion in a catchment basin.  The main
goal was to illustrate the ability of the combined extended
Saint-Venant and Hairsine-Rose mathematical model and of the
accompanying software to cope with various hydrological
processes.

The model ``affirms'' that the evolution of the hydrodynamic
$(h, \boldsymbol{v})$ and soil erosion
$\left.\left(\rho_{\alpha},m_{\alpha}\right)\right|_{{\alpha}=\overline{1,M}}$
variables, collectively denoted by $U$, is governed by
certain general principles of physics and by a set of
environmental ($z$, $\theta$) and structural parameters of
the soil and cover plant ($\alpha_s$, $\alpha_p$, $F$, $J$,
$\Omega_{cr}$, etc.), colectivelly denoted by $\Lambda$.
The variabilty of the dynamics of $U$ is due to the
variability of $\Lambda$.  In this paper, we have shown how
different values of $\Lambda$ can be chosen to model a given
hydrological process occurring in a given natural medium and
how the variation in $\Lambda$ is reflected in the variation
of $U$.

Before any conclusion, we must point out that the problem of
soil erosion on vegetated surfaces is far from being closed
by the mathematical model considered in this paper.  Plant
roots strongly influence the physico-chemical properties of
the soil, while plant stems interact with water flow so that
the parameters $J$ related to the energy of soil particle
detachment, $F$ related to the power stream, and
$m^{\star}_t$ related to the protection of the soil surface
from erosion needs more attention.

Based on the previously presented numerical results, let us
highlight several observations that are particularly
relevant to the practical application of the proposed model.

(a) The results of all three experiments demonstrate that
the presence of vegetation reduces the flow velocity,
thereby decreasing the rate of erosion, see
Fig.~\ref{fig_2Dcomparison_velocity} and
Fig.~\ref{fig_2Dcomparison_erosion}.

(b) The presence of plant or solid obstacles generates a
backward wave, see
Fig.~\ref{fig_water_dynamics1_P1_P4_RectangularDomain} and
Fig.~\ref{fig_water_distribution_snapshots_RectangularDomain}.

(c) The presence of an ``island of vegetation'' generates a
process of water accumulation in front of the island, see
Fig.~\ref{fig_water_dynamics1_P1_P4_RectangularDomain}.

(d) Catchment areas covered with vegetation of higher density
are less exposed to the flood and erosion hazards than those
covered with lower density, see
Figs.~\ref{fig_2Dcomparison_suspension},
\ref{fig_2Dcomparison_erosion}, and
\ref{fig_discharge_on_lipaia}.

From a mathematical point of view, several aspects of the
proposed model deserve particular attention.  The governing
equations (\ref{eq_swe1}-\ref{eq_eros2}) constitute a system
of nonlinear hyperbolic PDEs, for which analytical solutions
are generally difficult to obtain.  Nevertheless, we
successfully adapted the analytical solution presented in
\cite{sander} to the proposed model incorporating
vegetation, thereby providing a benchmark for validating the
developed numerical method.  Furthermore, we proved that the
equilibrium point of the numerical scheme converges to the
equilibrium point of the continuous model and established
the asymptotic stability of the equilibrium point of the
dynamical system associated with the discrete model.

For the non-stationary sediment evolution, some analytical
solutions were derived only in the case of a single sediment
size class.

These analytical solutions (stationary and non-stationary)
provide a basis for validating the proposed numerical scheme
for the sediment transport equations.

As a direction for future research, we intend to develop a
fully coupled Saint-Venant-Hairsine-Rose model.  Although
such a model is considerably more complex, it may be of
interest at least in situations involving high suspended
sediment concentrations or significant erosion-induced
changes in the soil geometry.

As a final remark, we stress that the use of models based on
physical processes in hydrology should be sustained and at
least two arguments can be invoked in this sense:

$\circ$ the increased power of mathematics to solve
complicated equations and 

$\circ$ it is not possible to do flash flood or dam break
experiments at a real scale.

We note the reader that the results of the numerical
simulations considered here are obtained with ASTERIX - an
open source software we have built using the ``Data
Porting'' process described in \cite{sds-ADataPortingTool}
- and the water flow model developed and studied in
\cite{sds_apnum}.
%==========================================================
%==========================================================

%==========================================================
%==========================================================
\section*{Author Declarations}

%----------------------------------------------------------
\subsection*{Conflict of Interest}
The authors have no conflicts to disclose.
%----------------------------------------------------------

%----------------------------------------------------------
\subsection*{CRediT authorship contribution statement}
All three authors contributed equally to this work:
Conceptualization, Methodology, Validation, Formal
analysis, Data Curation, Investigation, Writing - Original
Draft.
%----------------------------------------------------------

%----------------------------------------------------------
\subsection*{Data Availability Statement}
The data that supports the findings of this study are
available within the article.
%----------------------------------------------------------
%==========================================================
%==========================================================

%==========================================================
%==========================================================
%\nocite{*}
\bibliographystyle{elsarticle-num}
\bibliography{biblio_apnum}
%==========================================================
%==========================================================

\end{document}